\documentclass[12pt,preprint]{aastex}

\shortauthors{Lee et al.}
\shorttitle{Physical Parameters of LBGs from Broadband Photometry}

\begin{document}

\title{Biases and Uncertainties in Physical Parameter Estimates of
  Lyman Break Galaxies from Broad-band Photometry}

\author{Seong-Kook Lee}
\affil{Department of Physics and Astronomy, Johns Hopkins University, 
3400 North Charles Street, Baltimore, MD 21218-2686}
\email{joshua@pha.jhu.edu}

\author{Rafal Idzi}
\affil{Department of Physics and Astronomy, Johns Hopkins University, 
3400 North Charles Street, Baltimore, MD 21218-2686}

\author{Henry C. Ferguson}
\affil{Space Telescope Science Institute, 3700 San Martin Drive,
Baltimore, MD 21218}

\author{Rachel S. Somerville}
\affil{Space Telescope Science Institute, 3700 San Martin Drive,
Baltimore, MD 21218}

\author{Tommy Wiklind}
\affil{Space Telescope Science Institute, 3700 San Martin Drive,
Baltimore, MD 21218}

\and

\author{Mauro Giavalisco}
\affil{Astronomy Department, University of Massachusetts, 
Amherst, MA 01003}

\begin{abstract} 

We investigate the biases and uncertainties in estimates of physical
parameters of high-redshift Lyman break galaxies (LBGs), such as
stellar mass, mean stellar population age, and star formation rate (SFR), 
obtained from broad-band photometry. These biases arise from the simplifying
assumptions often used in fitting the spectral energy distributions
(SEDs).  By combining $\rm{\Lambda}$CDM hierarchical structure
formation theory, semi-analytic treatments of baryonic physics, and
stellar population synthesis models, we construct model galaxy
catalogs from which we select LBGs at redshifts $z$ $\sim$ 3.4,
4.0, and 5.0.  The broad-band photometric SEDs of these model LBGs are
then analysed by fitting galaxy template SEDs derived from stellar
population synthesis models with smoothly declining SFRs. 
We compare the statistical properties of LBGs' physical
parameters -- such as stellar mass, SFR, and stellar population age -- 
as derived from the best-fit galaxy templates with the intrinsic
values from the semi-analytic model.  We find some trends in these
distributions: first, when the redshift is known, 
SED-fitting methods reproduce the input distributions of 
LBGs' stellar masses relatively well, with a minor tendency to underestimate 
the masses overall, but with substantial scatter.  Second, there are large
systematic biases in the distributions of best-fit SFRs and mean ages, 
in the sense that single-component SED-fitting methods underestimate
SFRs and overestimate ages. We attribute these trends to the
different star formation histories predicted by the semi-analytic
models and assumed in the galaxy templates used in SED-fitting procedure, 
and to the fact that light from the current generation of
star-formation can hide older generations of stars. These biases,
which arise from the SED-fitting procedure, can significantly affect
inferences about galaxy evolution from broadband photometry.
 
\end{abstract}


\section{Introduction}

Over the last decade, astronomy has experienced a boom of panchromatic
surveys of high redshift galaxies -- such as the Great Observatories
Origins Deep Survey \rm{\citep[GOODS,][]{gia04a}} and the Cosmic
Evolution Survey \rm{\citep[COSMOS,][]{sco07}} -- thanks to the
developments of space telescopes, like \it{Hubble Space Telescope}
\rm{(HST)} and \it{Spitzer Space Telescope} \rm{($Spitzer$)}, and 
also of ground-based facilities, including Keck observatory, Very Large
Telescope (VLT), and Subaru telescope.  Spectral energy distributions
(SEDs) constructed from the photometric data of wide wavelength
coverage have been used to constrain galaxy parameters, such as
stellar masses, SFRs, and stellar population ages, of high redshift
galaxies via comparison with simple stellar population synthesis
models \citep[e.g.][]{pap01,sha01,sha05,shi07,sta07,ver07}.  Accompanied 
byselection techniques to separate high-redshift galaxy samples
efficiently via their broad-band colors and magnitudes, these surveys
have provided increasing information about the nature of high-redshift
galaxies.
 
However, despite recent advances in our knowledge of the properties 
of high redshift galaxies, more accurate estimation of 
physical parameters is necessary for addressing several important
issues in galaxy evolution and cosmology.  For example, it is as yet
uncertain whether the Lyman break galaxies
\rm{\citep[LBGs,][]{ste96,gia02}} -- high-redshift, star-forming
galaxies selected according to their rest-frame ultraviolet (UV)
colors -- will evolve into large ellipticals \citep{ade98,ste98}, 
or into smaller spheroids, such as small ellipticals/spiral 
bulges or subgalactic structures, which will later be merged to 
form larger galaxies at $z \sim 0$ \citep{low97,saw98,som01}.  
Accurate knowledge of physical parameters, such as stellar masses
and SFRs, serves as one of the important elements in discriminating
among the possible evolutionary descendants of LBGs.  Also, accurately
constraining LBG's stellar-population ages and stellar masses is
necessary for determining how much the LBG populations contributed to
cosmic reionization.  Better estimation of these parameters is also
crucial in comparing different galaxy populations at
different redshifts.
       
The stellar populations of LBGs as well as other populations of
high-redshift galaxies have been studied by several authors using
SED-fitting methods to compare the photometric SEDs of observed galaxies 
with various galaxy spectra from stellar population synthesis models.
\citet{pap01} investigated 33 spectroscopically-confirmed LBGs with a
redshift range, $2.0 \leq z \leq 3.5$. They found that the mean lower
limit of stellar masses of these LBGs is $\sim 6 \times 10^{9}$
$M_{\sun}$ with upper limits of $\sim$ 3-8 times larger and that the
mean age is $\sim$ 120 Myr (assuming solar metallicity and
\citet{sal55} IMF) with broad range between 30 Myr and 1 Gyr.  More
importantly, they concluded that the most robust parameter constrained
through SED-fitting is stellar mass, while stellar population age and
especially star formation rate (SFR) are only poorly constrained.
Also, they speculated that SED-fitting methods (single-component
fitting) can only give lower limits to galaxies' stellar masses
because the results from the SED-fitting methods are largely driven by
the light from the most massive, most recently formed stars.
\citet{sha01} studied the physical parameters of 74, $z\sim$
3 LBGs with spectroscopic redshifts.  
They confirmed that among various physical parameters, stellar mass 
is the most tightly constrained, and that constraints on other parameters 
such as dust extinction and age are weak.  
With the assumption of solar metallicity
and a Salpeter IMF, the median age (defined as time since the onset of
current star formation) was found to be $t_{sf} \sim$ 320 Myr with a
large spread from several Myr to more than 1 Gyr. 
The median stellar mass was 1.2$\times 10^{10} ~h^{-2} ~M_{\sun}$, 
which is higher than the \citet{pap01} value.  
However, the \citet{sha01} LBGs were generally brighter than LBGs in 
the \citet{pap01} sample. The two studies found similar stellar masses 
for LBGs with similar rest-frame UV luminosities.  
\citet{sha01} also found that about 20 $\%$ of the
LBGs have best-fit ages older than 1 Gyr, stellar masses larger than
$10^{10} ~M_{\sun}$, and SFR $\sim$ 30 $M_{\sun}$ $yr^{-1}$, and they
interpreted these as galaxies which have formed their stars over a
relatively long period in a quiescent manner.  
\citet{sha01} also noted that more luminous galaxies are dustier than 
less luminous ones and that younger galaxies are dustier and have higher SFR.

More recently, \citet{ver07} studied 21 $z\sim$ 5 LBGs, six of which
have confirmed spectroscopic redshifts and the remaining 15 of which
have photometric redshifts.  These LBGs were found to be moderately
massive with median stellar mass $\sim$ $2 \times 10^{9}$ $M_{\sun}$,
and to have high SFRs with a median of $\sim$ 40 $M_{\sun}
\rm{yr^{-1}}$.  They also found that the stellar mass estimates are
the most robust of all derived properties.  Best-fit ages have a large
spread with a median value $\sim$ 25 Myr, assuming a metallicity of
one-fifth solar (0.2 $Z_{\sun}$) and a Salpeter IMF.  
\citet{ver07} also compared their $z \sim$ 5 LBGs with \citet{sha01}'s 
$z \sim$ 3 LBGs, assuming the same IMF and metallicity (i.e. solar 
metallicity and Salpeter IMF), and concluded that these two
samples of LBGs with similar rest-frame UV luminosity are clearly
different. Specifically, $z\sim$ 5 LBGs are much younger ($\lesssim$
100 Myr) and have lower stellar masses ($\sim ~10^{9} ~M_{\sun}$) than
$z\sim$ 3 LBGs.  The fraction of young (age $<$ 100 Myr) galaxies is
$\sim$ $70 \%$ for $z \sim$ 5 LBGs while it is $\sim$ $30 \%$ for $z
\sim$ 3 LBGs.  They also concluded that these $z\sim$ 5 LBGs are the
likely progenitors of the spheroidal components of present-day massive
galaxies, based on their high stellar mass surface densities.
\citet{sta07} analysed 72 $z\sim$ 5 galaxies (not just LBGs) with
photometric redshifts. They performed an SED-fitting analysis on a
spectroscopically confirmed subset of 14 $z\sim$ 5 galaxies to derive
best-fit stellar masses ranging between $3 \times 10^{8}$ and $2
\times 10^{11}$ $M_{\sun}$, and ages from 1 Myr to 1.1 Gyr. Three out
of these 14 galaxies have stellar masses in excess of $10^{11}
~M_{\sun}$.  Using stellar mass estimates from both spectroscopically
confirmed and candidate galaxies, they calculated a stellar mass
density at $z\sim$ 5 of $6 ~\times ~10^{6} ~M_{\sun} ~\rm{Mpc^{-3}}$,
which is much larger than the integration of the star formation rate
density (SFRD) from $z\sim$ 10 to 5. They attributed this discrepancy
either to significant dust extinction or to undetected low-luminosity
star-forming galaxies at z $\gtrsim$ 5.
\citet{shi07} studied the properties of 1088 massive LBGs whose best-fit 
stellar masses are larger than $10^{11} ~M_{\sun}$ at $z\sim$ 3. 
They derived stellar masses of these massive LBGs through SED-fitting with 
the assumption of a Salpeter IMF, and noted that LBGs which are detected 
in mid-infrared wavelength are the ones with large stellar mass and severe 
dust extinction among LBG population.

The SED-fitting method is also used to constrain physical
parameters for high-z Lyman-$\alpha$ emitting galaxies
\citep[LAEs,][]{fink07} and for submillimetre galaxies \citep{dye08}.
\citet{fink07} analysed $z \sim 4.5$ LAEs through SED fitting and
found their ages range from 1 Myr to 200 Myr (assuming a constant star
formation history), and stellar masses from $2 \times 10^{7} - 2
\times 10^{9}$ $M_{\sun}$.  \citet{dye08} used SED fitting to study
physical parameters of SCUBA (Submillimetre Common-User
Bolometer Array) sources, finding an average stellar mass of $\sim
10^{11.8 \pm 0.1} M_{\sun}$.

However, the estimated parameter values and implications derived from
these values are prone to several sources of error including errors
inherent in SED-fitting methods.  It is unclear whether or not
SED-fitting methods deliver biased estimates of parameters such as
stellar mass, age, and SFR.  It is also unclear how much worse the
parameter estimates become if spectroscopic redshifts are unavailable
and the photometric data must be used to constrain not only the star
formation history, but also the redshift.  The critical question is
how far we can trust the physical parameters derived using mainly
photometric data under the assumption of simple star-formation
histories (SFHs), when real galaxies are expected to have more complex
formation/evolution histories with several possible episodes of
star-formation.

In this paper, we address this issue by comparing the statistical
distributions of intrinsic physical parameters of model Lyman break
galaxies (LBGs) from semi-analytic models (SAMs) of galaxy formation,
with the estimates of those same parameters derived from stellar
population synthesis models through the commonly used SED-fitting
method.

There have been several studies which tried to constrain the stellar
populations of LBGs or of high-redshift galaxies by comparing them with
predictions from theoretical models, such as semi-analytic models
\citep[e.g.][]{som01,idz04} or cosmological hydrodynamic simulations
\citep[e.g.][]{nag05,nig06,fin07}.  \citet{som01} compared $z \sim 3$
and $z \sim 4$ observed LBGs with semi-analytic models to investigate
possible scenarios for LBGs, and \citet{idz04} put constraints on the
properties of $z \sim 4$ LBGs through comparison of colors of observed
LBGs and SAMs model LBGs.  \citet{nag05} and \citet{nig06} used
cosmological simulations to place constraints on properties of
UV-selected, $z \sim 2$ galaxies and $z \sim 4-6$ LBGs.  \citet{fin07}
constructed a set of galaxy templates from hydrodynamical simulations
and used them to place constraints on properties of 6 $z \leq 5.5$
observed galaxies. Our approach is distinct from these studies.
Here, instead of comparing model galaxies with observed ones, we
perform SED fitting on model galaxies, derive best-fit parameters, and
compare them with the intrinsic values given in the model, trying to
understand quantitatively as well as qualitatively the biases and
uncertainties of SED-fitting methods in constraining the physical
parameters of LBGs.

To accomplish this, we construct galaxy formation histories by
combining $\rm{\Lambda}$CDM hierarchical structure formation theory
with semi-analytic treatments of gas cooling, star formation,
supernova feedback, and galaxy mergers. Then, using the stellar
population synthesis models of Bruzual $\&$ Charlot 
\citep[][hereafter BC03]{bru03}, and a simple model for dust extinction, 
we derive the photometric properties of these model galaxies. 
After selecting LBGs through appropriate color criteria, we compare the 
photometric spectral energy distributions (SEDs) in observed-frame optical
(HST/ACS) through mid-infrared ($Spitzer$/IRAC) passbands of these
model LBGs with those of SED templates from stellar population
synthesis model of BC03 using a $\chi^{2}$-minimization method. 
In this SED-fitting procedure, a large multi-parameter space is explored 
to minimize any prior. Each parameter -- such as star-formation time scale, 
age, dust-extinction, and metallicity -- in this parameter space spans a 
broad as well as physically realistic range. 
The distributions of physical parameters of these model LBGs derived 
from this SED-fitting method are compared with the input distributions 
of these parameters from the semi-analytic model.

Various parameters in the semi-analytic model are calibrated to
reproduce the colors of observed LBGs in the southern field of the
GOODS (GOODS-S).  The GOODS \citep{gia04a} is a deep, multiwavelength
survey which covers a combined area of $\sim 320$ $\rm{arcmin^{2}}$ in
two fields -- GOODS-N centered on the Hubble Deep Field-North (HDF-N),
and GOODS-S centered on the Chandra Deep Field-South (CDF-S).  The
extensive wavelength coverage of the photometric data in the GOODS
fields including HST/ACS (${\it Advanced~ Camera~ for~ Surveys}$) and
$Spitzer$/IRAC (${\it Infrared~ Array~ Camera}$) is beneficial for the
derivation of various physical parameters of galaxies including
stellar mass, SFR, and age.  Deep observations (reaching $z_{850} \sim
26.7$) in the GOODS fields can probe the high redshift universe in a
comprehensive and statistically meaningful manner.  This has enabled
many authors to investigate high-redshift LBG populations in the GOODS
fields \citep[e.g.][]{gia04b,idz04,pap04,lee06,rav06,yan06,sta07,ver07}. 
In this work, we use bandpasses for the HST/ACS filters F435W 
($B_{435}$), F606W ($V_{606}$), F775W ($i_{775}$), F850LP ($z_{850}$),
VLT/ISAAC (Infrared Spectrometer And Array Camera) $J$, $H$,
$Ks$ bands, and $Spitzer$/IRAC 3.6, 4.5, 5.8, 8.0 $\mu m$ channels for
SED-fitting.  The CTIO (Cerro Tololo Inter-American Observatory)
MOSAIC U band is used for selecting U-dropouts.

A description of the semi-analytic model of galaxy formation used in
this work is given in \S~2, with the Lyman break galaxy sample
selection and the SED-fitting procedures explained in \S~3.  The
statistical properties of the derived physical parameters for model
LBGs are shown in \S~4. We analyse, in detail, the effects of various
factors on the biases in SED-fitting in \S~5, and the bias in estimating 
SFR from rest-frame UV magnitude in \S~6. We discuss the effects
of these biases on the galaxy evolution studies in \S~7 and a summary and
our conclusions are given in \S~8.  
Throughout the paper, we adopt a flat $\rm{\Lambda}$CDM cosmology, 
with ($\Omega_{m}, \Omega_{\Lambda}$) = (0.3,0.7), and $H_{0}$ = 
100$h$ $km$ $s^{-1}$ $Mpc^{-1}$, where $h$=0.7, and all magnitudes 
are expressed in the AB magnitude system \citep{oke74}.

\section{Semi-Analytic Models of Galaxy Formation} \label{model} 

Semi-analytic models (SAMs) of galaxy formation are embedded within
the framework of a $\Lambda$CDM initial power spectrum and the theory
of the growth and collapse of fluctuations through gravitational
instability.  The models include simplified physical treatments of gas
cooling, star formation, supernova feedback, dust extinction, and
galaxy merging.  In this work, we use the SAMs model run which was
constructed and used in \citet{idz07}.  The \citet{idz07} work was
based on the \citet{som99b} model, which is also used by \citet{som01}
and \citet{idz04}.  Specifically, among various model runs in
\citet{idz07}, we use here the model run which showed the closest
match to the rest-frame UV-continuum and UV-optical
colors\footnote{$V_{606}$ - $i_{775}$ and $V_{606}$ - IRAC 3.6 $\mu$m 
  colors are used for U-dropouts, and $i_{775}$ - $z_{850}$ and 
  $z_{850}$ - IRAC 4.5 $\mu$m colors are used for B-dropouts as 
  diagnostics to determine best-fit model run.} 
of observed U- $\&$ B-dropouts in the GOODS-S field. 
The details of the model can be found in the above references, and here 
we briefly highlight the most important aspects of the model (including 
some changes from the models described in the original papers).

\subsection{Description of Semi-Analytic Models}

The semi-analytic model used in this work implements the method of
\citet{som99a} to build merging histories of dark matter halos.  After
dark matter merging histories are constructed, the semi-analytic
treatments for various physical processes, such as gas cooling,
mergers, star formation (both in a merger-induced burst mode and in a
quiescent mode), supernovae feedback, chemical evolution, and dust
extinction, are applied to realize predictions of the formation
history of a statistical ensemble of galaxies.

A newly formed dark matter halo (residing at top of the tree) contains
pristine shock-heated hot gas at the virial temperature. When a halo
collapses or undergoes a merger with a larger halo, the associated gas
is assumed to be shock-heated to the virial temperature of the
halo. This gas then radiates energy and consequently cools.  For small
halos at high redshift, cooling is limited by the accretion rate,
since the amount of gas that can cool at any given time cannot exceed
the amount of hot gas contained within the halo's virial radius.

Once halos merge, the galaxies within them remain distinct for some
time. The central galaxy of the largest progenitor halo is set as the
central galaxy of the merged dark matter halo. All the other galaxies
become satellites, which then fall in towards the central galaxy due 
to dynamical friction. Unlike in \citet{som01}, collisions between 
satellite galaxies are neglected here.

Both quiescent and merger-driven modes of star formation are included
in the model.  
We assume that every ``major'' galaxy-galaxy merger above a certain
mass-ratio threshold (1:4) triggers a starburst, which converts 100 \%
of the available gas into stars in a tenth of a halo dynamical
time. Quiescent star formation is modeled with a Kennicutt-like law,
such that the star formation rate is proportional to the mass of cold
gas in the disk divided by the dynamical time. We also scale the SFR
by a power-law function of the galaxy circular velocity, such that
star formation is less efficient in low-mass galaxies. This mimics the
effect of a SF threshold, as implemented in more recent semi-analytic
models \citep[e.g.][]{som08}.  The total star formation rate of a galaxy
is the sum of the burst and quiescent modes.

Supernova feedback is also modeled via a simple recipe, where the rate
of reheating of cool gas by supernovae is proportional to the SFR
times a power-law function of circular velocity, such that the
reheating is more efficient in smaller mass galaxies. If the halo's
virial velocity is less than a preset ejection threshold (set to 100
$km$ $s^{-1}$ in our model), then all of the reheated gas is ejected
from the halo; otherwise the reheated gas is placed in the hot gas
reservoir within the halo.  The gas and metals that are ejected from
the halo are distributed outside of the halo with a continuation of
the isothermal $r^{-2}$ profile that we assumed inside the halo.  This
material falls in gradually as the virial radius of the halo increases
due to the falling background density of the Universe.

Chemical evolution is treated by assuming a constant mean mass of
metals produced per mass of stars. The metals produced in and ejected
from stars are first deposited into the surrounding cold gas, at which
point they may be ejected from the disk and mixed with the hot halo
gas.  The metallicity of any newly formed stars is set to equal the
metallicity of the ambient cold gas at the time of formation.

For each galaxy, the SAM predicts the two-dimensional distribution of
the mass in stars of a given age and metallicity. We convolve this
distribution with the SSP models of BC03 (using the Padova 1994
isochrones) to create a synthetic SED. We adopted a \citet{cha03} IMF
with lower and upper mass cutoffs of $m_{L} ~= ~0.1 ~M_{\sun}$ and
$m_{U} ~= ~100 ~M_{\sun}$. We then use the model of \citet{mad95} to
account for the opacity of intervening HI in the intergalactic medium
as a function of redshift, and convolve the SEDs with the response
functions appropriate to the ACS, ISAAC, IRAC, and MOSAIC photometric
bands.

Dust extinction is treated assuming that the face-on optical depth in
the V-band is $\tau_{V,0} = \tau_{dust,0} \times
(\dot{m_{*}})^{\beta_{dust}}$ where $\tau_{dust,0}$ and $\beta_{dust}$
are free parameters set as 1.2 and 0.3, to match the observations in
the GOODS-S. This choice is motivated by the observational results of
\citet{hop01}. The dependence of the extinction on wavelength
(the attenuation curve) is calculated using a Calzetti attenuation
curve \citep{cal00}.

In summary, we use the semi-analytic models to build galaxies that
have rich and varied star formation histories that are motivated by
the hierarchical galaxy formation picture.  Star formation is bursty
and episodic. Stars have a distribution of metals consistent with the
assumed star formation history, and the dust content is plausible.  In
contrast, the simple SED templates which we will use to $analyze$
these model galaxies have uniform metallicity and monotonically
declining star formation rates.

\section{Lyman Break Galaxy Samples and Spectral Energy Distribution Fitting}

\subsection{Model Galaxy Catalog and Lyman Break Color Selection} \label{modelLBG}

We created a suite of model runs by varying the uncertain model
parameters controlling the burstiness of star formation and dust
extinction. As stated in \S~\ref{model}, the rest-frame UV-continuum
and UV-optical colors of model LBGs from each semi-analytic model run
were compared with the colors of LBGs observed in the GOODS-S field,
and the best-fit model run was selected based on these
comparisons. From this best-fit model run, a model galaxy catalog,
which contains a total of 44281 galaxies within a redshift range 2.3
$\leq z \leq$ 5.7, has been constructed.  This catalog carries the
various physical parameters such as stellar mass and SFR, along with
the (dust-extinguished) broad band photometry for several bandpasses
used in the GOODS observations.  The photometric data in
this catalog are derived by combining the SAM's star formation and
enrichment histories with BC03 stellar population synthesis models,
and including models for dust extinction and absorption by the IGM, as
described in \S~\ref{model}.

These photometric data are used to select high-redshift, star-forming
galaxies through the Lyman-break color selection technique.  The
Lyman-break color selection technique -- which uses the `Lyman-break'
feature at $\lambda \sim 912$ $\rm{\AA}$ in galaxy spectra and the
Lyman-$\alpha$ forest flux deficit between 912 $\rm{\AA}$ and 1216
$\rm{\AA}$ together with blue colors at longer rest-frame UV
wavelengths to identify star-forming galaxies located at high redshift
-- has been shown to be an effective way to construct large samples of
high-redshift, star-forming galaxies from optical photometric data
sets \citep{mad96, ste03}.  Spectroscopic follow-ups for the LBG
samples have verified the robustness and efficiency of this technique
in building up high-$z$ galaxy samples \citep{ste96,ste99,van06}

The color criteria used in this work to select U-, B-, $\&$ V-dropout
model LBGs are as follows:

\begin{equation} \label{Udrop1}
(U~-~B_{435})~\geq ~ 0.62 ~+~ 0.68 ~\times ~ (B_{435} ~-~ z_{850}) ~ \wedge 
\end{equation}
\begin{equation} \label{Udrop2}
(U~-~B_{435})~\geq ~ 1.25 ~ \wedge 
\end{equation}
\begin{equation} \label{Udrop3}
(B_{435}~-~z_{850}) ~\leq ~ 1.93
\end{equation}

for U-dropouts (z$\sim$3.4),

\begin{equation} \label{Bdrop1}
(B_{435} ~-~ V_{606}) ~> ~ 1.1 ~+~ (V_{606} ~-~ z_{850}) ~ \wedge  
\end{equation}
\begin{equation} \label{Bdrop2}
(B_{435} ~-~ V_{606})~> ~ 1.1 ~ \wedge 
\end{equation}
\begin{equation} \label{Bdrop3}
(V_{606}~-~z_{850}) ~< ~ 1.6 
\end{equation}

for B-dropouts (z$\sim$4), and

\begin{equation} \label{Vdrop1}
((V_{606} ~-~ i_{775}) ~> ~ 1.4667 ~+~ 0.8889 ~\times ~ (i_{775} ~-~ z_{850})) ~ \vee ((V_{606} ~-~ i_{775}) ~> ~2.0) ~\wedge 
\end{equation}
\begin{equation} \label{Vdrop2}
(V_{606} ~-~ i_{775}) ~> ~ 1.2 ~ \wedge 
\end{equation}
\begin{equation} \label{Vdrop3}
(i_{775}~-~z_{850}) ~< ~ 1.3
\end{equation}

for V-dropouts (z$\sim$5).

Here, $\rm{\wedge}$ means logical `AND', and $\rm{\vee}$ is logical `OR'.

The criteria for B- $\&$ V-dropouts used here are the same ones which
have been tested and used in previous works on GOODS survey fields
\citep[e.g.,][]{gia04b,idz04,pap04,idz07}.  The criteria for
U-dropouts are set (1) to select $z\sim$ 3 star-forming galaxies, (2)
to effectively avoid contamination from foreground stars, and (3) to
minimize the overlap with B-dropouts.  These color criteria and
additional magnitude cuts for the ACS z-band ($z_{AB} \leq 26.6$) and
IRAC channel 1 or channel 2 ($m_{3.6 \micron} \leq 26.1$ or $m_{4.5
  \micron} \leq 25.6$), which correspond to detection magnitude limits
for each bands in GOODS-S, give samples of LBGs containing 2729
objects for U-dropouts with redshift range $3.1 \leq z \leq 3.6$, 2638
objects for $\rm{B}_{435}$-dropouts with $3.6 \leq z \leq 4.6$, and
808 for $\rm{V}_{606}$-dropouts with $4.6 \leq z \leq 5.6$.

\begin{figure}
\plotone{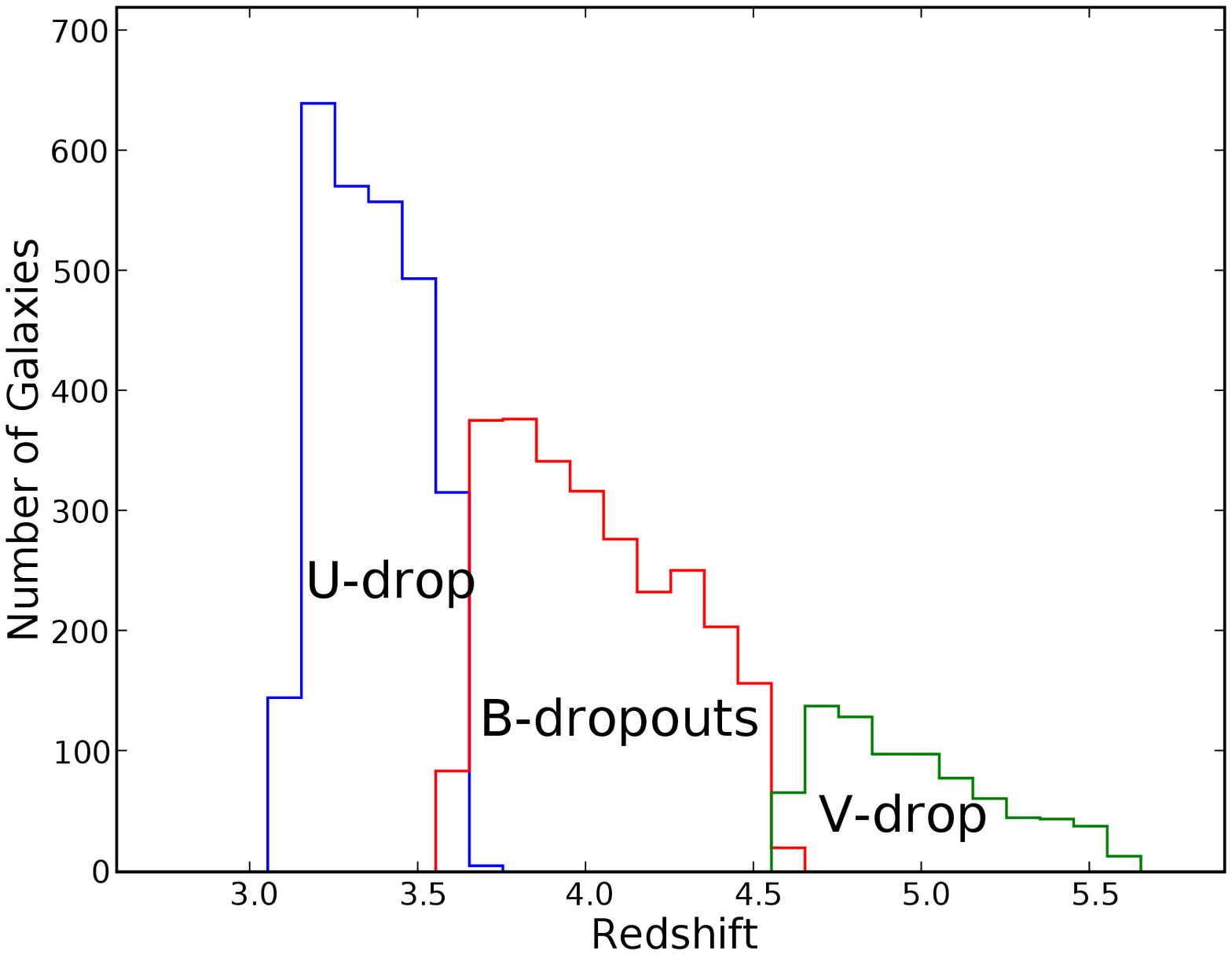}
\caption{Redshift distributions of model U-, B-, $\&$ V-dropout galaxies selected 
through Lyman break color criteria as explained in \S~\ref{modelLBG}. \label{zdist_tot}}
\end{figure}

Figure~\ref{zdist_tot} shows the redshift distributions of model U-,
B-, and V-dropout galaxies selected with the above criteria.  The
color selection criteria effectively isolate three distinct redshift
intervals.  Because the models include a realistic distribution of
galaxy luminosities, the galaxies preferentially lie toward the lower
redshift boundary of each redshift interval, as they do for real
surveys.

\subsection{Spectal Energy Distribution Fitting} \label{SED}

The goal of this work is to compare the statistical distributions of
various physical parameters of model Lyman break galaxies
with the ones derived from the SED-fitting method.  To do this, we
performed $\chi^{2}$-minimization to find the best-fit stellar
population synthesis model templates and best-fit parameters derived
from them.  Here, we use as the stellar population synthesis model the
BC03 model with \citet{cha03} IMF and Padova 1994 evolutionary
tracks. This is the same set of SSP models used to construct the SEDs
for the SAM galaxies.  
Only few discrete values of metallicities are available in the BC03 
model library; among these, two sub-solar (0.2 $Z_{\sun}$ and 0.4 
$Z_{\sun}$) and solar metallicities were used.  
Studies of high redshift LBG spectra have shown that their 
metallicities are subsolar or solar ($Z \sim 0.2 ~-~1.0 ~Z_{\sun}$) 
\citep{pet01,sha03,and04}.  For internal dust
extinction, the Calzetti extinction law \citep{cal00} is used with
0.000 $\leq$ $E(B-V)$ $\leq$ 0.950 with a step size 0.025, and
intergalactic absorption by neutral hydrogen according to
\citet{mad95} is included.  The star formation histories are
parameterized as $\varpropto e^{-t/\tau}$ with the $e$-folding time,
$\tau$ ranging between 0.2 and 15.0 Gyr, and the time since onset of
star formation, $t$ spanning from 10 Myr to 2.3 Gyr, with limiting $t$
being smaller than the age of the universe at each corresponding
redshift.  The parameter values used in our SED-fitting are summarized
in table.~\ref{paraset}.  From the observed LBGs in GOODS-S, mean
errors for different magnitude bins at each passband of ACS, ISAAC and
IRAC are calculated, and assigned to each SAM galaxy
photometric value according to their magnitudes in the calculation of
$\chi^{2}$.  Fluxes of the SAM galaxies in the ACS
$B_{435},~V_{606},~i_{775},~z_{850}$ bands ($B_{435}$ band photometry
are not used in fitting for $V_{606}$-dropouts), ISAAC $J,~H,~Ks$
bands, and IRAC 3.6, 4.5, 5.8, 8.0 $\mu$m channels are compared with
those of BC03 model templates\footnote{U-band data is used only for
  selecting U-dropout samples, and is not used in the SED-fitting
  procedure.}, and based on the calculated $\chi^{2}$ values, the
best-fit galaxy template is determined for each model LBG.  The
best-fit stellar mass is calculated by multipling the mass-to-light
ratio and bolometric luminosity of the best-fit template.  

The best-fit stellar mass equals to the integration of the SFR 
over time, correcting for stellar mass recycling. 

\begin{equation} \label{SFRcalc1}
M_{*} = A \times [ \int_{0}^{t_{0}} \Psi (t',\tau) dt' - \Gamma (t_{0},\tau) ] 
\end{equation}

Here, $t_{0}$ and $\tau$ are the best-fit $t$ and $\tau$ respectively, 
$M_{*}$ is the best-fit stellar mass, and $A$ is a normalization factor. 
$\Psi (t,\tau)$ and $\Gamma (t,\tau)$ are unnormalized SFR and stellar 
mass recycling fraction respectively, and can be known from the best-fit 
$t$ and $\tau$. 
The best-fit instantaneous SFR is then, 

\begin{equation} \label{SFRcalc2}
SFR(t_{0},\tau) = A \times \Psi (t_{0},\tau),
\end{equation}

where $\Psi (t_{0},\tau) = e^{-t_{0} / \tau}$.

Not correcting for recycling would result in a $\sim$ 29 $\%$ 
underestimation of the correct SFR and 8 $\sim$ 9 $\%$
overestimation of the mean age\footnote{In some studies, SFR is
derived first from the best-fit t and $\tau$, then stellar mass is
calculated by integrating the SFR over time. In these cases, failure
to correct for stellar mass recycling results in an overestimation
of stellar mass.}. 
The statistical behaviors of various physical parameters, such as
stellar mass, SFR, and mean age, derived from these best-fit templates
are compared with the intrinsic distributions from the SAMs in the
following sections.

\section{Constraining Physical Parameters using the SED-Fitting Method} \label{sedfit}

\subsection{Physical Parameters of Model Lyman Break Galaxies} \label{statdist}

In the ideal case, in which the SED-fitting methods can determine
various physical parameters accurately, the one-to-one comparison
between intrinsic SAM parameters and best-fit parameters from
SED-fitting should show no deviation from a straight line with a slope
of unity.  However, there are several factors that introduce errors in
the parameters derived from SED-fitting, which include the complex
star formation/merging histories of individual galaxies and photometric
errors.  So, the interesting question is how well the SED-fitting
methods can recover the intrinsic or true statistical distributions of
various physical parameters -- such as stellar mass, star formation rate
(SFR), mean stellar population age, and so on -- which are used to
constrain galaxies' formation histories.

The star formation rate of a galaxy is an ill-defined
quantity. Strictly speaking, it is always zero except at those
instants in time when a collapsing ball of gas begins to generate
energy through nuclear fusion. It is common in the literature to
implicitly average over some time interval $\Delta t$ without
specifying $\Delta t$. In our view, it is important to be specific
about the timespan for this averaging.  Henceforth in this paper, we
define the SFR to be the mass in stars formed in the past 100 Myr.
Our reason for this choice is as follows: unlike the BC03 model galaxy
templates whose SFRs are assumed to decrease exponentially with time,
the star formation activity in semi-analytic model galaxies -- and
probably in real LBGs also \citep[e.g.][]{pap05} -- is more episodic
and complex.  So, instantaneous SFRs or SFRs averaged over very small
$\Delta t$ are less meaningful for SAM LBGs.

In this work, we repeat our SED-fitting experiment with several
different choices of conditions: (1) holding redshift fixed and
fitting redshift as an additional free parameter, (2) using various
combinations of passbands of $ACS$, $ISAAC$, and $IRAC$, and (3) with
limited $\tau$ ranges.  We also try two-component templates as well as
single-component templates.

In this section, we show the SED-fitting results in cases where we use
all the passbands from $ACS$, $ISAAC$, and $IRAC$: first, assuming the
redshift is known, and second, fitting redshift as an additional free
parameter.

\begin{figure}
\plotone{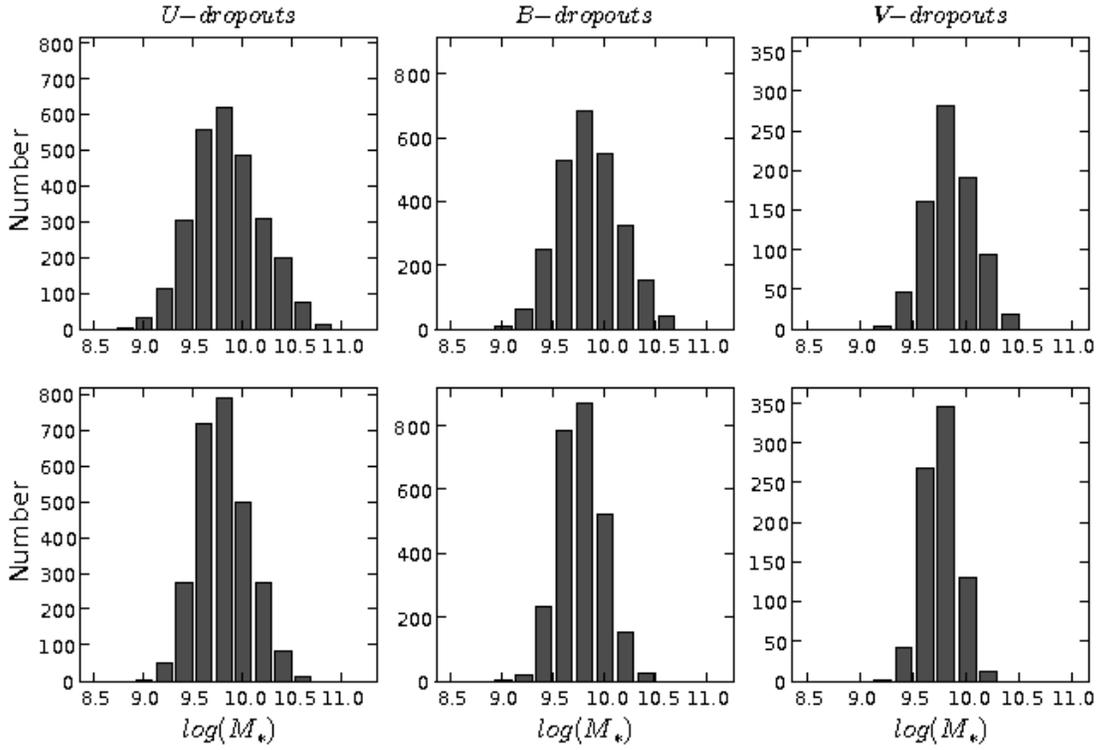}
\caption{Distribution of the logarithm of stellar masses for U-dropouts (left column), 
B-dropouts (middle column), and V-dropouts (right column) when galaxy redshifts are 
fixed and all of the $ACS$/$ISAAC$/$IRAC$ passbands are used in the SED-fitting procedure. 
Figures in the top row are the intrinsic distributions from the semi-analytic models 
and figures in the bottom row are the distributions of best-fit values from SED-fitting. 
Stellar masses are given in $M_{\sun}$. \label{lmsdstzfix}}
\end{figure}

\begin{figure}
\plotone{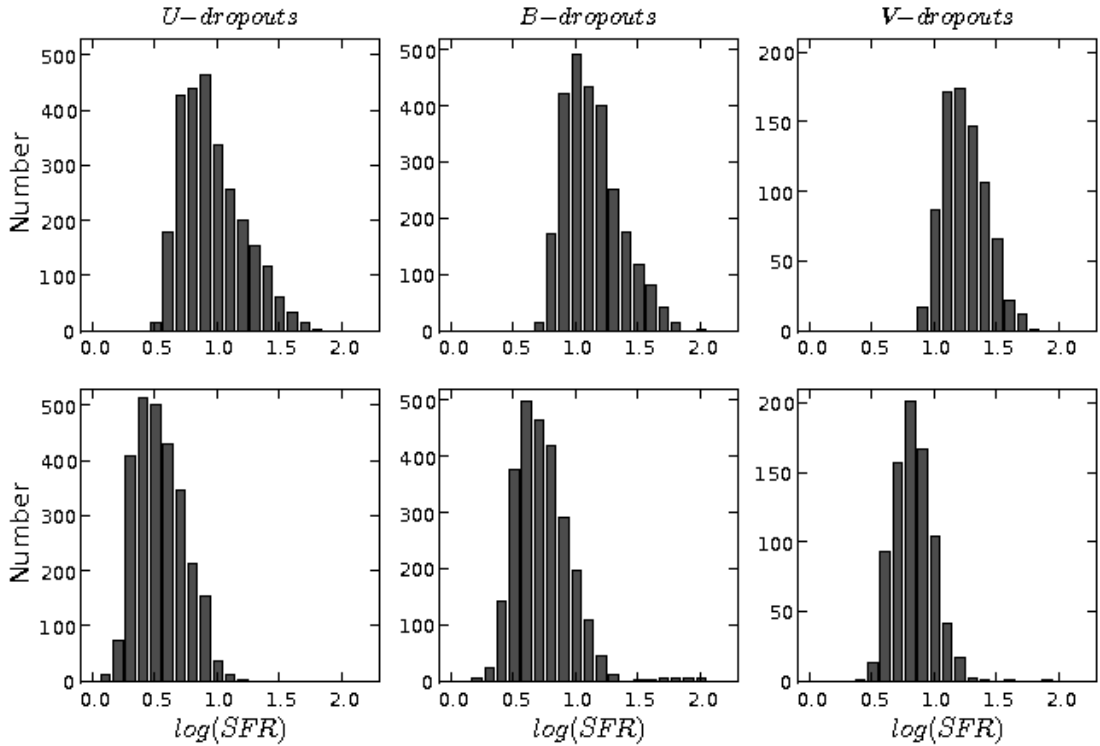}
\caption{Distribution of the logarithm of SFRs averaged over last 100 Myr (in $M_{\sun}$ $yr^{-1}$) 
for U-dropouts (left column), B-dropouts (middle column), and V-dropouts (right column) 
when galaxy redshifts are fixed and all of the $ACS$/$ISAAC$/$IRAC$ passbands are used. 
The top row shows the intrinsic distributions from the semi-analytic models and the bottom row 
shows the distributions of best-fit values from SED-fitting. \label{sfrdstzfix}}
\end{figure}

\begin{figure}
\plotone{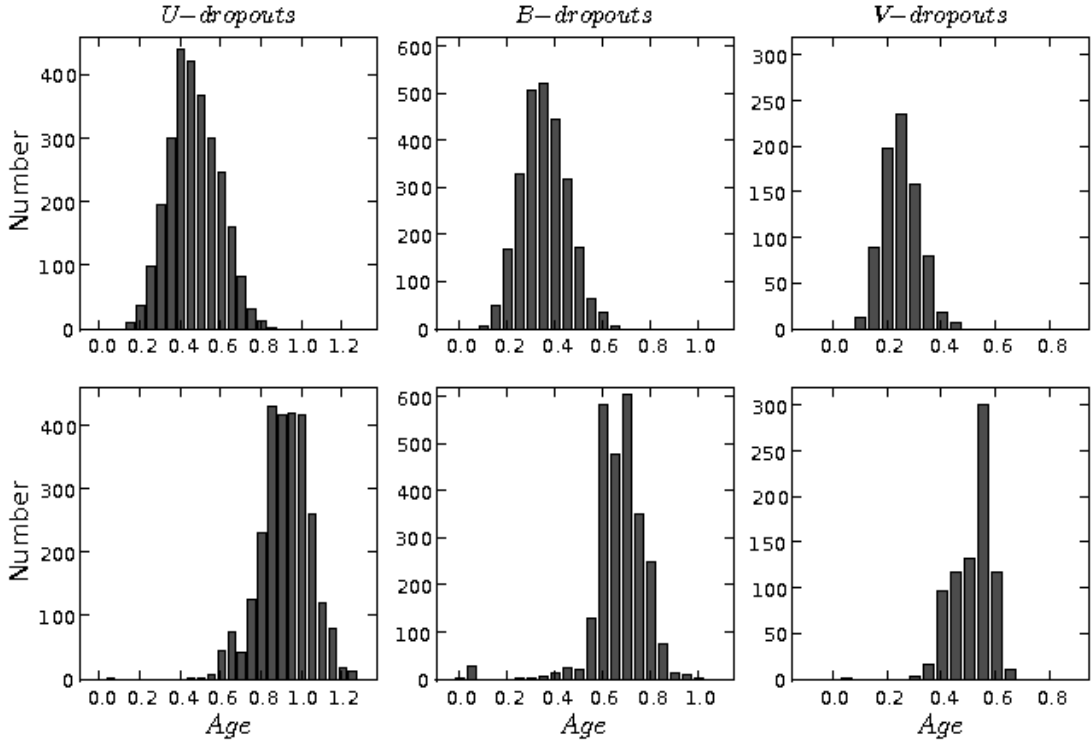}
\caption{Distribution of the stellar-mass weighted mean stellar-population ages (given in Gyr) 
for U-dropouts (left column), B-dropouts (middle column), and V-dropouts (right column) 
when galaxy redshifts are fixed and all of the $ACS$/$ISAAC$/$IRAC$ passbands are used 
for SED fitting. 
Figures in the top row are the intrinsic distributions from the semi-analytic models and figures 
in the bottom row are the distributions of best-fit values from SED-fitting. \label{agedstzfix}}
\end{figure}

Figures~\ref{lmsdstzfix}, \ref{sfrdstzfix}, and \ref{agedstzfix} compare
the intrinsic distributions to the fitted values of stellar mass,
SFR (averaged over the past 100 Myr), and (mass-weighted) mean
stellar population age for $z$ $\sim$ 3.4 (U-dropouts), 4.0 (B-dropouts), 
and 5.0 (V-dropouts) model LBGs, with the redshift fixed to the actual 
value in the SAM model catalog.

Figure~\ref{lmsdstzfix} shows that the SED-fitting method recovers
relatively well the input distributions of stellar masses in spite of
the very different star formation histories (SFHs) in the
semi-analytic model and in the templates from the BC03 stellar
population synthesis models.  The mean values of the SED-derived
stellar masses of U-, B-, and V-dropouts differ from the intrinsic
mean values by $\sim$ 19 $\%$ (U-dropouts), 25 $\%$ (B-dropouts), and
25 $\%$ (V-dropouts) of the intrinsic mean values, in the sense that
the SED-fitting method systematically underestimates the mean values
of stellar masses for all of three dropout samples.  
This trend may be attributed to the fact that light from a recent 
generation of star formation can easily mask (some portion of) the 
presence of an older stellar population. 
Our results are consistent with the earlier arguments that the stellar 
masses of LBGs are underestimated when derived from (single-component) 
SED-fitting methods \citep[e.g.][]{pap01,sha05}.

However, figures~\ref{sfrdstzfix} and \ref{agedstzfix} show that there
are larger biases in constraining SFRs and mean ages of galaxies using
the SED-fitting method.  As can be seen in these figures, the mean
SFRs derived from the SED-fitting method systematically underestimate
the intrinsic mean values by $\sim$ 65 $\%$ for U-dropouts, $\sim$ 58
$\%$ for B-dropouts and $\sim$ 62 $\%$ for V-dropouts, while the
stellar population mean ages are overestimated by about factors of 
two for all three sets of dropouts.

\begin{figure}
\plotone{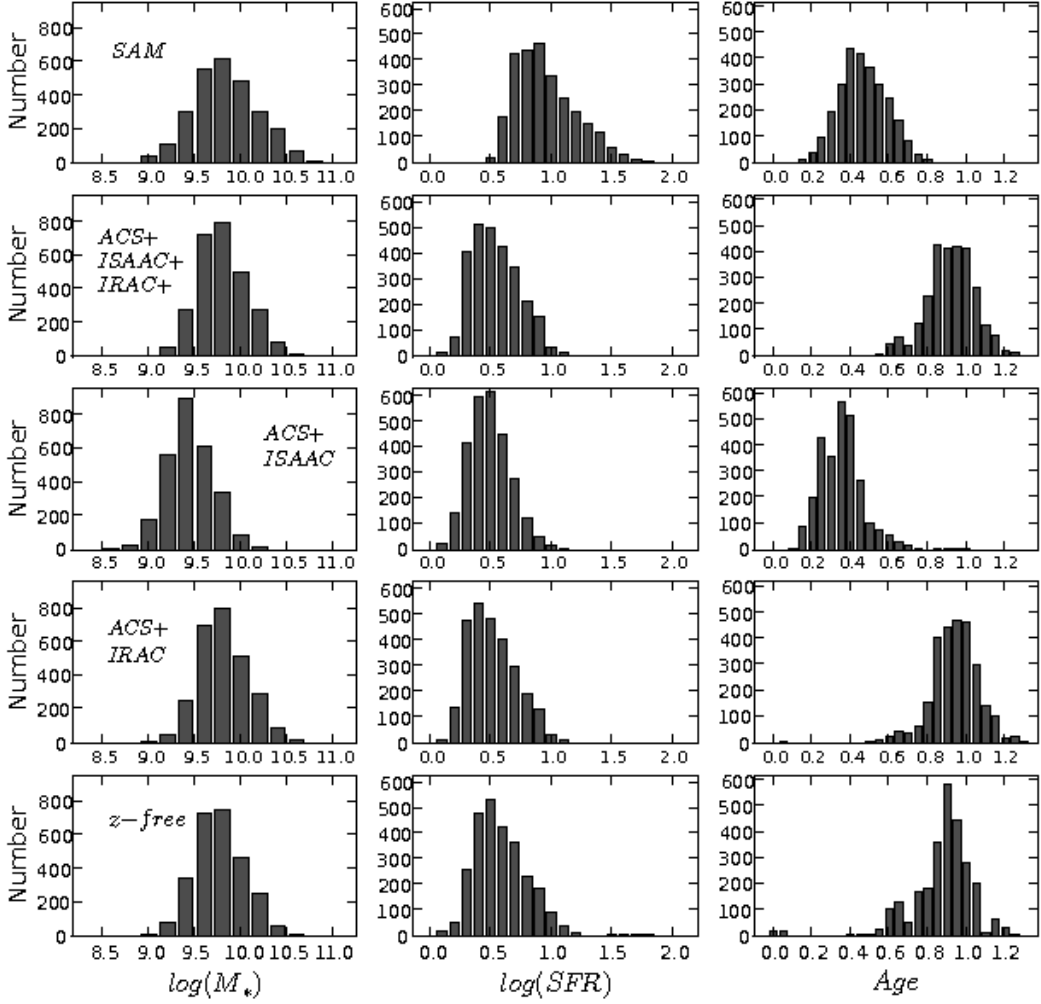}
\caption{Comparison of the statistical distributions of stellar masses (left column), 
SFRs (middle column), and mean ages (right column) of U-dropouts when we use 
$ACS$, $ISAAC$, $\&$ $IRAC$ (the second row), when we use $ACS$ $\&$ $ISAAC$ only 
(i.e. without using $IRAC$ bands, the third row), and when we use $ACS$ $\&$ $IRAC$ 
only (i.e. without $ISAAC$, the fourth row), all with the redshift held fixed.
The bottom row shows the case when we vary redshift as a free parameter. 
Intrinsic distributions are shown in the top row. 
Stellar masses are given in $M_{\sun}$, SFRs are given in $M_{\sun}$ $yr^{-1}$, 
and mass-weighted mean ages are given in $Gyr$. \label{udropdsts}}
\end{figure}

\begin{figure}
\plotone{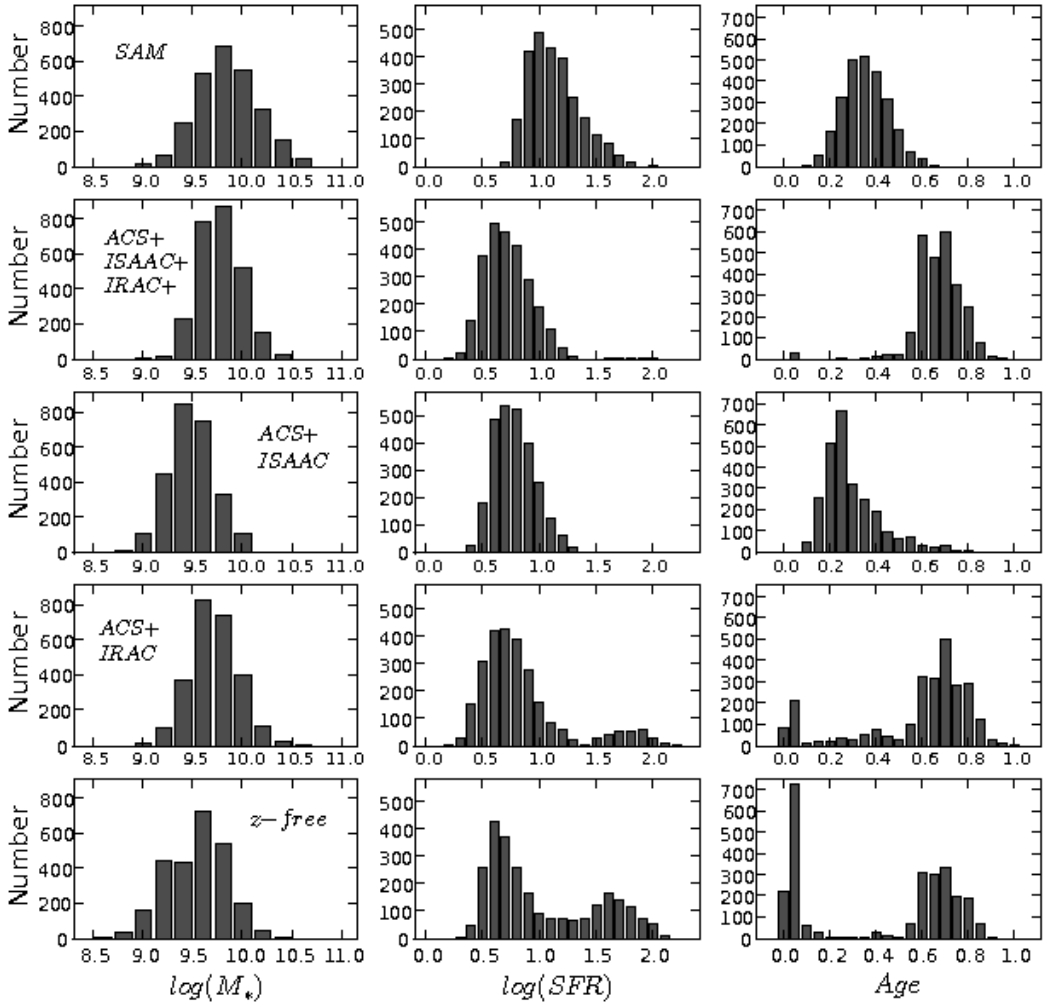}
\caption{Same as figure 5, but for B-dropouts. \label{bdropdsts}}
\end{figure}

\begin{figure}
\plotone{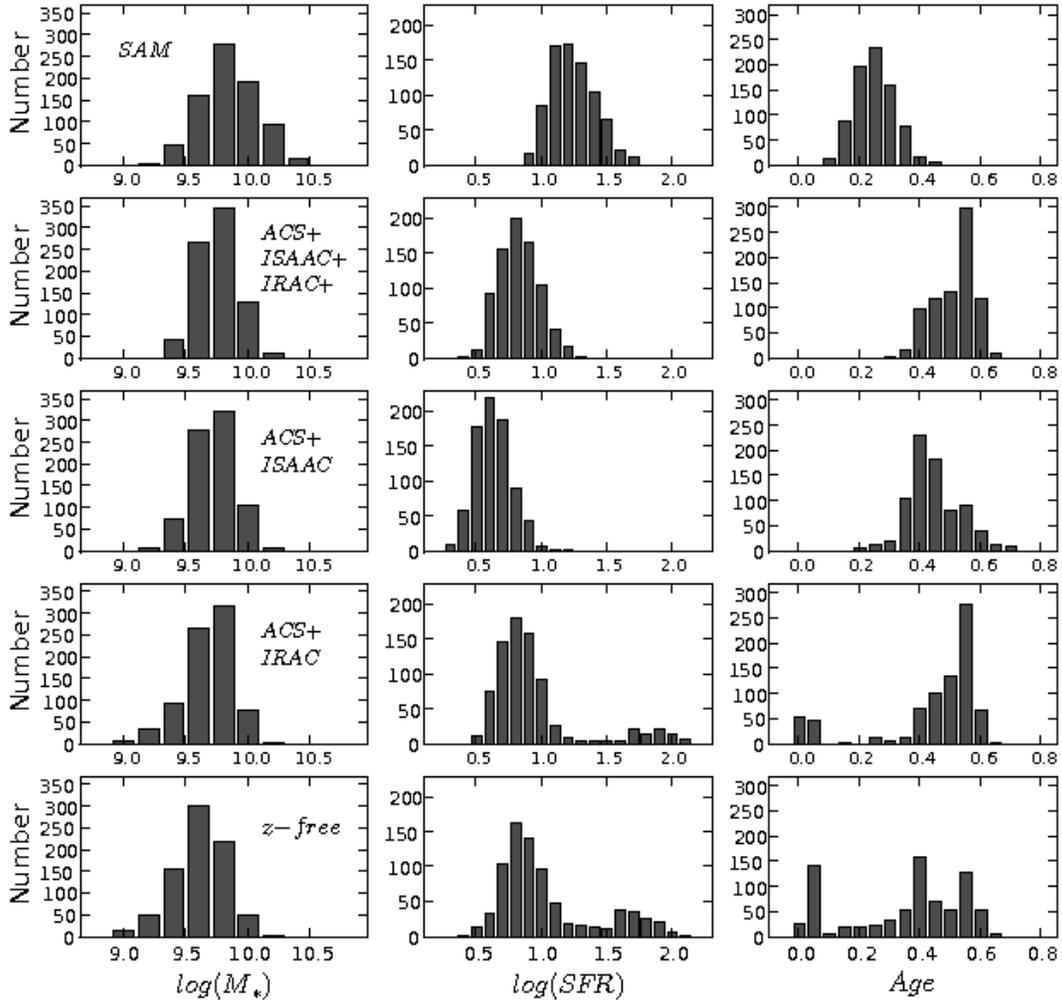}
\caption{Same as figures 5 and 6, but for V-dropouts. \label{vdropdsts}}
\end{figure}

The bottom rows of figure~\ref{udropdsts}, \ref{bdropdsts}, and
\ref{vdropdsts} show the results when the redshift is allowed to float
as an additional free parameter in the fit -- which is analogous to
the case of color-selected LBG samples without spectroscopic (or
pre-calculated photometric) redshift information.

The bottom rows of the first columns of figure~\ref{udropdsts},
\ref{bdropdsts}, and \ref{vdropdsts} show that when we allow redshift
to vary as an additional free parameter, the differences between the
distributions of SAM intrinsic and SED-derived stellar masses
increase especially for B- and
V-dropouts.  The mean values of the SED-derived stellar masses are
underestimated by 25 $\%$ (U-dropouts), 51 $\%$ (B-dropouts), and 
43 $\%$ (V-dropouts) when redshifts were allowed to float.

More interestingly, bimodalities in the distributions of SED-derived
SFRs and mean ages become more significant, as can be seen in the
second and third columns of the bottom row of figures~\ref{udropdsts},
\ref{bdropdsts}, and \ref{vdropdsts}. The change is more significant
for B- and V-dropouts.  When we fix the redshift, we can only see
hints of the existence of this bimodality in the SFR and age 
distributions of B-dropouts.  
The existence of these bimodalities, which are absent in
the intrinsic distributions, indicates that there are sub-populations
of LBGs whose behavior in the SED-fitting procedure is distinct from
others, and the different behaviors of these subpopulations are
exaggerated when we do not fix the redshifts.  The directions of these
bimodal distributions show that for this sub-population of LBGs, the
SED-fitting method does not underestimate (or even overestimates) SFRs
and underestimates mean ages. The characteristic SFHs of this
sub-population of galaxies are discussed in \S~\ref{subpop}

SED-fitting generally underestimates redshifts slightly and the ranges
of redshift discrepancies, ($z_{SED} - z_{SAM}$)/(1+$z_{SAM}$) are
$\sim$ -0.137 $-$ 0.043 for U-dropouts, $\sim$ -0.115 $-$ 0.044 for
B-dropouts, and $\sim$ -0.052 $-$ 0.010 for V-dropouts.  For galaxies
whose redshifts are severely underestimated, stellar masses and ages
are severely underestimated and SFRs are overestimated.  The means of
$\vert z_{SED} - z_{SAMs} \vert$/(1+$z_{SAMs}$) are 0.022, 0.032, and
0.014 for U-, B-, and V-dropouts.

Table.~\ref{means} summarizes the main results of the SED-fitting in
the case where we fix the redshifts to the values in the SAM model
galaxy catalog (analogous to observations with spectroscopic
redshifts) as well as in the case where we allow redshift to float as
an additional free parameter (analogous to observations without
spectroscopic redshifts).  The contents of table.~\ref{means} are the
mean values of SAM intrinsic and SED-derived stellar masses, SFRs, and
mean ages as well as redshifts for each set of dropouts.

\subsection{Biases in Constraining Physical Parameters} \label{bias}

In section \ref{statdist}, we showed the statistical behavior of
various physical parameters derived from the SED-fitting compared with 
the intrinsic distributions. In this and the next section, we investigate
more thoroughly the biases in the statistical properties of physical
parameters derived from SED-fitting, focusing in particular on the
dependencies of the offsets on various galaxy parameters.

\begin{figure}
\plotone{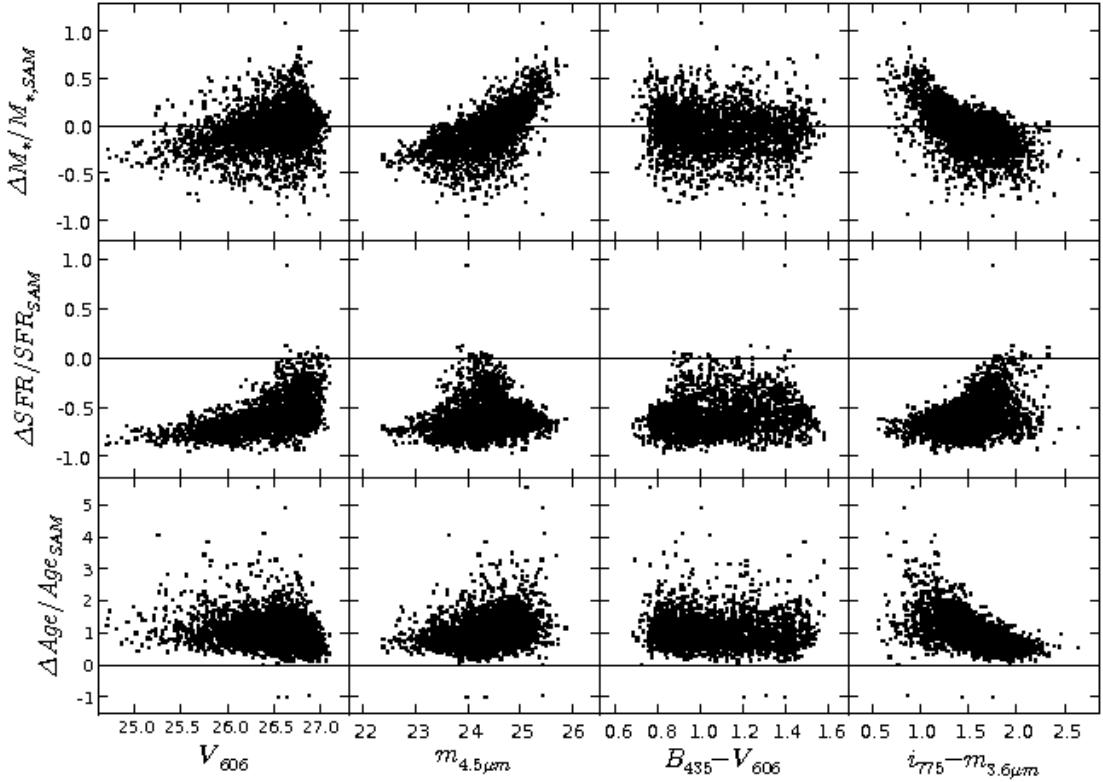}
\caption{Relative errors of stellar masses, SFRs, and mean ages vs. rest-frame UV 
(ACS $V_{606}$) and optical (IRAC 4.5$\mu$m) magnitudes, and vs. rest-framge UV 
($B_{435}-V_{606}$) and UV-optical ($i_{775} - m_{3.6 ~\mu m}$) colors 
of U-dropout LBGs in case when $ACS$/$ISAAC$/$IRAC$ passbands are used and 
galaxy redshifts are fixed at the input values during the SED fitting. 
One object with $\Delta SFR / SFR_{SAM}$ larger than 3 is excluded in figures 
in the middle row for visual clarity. \label{umgclrzfix}}
\end{figure}

\begin{figure}
\plotone{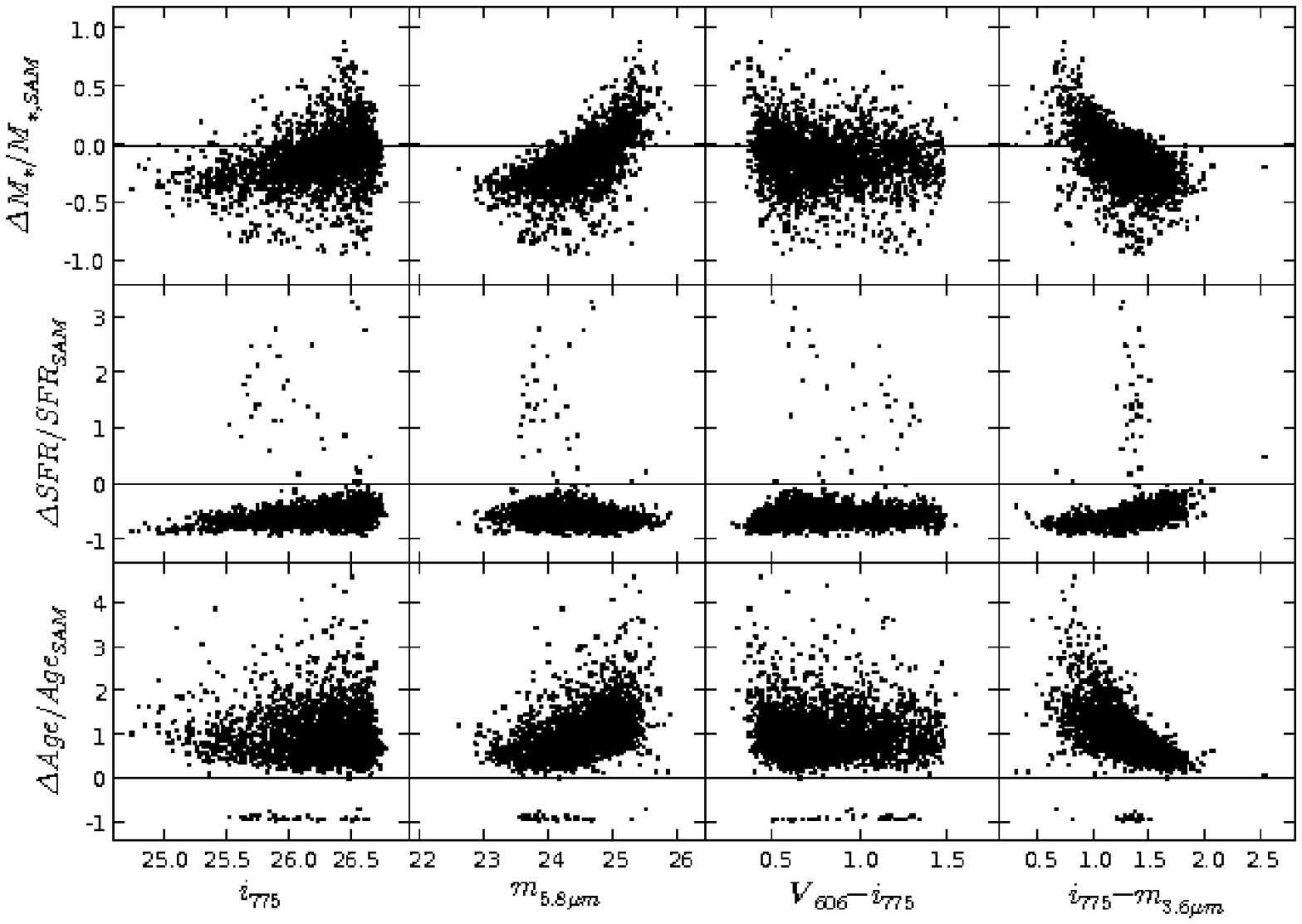}
\caption{Relative errors of stellar masses, SFRs, and mean ages vs. rest-framge UV 
(ACS $i_{775}$) and optical (IRAC 5.8$\mu$m) magnitudes, and vs. rest-frame UV 
($V_{606} - i_{775}$) and UV-optical ($i_{775}-m_{3.6 ~\mu m}$) colors  
of B-dropout LBGs in case when $ACS$/$ISAAC$/$IRAC$ passbands are used and 
galaxy redshifts are fixed at the input values. \label{bmgclrzfix} }
\end{figure}

\begin{figure}
\plotone{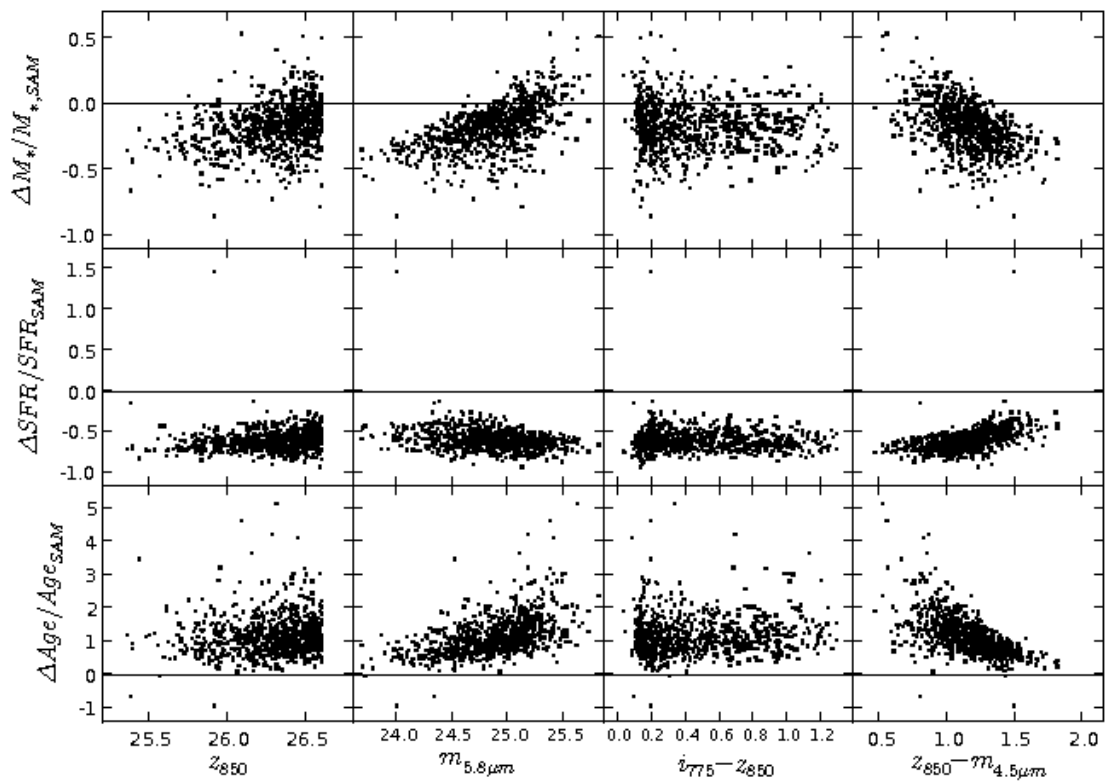}
\caption{Relative errors of stellar masses, SFRs, and mean ages vs. rest-frame UV 
(ACS $z_{850}$) and optical (IRAC 5.8$\mu$m) magnitudes, and vs. rest-frame UV 
($i_{775} - z_{850}$) and UV-optical ($z_{850} - m_{5.8 ~\mu m}$) colors  
of V-dropout LBGs in case when $ACS$/$ISAAC$/$IRAC$ passbands are used and 
galaxy redshifts are fixed. \label{vmgclrzfix} }
\end{figure}

\begin{figure}
\plotone{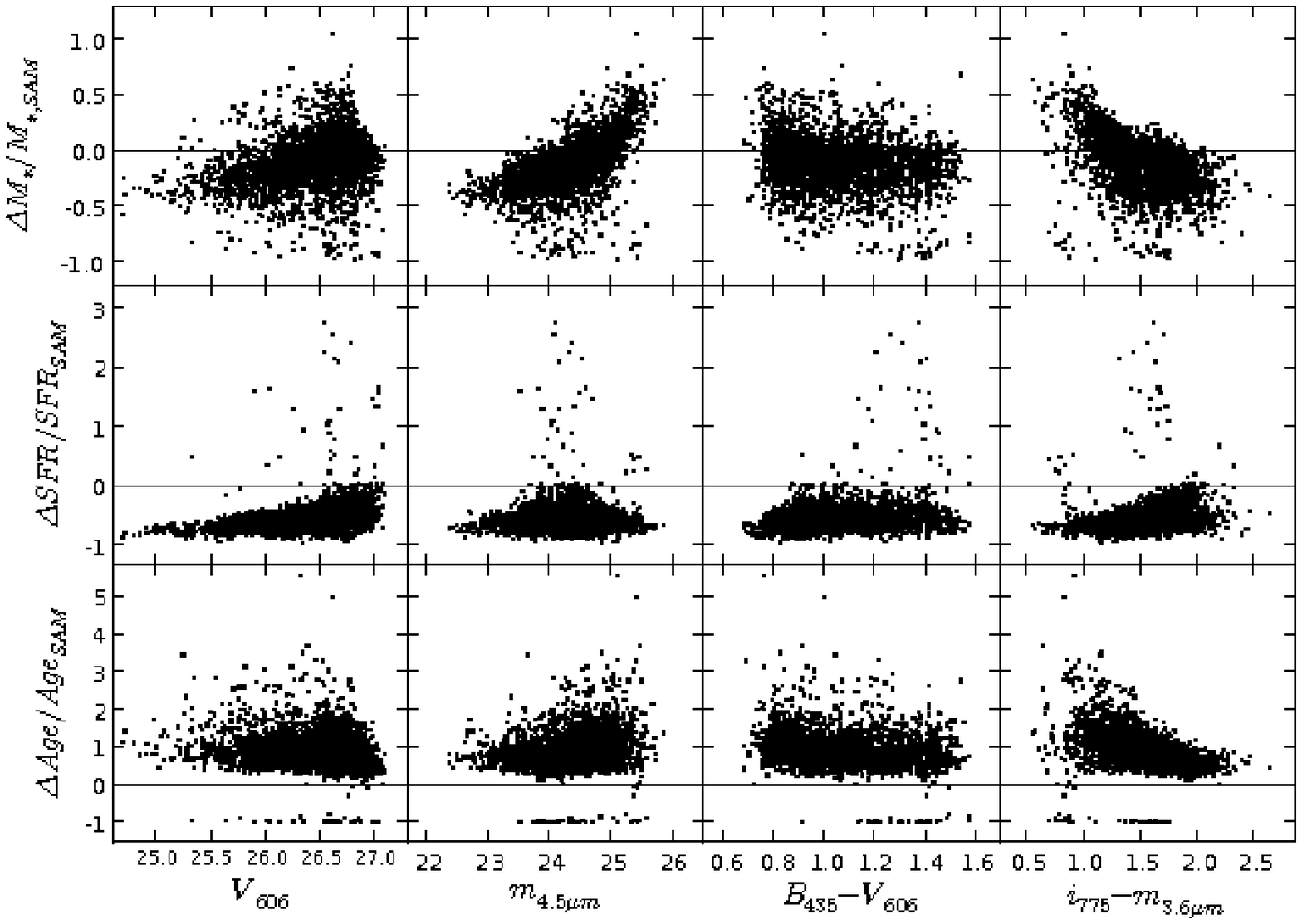}
\caption{Relative errors of stellar masses, SFRs, and mean ages vs. rest-frame UV 
(ACS $V_{606}$) and optical (IRAC 4.5$\mu$m) magnitudes, and vs. rest-framge UV 
($B_{435} - V_{606}$) and UV-optical ($i_{775} - m_{3.6 ~\mu m}$) colors 
of U-dropout LBGs in case when $ACS$/$ISAAC$/$IRAC$ passbands are used and 
galaxy redshifts are allowed to vary as an additional free parameter. \label{umgclrzfree} }
\end{figure}

\begin{figure}
\plotone{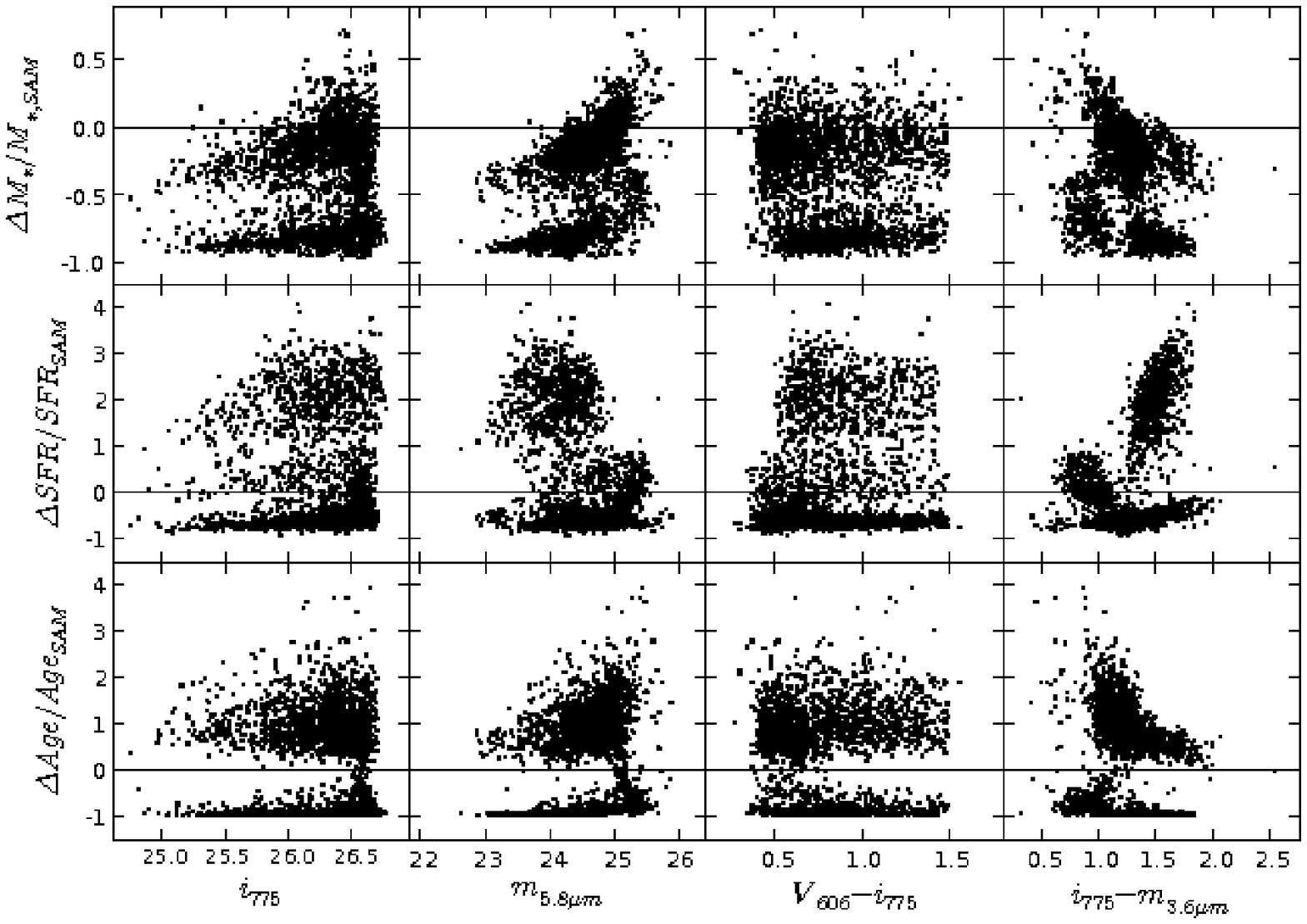}
\caption{Relative errors of stellar masses, SFRs, and mean ages vs. rest-frame UV 
(ACS $i_{775}$) and optical (IRAC 5.8$\mu$m) magnitudes, and vs. rest-frame UV 
($V_{606} - i_{775}$) and UV-optical ($i_{775} - m_{3.6 ~\mu m}$) colors  
of B-dropout LBGs in case when $ACS$/$ISAAC$/$IRAC$ passbands are used and 
galaxy redshifts are allowed to vary as an additional free parameter. \label{bmgclrzfree}}
\end{figure}

\begin{figure}
\plotone{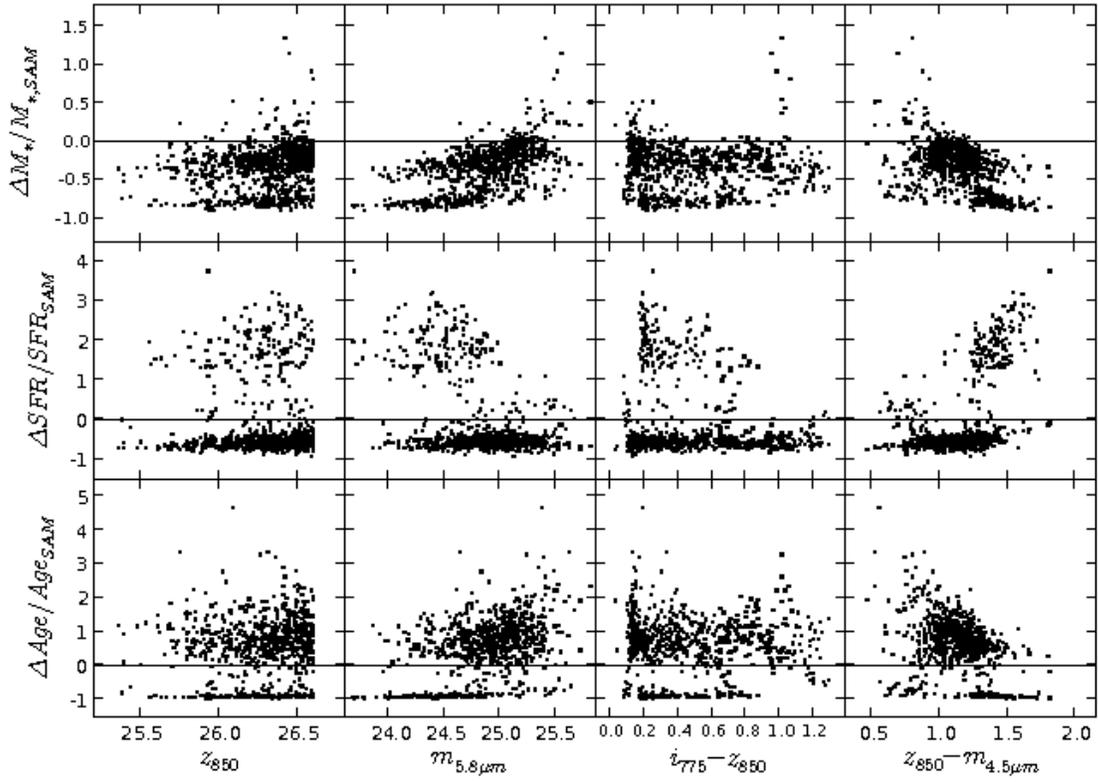}
\caption{Relative errors of stellar masses, SFRs, and mean ages vs. rest-frame UV 
(ACS $z_{850}$) and optical (IRAC 5.8$\mu$m) magnitudes, and vs. rest-frame UV 
($i_{775} - z_{850}$) and UV-optical ($z_{850} - m_{4.5 ~\mu m}$) colors  
of V-dropout LBGs in case when $ACS$/$ISAAC$/$IRAC$ passbands are used and 
galaxy redshifts are allowed to vary as an additional free parameter. 
One object with $\Delta M_{*} / M_{*,SAM}$ larger than 2.5 is excluded from the figures 
in the top row for visual clarity. \label{vmgclrzfree}}
\end{figure}

First, in this section, we show how the discrepancies in LBGs' stellar
mass, SFR, and mean age depend on the magnitudes and colors of LBGs.
Specifically, we investigate the behavior of biases as functions of
rest-frame UV and optical magnitudes (i.e. ACS $V_{606}$ and IRAC
$m_{4.5 \micron}$ for U-dropouts, ACS $i_{775}$ and IRAC $m_{5.8
  \micron}$ for B-dropouts, and ACS $z_{850}$ and IRAC $m_{5.8
  \micron}$ for V-dropouts), and rest-frame UV and UV-optical colors
($B_{435}-V_{606}$ and $i_{775}- m_{3.6 \micron}$ for U-dropouts,
$V_{606}-i_{775}$ and $i_{775} - m_{3.6 \micron}$ for B-dropouts, and
$i_{775}-z_{850}$ and $z_{850} - m_{4.5 \micron}$ for V-dropouts) in
figures~\ref{umgclrzfix}-\ref{vmgclrzfix} for the case of fixed
redshifts and in figures~\ref{umgclrzfree}-\ref{vmgclrzfree} for the
case of allowing redshift as an additional free parameter.  In these
figures, relative errors of stellar mass, SFR, and mean age are defined 
as $\Delta^{r} M_{*} = (M_{*,SED} - M_{*,SAM})/M_{*,SAM}$, 
$\Delta^{r} SFR = (SFR_{SED} - SFR_{SAM})/SFR_{SAM}$, and 
$\Delta^{r} Age = (Age_{SED} - Age_{SAM})/Age_{SAM}$, respectively. 
Here, $value_{SED}$ is stellar mass, SFR, or mean age derived from 
SED-fitting, and $value_{SAM}$ is the intrinsic stellar mass, SFR, or 
mean age of each galaxy.  
These figures clearly show that the SFRs of most galaxies are 
underestimated and mean ages are almost always overestimated.

For both the case of fixed redshifts and where we vary redshift as a
free parameter, the magnitude- and color-dependent behaviors are
similar; however, the plots for B-dropouts when redshifts are fixed
(figure \ref{bmgclrzfix}) and for U-, B-, and V-dropouts when
redshifts are allowed to vary freely (figures
\ref{umgclrzfree}-\ref{vmgclrzfree}) reveal subpopulation(s) of LBGs
whose behaviors in SED-fitting are distinguished from the majority of
LBGs in each dropout sample even in stellar mass estimation.  The
bimodality which is clearly seen in figure \ref{bmgclrzfix} is not
evident in the mass distributions shown in section \ref{statdist}.
According to figure \ref{bmgclrzfix} and figures
\ref{umgclrzfree}-\ref{vmgclrzfree}, galaxies which are belong to this 
subpopulation show distinguished pattern of biases from the majority 
of galaxies: (1) Their stellar masses are more severely underestimated 
than the majority of galaxies. (2) SFRs are less severely 
underestimated or even overestimated. (3) Mean ages are underestimated 
for these galaxies in the subpopulation while they are overestimated 
for the majority of galaxies. Also, this subpopulation of
galaxies is more likely to reside on the fainter side of the
rest-frame UV magnitude distribution.  
A more detailed investigation of this subpopulation is given in 
section \ref{subpop}. It should also be noted that figure 
\ref{bmgclrzfree} indicates that there are actually (at least) two 
subpopulations whose behavior in the SED-fitting procedure are 
distinct from the majority of LBGs.

The offsets in stellar mass, SFR, and mean age show relatively clear
dependencies on rest-frame UV-optical color. (1) The stellar mass is 
more likely underestimated for redder LBGs, (2) the bluer the LBG is 
in rest-frame UV-optical color, the more the SFR is underestimated and 
the more the mean age is overestimated. 
Rest-frame UV-optical color is considered to be a crude indicator of a 
galaxy's stellar population age, so dependencies of offsets on 
rest-frame UV-optical color may indicate that (one of) the main cause(s) 
of the biases in SED-fitting is the mean age (or SFH), in the sense 
that systematic offsets between intrinsic and best-fit SFRs and mean 
ages increase for younger galaxies.

Interestingly, stellar masses tend to be overestimated for galaxies
that are blue in rest-frame UV-optical color.  These are also the
galaxies whose biases in SFR and mean age are largest,
which suggests that the mass-overestimation and age-overestimation
share the same origin.

The parameter that shows the clearest dependence on rest-frame optical
magnitude is stellar mass. Stellar mass is more likely to be
underestimated for LBGs which are brighter in rest-frame optical
magnitudes, while stellar masses of fainter LBGs are more likely to be
overestimated.  At both faint and bright rest-frame optical magnitude,
SFRs are clearly underestimated, while relative errors are distributed
widely.  This indicates that several different factors may contribute
to the biases in SFR estimates.

The offsets show almost no dependence on rest-frame UV-color, and
relatively weak dependence on rest-frame UV magnitude. The dispersions
of offsets are larger for galaxies whose rest-frame UV magnitudes are
fainter, and SFRs are more likely underestimated for galaxies with
bright rest-frame UV magnitudes.

These color/magnitude dependencies become more complex when we do not
fix the redshifts.  More specifically, for some galaxies with red
rest-frame UV-optical color, stellar masses are greatly
underestimated, SFRs are severely overestimated, and ages are severely
underestimated. These trends are similar for the subpopulation of
B-dropouts shown in figure~\ref{bmgclrzfix}.

Also, there are also some hints of galaxies with blue rest-frame
UV-optical color, whose stellar masses are underestimated, SFRs are
roughly correct, and ages are underestimated.

Table.~\ref{offsets} lists the means and standard deviations of
relative offsets for physical parameters such as stellar masses, SFRs,
mean ages, and redshifts for each set of dropouts.  Here, the relative
offset for each parameter is defined as ($Value_{SED} -
Value_{SAM}$)/($Value_{SAM}$).  For redshift, the relative offset is
defined slightly differently as ($z_{SED} - z_{SAM}$)/($1+z_{SAM}$),
following the convention used in the literature.

\subsection{Origins of the Biases of Galaxy Population Parameters} \label{intrinsic}

In this section, we investigate the dependencies of the fitting
discrepancies on intrinsic properties of the SAM model galaxies, such
as stellar mass, SFR, age, and specific SFR (SSFR; defined as SFR per
unit stellar mass) to investigate the causes of these biases in SED fitting.

Figures~\ref{uersamzfix}, \ref{bersamzfix}, and \ref{versamzfix} show
how relative errors (as defined in previous section) in stellar mass,
SFR, and mean age are correlated with intrinsic stellar mass, SFR,
mean age and SSFR for U-, B-, and V-dropouts, with redshift fixed.
These correlations shed light on the origins of the biases found in
sections \ref{statdist} and \ref{bias}.

\subsubsection{Origin of Bias in Age Estimation} \label{ageest}

\begin{figure}
\plotone{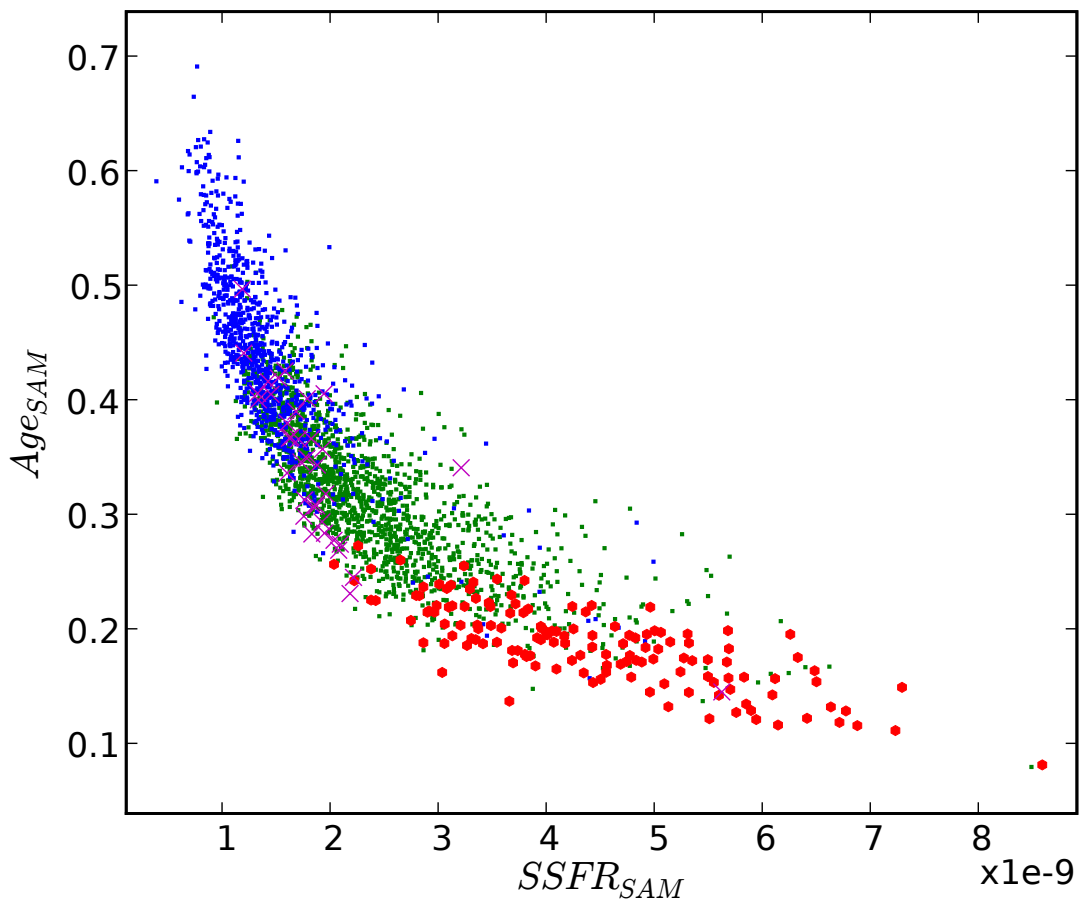}
\caption{Intrinsic age vs. specific SFR (SSFR) for B-dropout galaxies. 
Blue dots represent galaxies with $0.0 \leq \Delta Age/Age_{SAM} \leq 0.75$. 
Green dots are for galaxies with $0.75 < \Delta Age / Age_{SAM} \leq 2.0$, and 
larger, red dots are for ones with $\Delta Age / Age_{SAM} > 2.0$.
Purple crosses show age and SSFR for galaxies whose mean ages are underestimated, 
unlike the majority of B-dropouts. Ages are given in $Gyr$, and SSFRs are given 
in $yr^{-1}$. \label{subagessfr}}
\end{figure}

\begin{figure}
\plotone{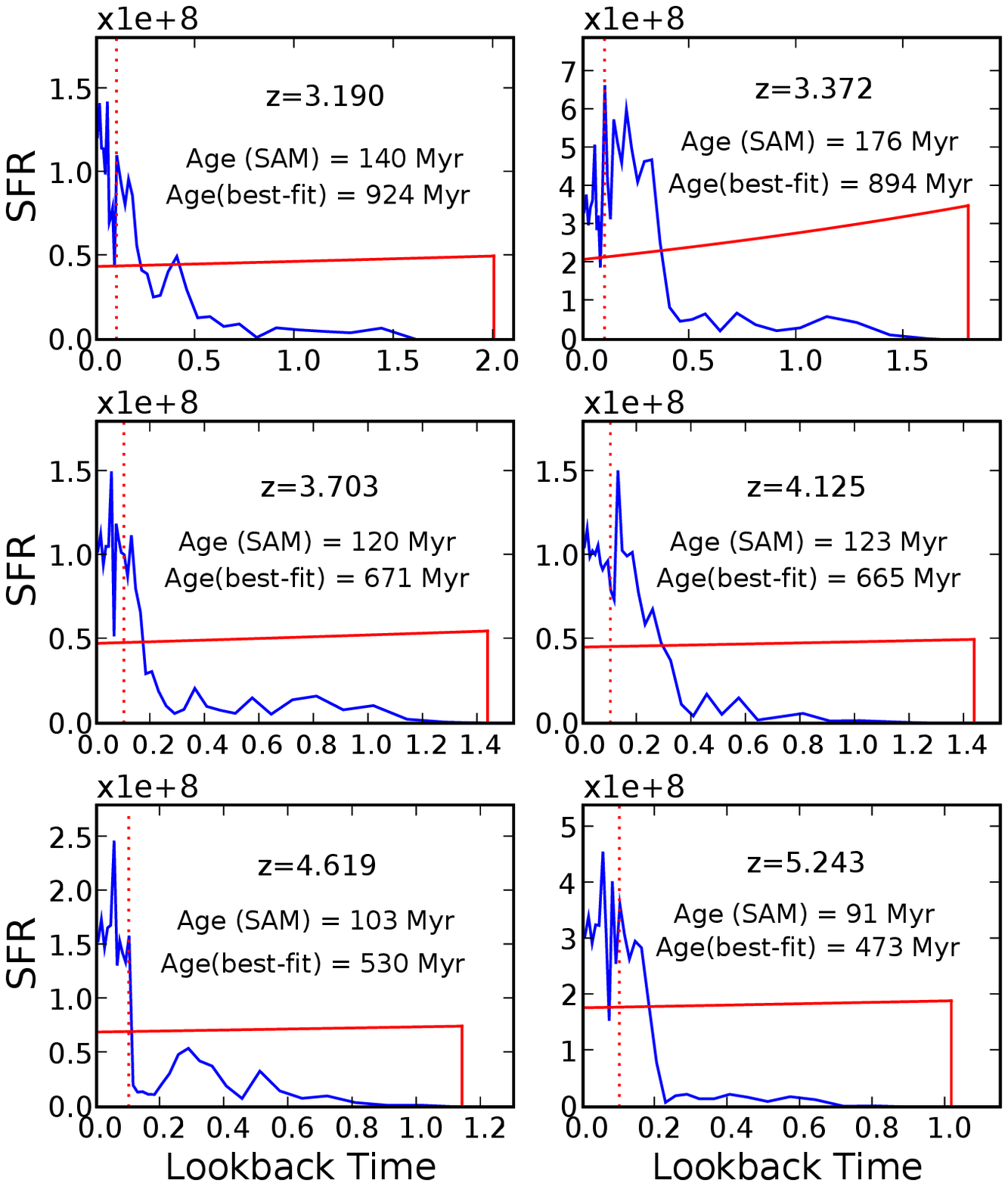}
\caption{Intrinsic star formation histories of typical dropout galaxies among the 
galaxies with the largest overestimation of mean stellar population ages (blue line), 
along with star formation histories of the best-fit BC03 template (red line). 
These galaxies are the ones with type-2 SFH (see text). 
The red, dotted vertical line shows the point where lookback time is 100 $Myr$. 
SFR are measured over past 100 $Myr$ timespan in this work. 
Lookback time is given in $Gyr$. \label{sfhrdaglg}}
\end{figure}

The clearest correlations are seen in `$\Delta^{r} M_{*}$ 
vs. $M_{*,SAM}$', `$\Delta^{r} Age$ vs. $Age_{SAM}$', and
`$\Delta^{r} Age$ vs. $SSFR_{SAM}$'. The tight correlation of 
relative mean age errors with intrinsic ages is not unexpected from
the correlation between relative mean age errors and rest-frame
UV-optical colors.  This correlation and the one between relative mean
age errors and intrinsic SSFRs are a strong indication that mean ages
and/or SSFRs are the main cause of bias.  
The sense of bias is that the stellar population mean age overestimates 
are worse for galaxies with the youngest intrinsic ages and/or largest SSFRs
(i.e. galaxies whose current SF activity is strong compared with the past
SF activity), as can be seen in figure \ref{subagessfr}.  In this
figure, which shows the intrinsic mean ages (y-axis) and SSFRs (x-axis) of
B-dropout galaxies, blue dots represent galaxies whose relative age
errors are $0.0 \leq \Delta^{r} Age \leq 0.75$, i.e. galaxies with the 
smallest $\Delta^{r} Age$. Green dots are for galaxies with 
$0.75 < \Delta^{r} Age \leq 2.0$, and large red dots are for galaxies 
with the largest $\Delta^{r} Age$ ($ > 2.0$).  This
figure clearly shows that galaxies with very large age overestimates 
are relatively young galaxies with high SSFRs. The Young mean ages and
large SSFRs of these galaxies indicate that they have experienced a 
relatively high level of SF activity recently.  
The mismatch between the SFHs predicted by the semi-analytic galaxy 
formation model and the simple SED templates from the BC03 stellar 
population model is expected to be largest for galaxies with this type of
SFH. In the BC03 templates, SFRs are assumed to decrease exponentially, 
so the strongest SF activity always occurs at early times, which is
nearly the opposite of the SFHs of these galaxies.  This difference
makes ages overestimated severely for these galaxies.  Figures
\ref{sfhrdaglg} and \ref{sfhrdagsm}, which show SF activity as a
function of lookback time for individual galaxies, support this
speculation.  Figure \ref{sfhrdaglg} shows the typical SFHs (we 
will refer this type of SFH as `type-2' from now on) of galaxies whose
age-overestimation is largest. The star formation histories of these 
galaxies are clearly distinct from the SFHs of galaxies whose 
age-overestimation is smallest. The typical SFHs of galaxies with the 
smallest age-overestimation are shown in figure \ref{sfhrdagsm} 
(`type-1' SFHs from now on).  
In figure \ref{subagessfr}, there is a dearth of galaxies with old age 
and high SSFR and galaxies with young age and small SSFR.  
Very young galaxies with low SSFR would not pass our color 
selection criteria and/or magnitude limits or would not be detected in 
real galaxy samples. 
Old galaxies with high SSFR, in contrast, would be probably detected. 
To have a high SSFR and an old mass-weighted stellar population age, 
galaxies would need to have had a very extreme SFH -- for example, 
two strong, very short, widely separated bursts.

\subsubsection{Origins of Bias in Stellar Mass Estimation} \label{msest}

How does this age overestimation affect other derived physical 
parameters, such as stellar mass and SFR? 
When the mean stellar age is overestimated, some portion of the 
galaxy's luminosity will be attributed to older stars, with 
consequently higher mass-to-light ratios, than would be the case 
for the true SFH.
This leads to an overestimation of the stellar mass.

\begin{figure}
\plotone{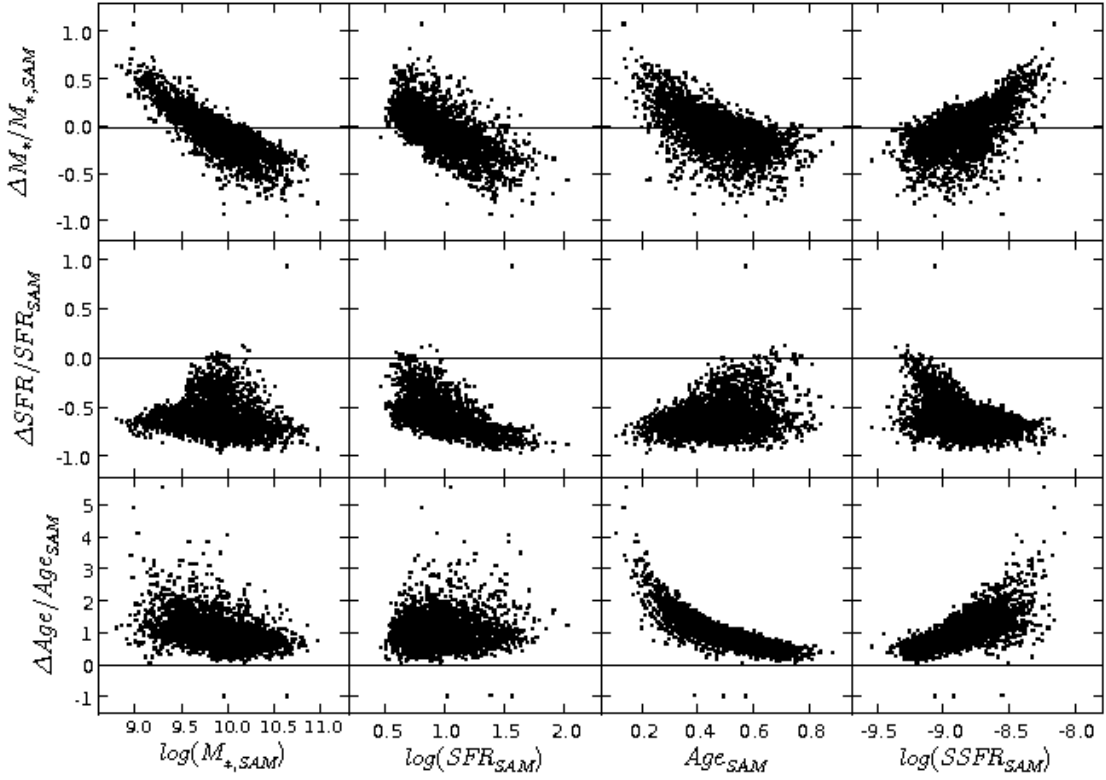}
\caption{Relative errors of stellar masses, SFRs, and mean ages vs. 
intrinsic stellar masses, SFRs, mean ages, and SSFRs for U-dropout LBGs. 
$ACS$/$ISAAC$/$IRAC$ passbands are used and redshifts are fixed. 
One object with $\Delta SFR / SFR_{SAM}$ larger than 3 is excluded from the figures 
in the middle row for visual clarity. \label{uersamzfix}}
\end{figure}

\begin{figure}
\plotone{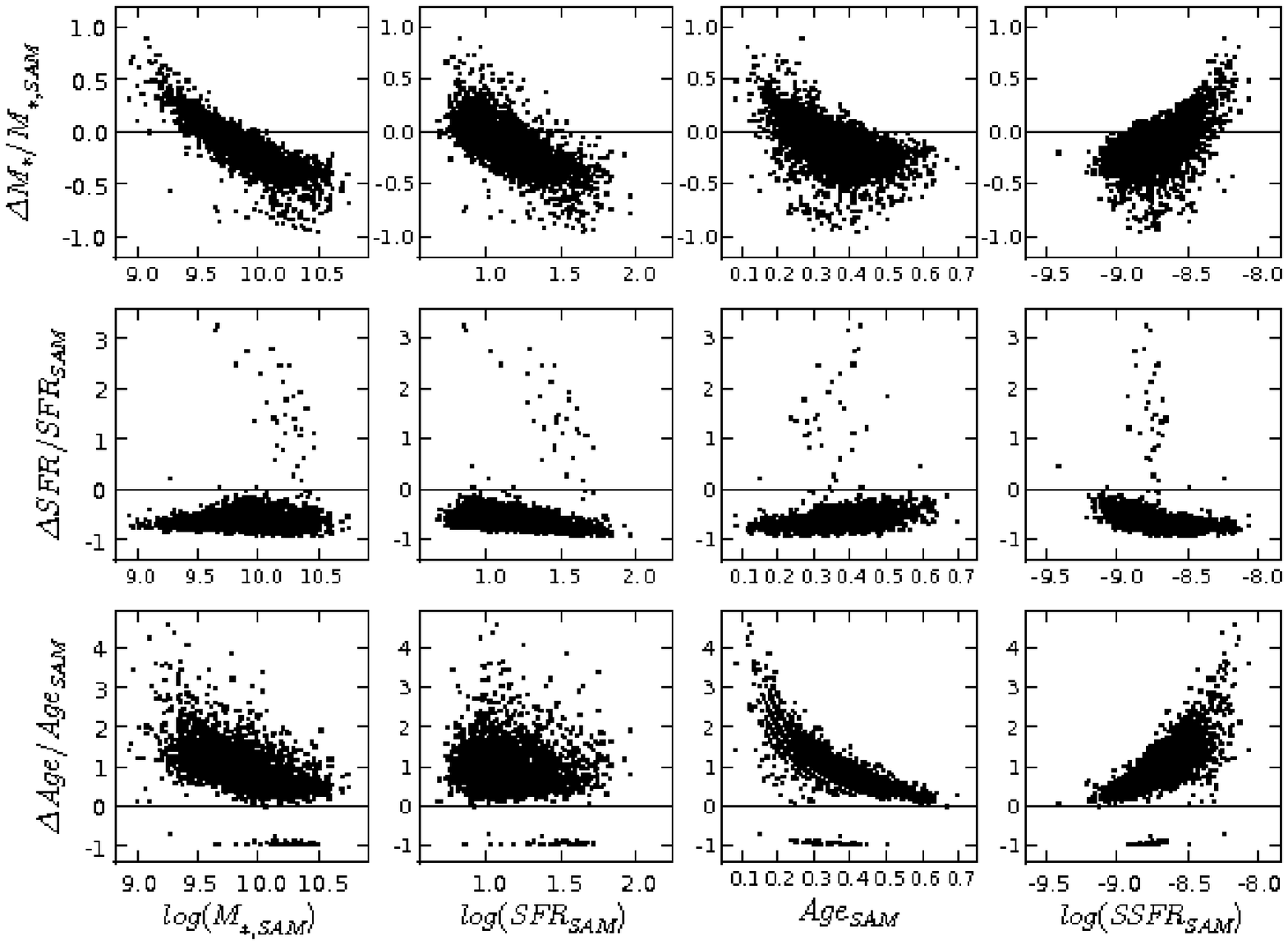}
\caption{Relative errors of stellar masses, SFRs, and mean ages vs. 
intrinsic stellar masses, SFRs, mean ages, and SSFRs for B-dropout LBGs. 
$ACS$/$ISAAC$/$IRAC$ passbands are used and redshifts are fixed. \label{bersamzfix}}
\end{figure}

\begin{figure}
\plotone{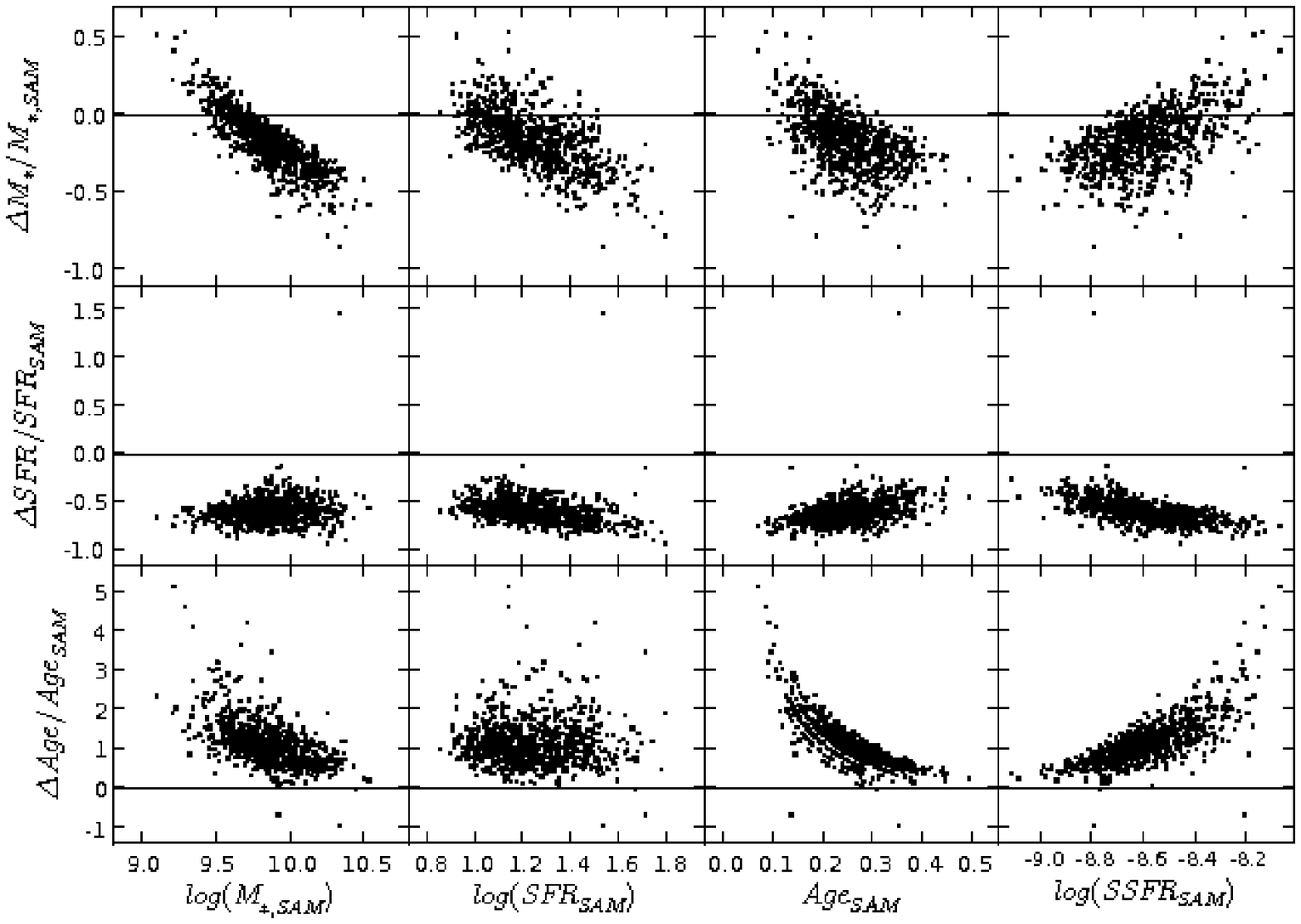}
\caption{Relative errors of stellar masses, SFRs, and mean ages vs. 
intrinsic stellar masses, SFRs, mean ages, and SSFRs for V-dropout LBGs. 
$ACS$/$ISAAC$/$IRAC$ passbands are used and redshifts are fixed. \label{versamzfix}}
\end{figure}

In previous studies, it has been suspected that the stellar masses 
of galaxies are underestimated through single-component SED-fitting, 
since light from the recent star formation can easily mask some 
portion of the older stellar population. 
This is confirmed through the experiments done with the composite 
BC03 templates in section \ref{test}. 
As stated in section \ref{statdist}, the mean values of stellar 
masses are underestimated by $\sim$ 19-25 $\%$, which is in 
qualitative agreement with the arguments of previous studies. 
However, figures \ref{umgclrzfix}-\ref{vmgclrzfix} and 
\ref{uersamzfix}-\ref{versamzfix} show that the stellar mass 
is not always underestimated. 
For some galaxies with very small stellar masses and/or relatively 
young ages, the stellar mass can be overestimated. 
These galaxies are the ones for which the age overestimation is large. 
We speculate that the stellar mass estimation of LBGs 
is affected by two different factors. 
One factor is the fact that the recent generation of star formation 
can dominate the broadband SED, leading to the underestimation of 
stellar mass. 
The other is the SFH difference between SAM galaxies and BC03 
templates. If the age is overestimated due to the SFH difference, 
the mass-to-light ratio is overestimated, which results in the 
stellar-mass overestimation. 

In figures \ref{uersamzfix}-\ref{versamzfix}, it can be seen 
that the stellar masses are most likely underestimated for the 
galaxies with the oldest ages and/or smallest SSFRs, for which 
the age-underestimation is minimal. 
More clearly, (a)-(c) in figure \ref{errvserr} show that 
the stellar masses are underestimated when the age estimation 
is nearly correct. 
This confirms that the stellar mass is underestimated in 
single-component SED-fitting when the effect of the other 
origin of bias -- SFH difference between the 
SAM galaxies and the BC03 templates -- is minimal     
 
In figure \ref{errvserr} (a)-(c), we can also see that the age 
discrepancies and the stellar-mass discrepancies are correlated. 
Stellar masses tend to be underestimated for galaxies with the 
smallest $\Delta^{r} Age$, while they are overestimated for most 
galaxies with the largest $\Delta^{r} Age$.
This supports the speculation that the age-overestimation (due to 
the SFH mismatch between the SAM galaxies and the BC03 templates) 
causes the mass-overestimation through mass-to-light ratio 
overestimation.  
Evidently, in the stellar mass estimation, two sources of bias are 
compensating with each other.  
While SFH differences between the SAM galaxies and the BC03 templates 
tend to make stellar masses to be overestimated in SED-fitting, 
recent SF activity can easily mask older stellar population causing 
the stellar masses to be underestimated.  
The compensating effects of these two biases in the estimation of  
stellar mass explains why the stellar mass distributions are recovered
better than other parameters through the SED-fitting, and also provides 
a clue as to why the stellar mass has turned out to be the most robust 
parameter in earlier studies based on the SED-fitting 
\citep[e.g.][]{pap01,sha01,sha05}.

\subsubsection{Origins of Bias in Star Formation Rate Estimation} \label{sfrest}

\begin{figure}
\plotone{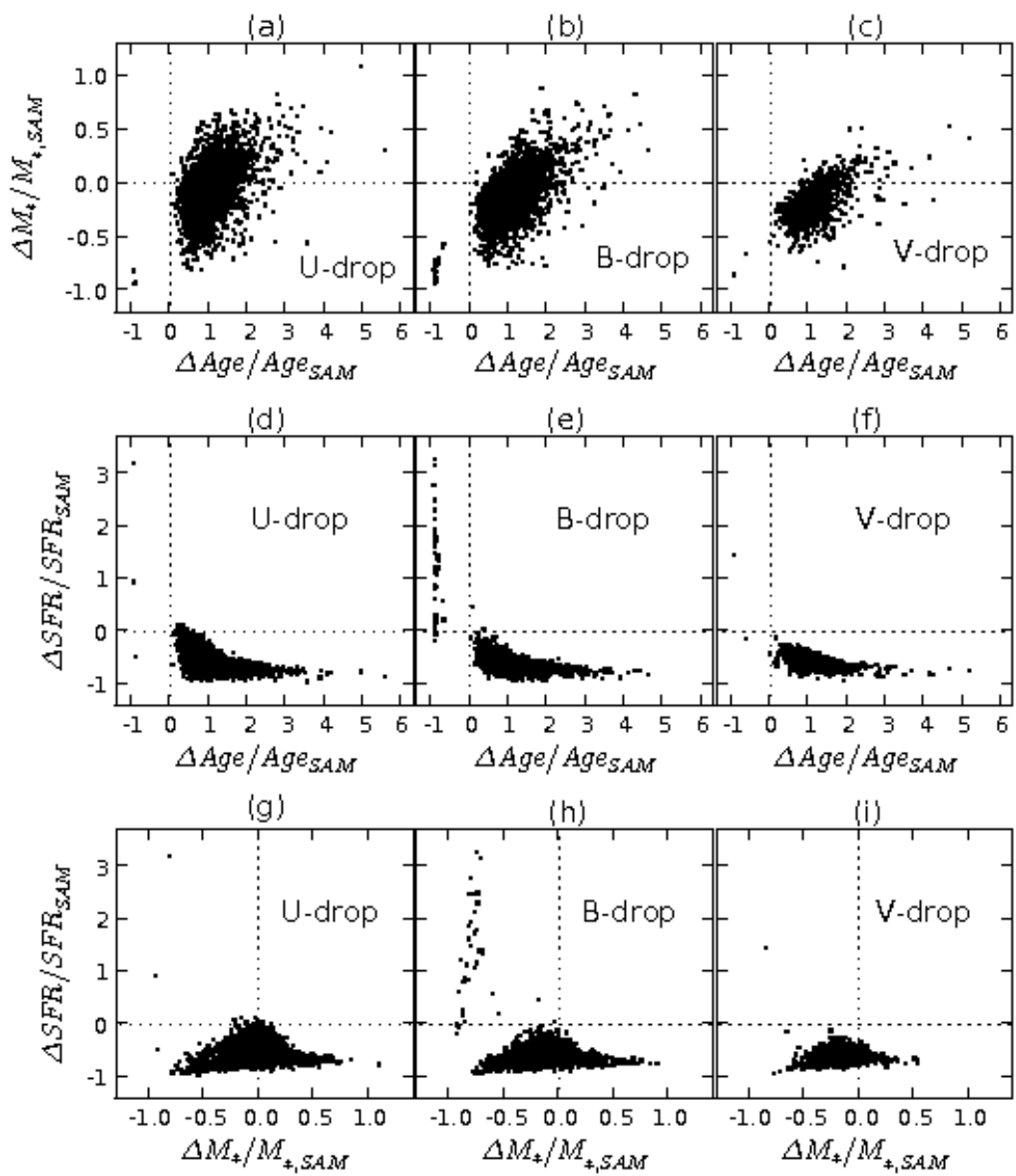}
\caption{Correlations among relative errors of stellar masses, SFRs, and 
mean ages for U-,B-,V-dropout galaxies. $ACS$/$ISAAC$/$IRAC$ passbands are 
used and redshifts are fixed in the SED-fitting. \label{errvserr}}
\end{figure}

Let us now consider the sources of bias in the SFR estimates. 
First, when ages are overestimated, the SED fitting erroneously 
assigns some portion of the luminosity to older stars instead of 
stars that are just forming.
Figures (d)-(f) of figure \ref{errvserr} show that the 
SFRs are most severely underestimated for galaxies with 
the largest $\Delta^{r} Age$.

However, the SFR still tends to be underestimated even when 
the age estimates are nearly correct.
Even some of the most extreme underestimates can be found 
for galaxies whose SED-derived age is correct to within 
a factor of two. This indicates that there is another source 
of bias in the SFR estimation. 
 
Interestingly, the correlation between the relative SFR 
error and the relative stellar mass error shown in (g)-(i) 
of figure \ref{errvserr} reveal distinct behaviors of the upper 
envelopes in this correlation between galaxies with the 
positive relative stellar mass errors and galaxies with the 
negative errors, providing another indication that there are 
two sources of bias in the SFR estimation.
For galaxies whose stellar masses are underestimated -- 
i.e. galaxies whose age-overestimation is small due to 
the little SFH difference or galaxies with type-1 SFH --  
the SFR discrepancy is proportional to the stellar mass 
discrepancy.
As stated in \S~\ref{SED}, the SFR is calculated from the 
estimated stellar mass accumulated over 100 $Myr$. 
Thus, for galaxies with the smallest age discrepancy (with 
the type-1 SFH), the underestimation of stellar mass results 
in the SFR underestimation.

However, unlike in the stellar mass estimation, both of these 
two origins of bias -- the age-overestimation due to the SFH 
mismatch and the hidden old stellar population by the recent 
SF activity -- cause the SFRs to be underestimated. 
This results in the overall underestimation of the SFR distributions 
and the large offsets in the mean SFRs. 

In addition to these two origins, the well-known `age-extinction 
degeneracy' leads to more significant SFR underestimation. For 
galaxies with large (positive) $\Delta^{r} Age$, dust extinctions 
tend to be underestimated due to the `age-extinction degeneracy'. 
This further deflates estimated SFRs for these galaxies. 

In summary, the main origins of biases in estimating physical 
parameters, such as stellar mass, SFR, and mean age, are: 
(1) the differences in the assumed SFHs in the SAM galaxies and 
in the BC03 stellar population templates, (2) the effects of the 
recent SF activity hiding some portion of old generations of 
stellar population, and (3) the age-extinction degeneracy. 
In the stellar mass estimation, issues (1) and (2) compete with 
each other, resulting in the best-fit stellar mass distributions 
that resemble the intrinsic distributions.
For the SFRs, all of these issues work in the same direction, leading 
to the large offsets in the distributions and in the mean values. 
The mass-weighted mean ages are mostly affected by issue (1).

\begin{figure}
\plotone{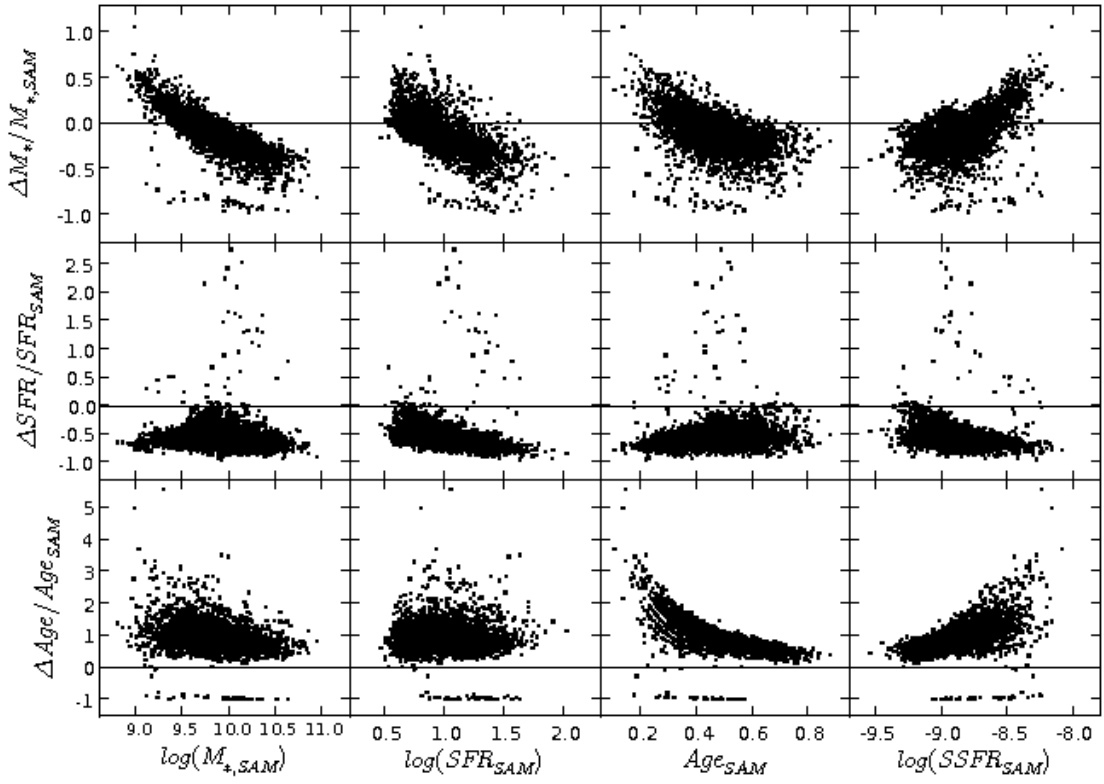}
\caption{Relative errors of stellar masses, SFRs, and mean ages vs. 
intrinsic stellar masses, SFRs, mean ages, and SSFRs for U-dropout LBGs. 
$ACS$/$ISAAC$/$IRAC$ passbands are used and redshifts are allowed to vary 
as a free parameter. \label{uersamzfree}}
\end{figure}

\begin{figure}
\plotone{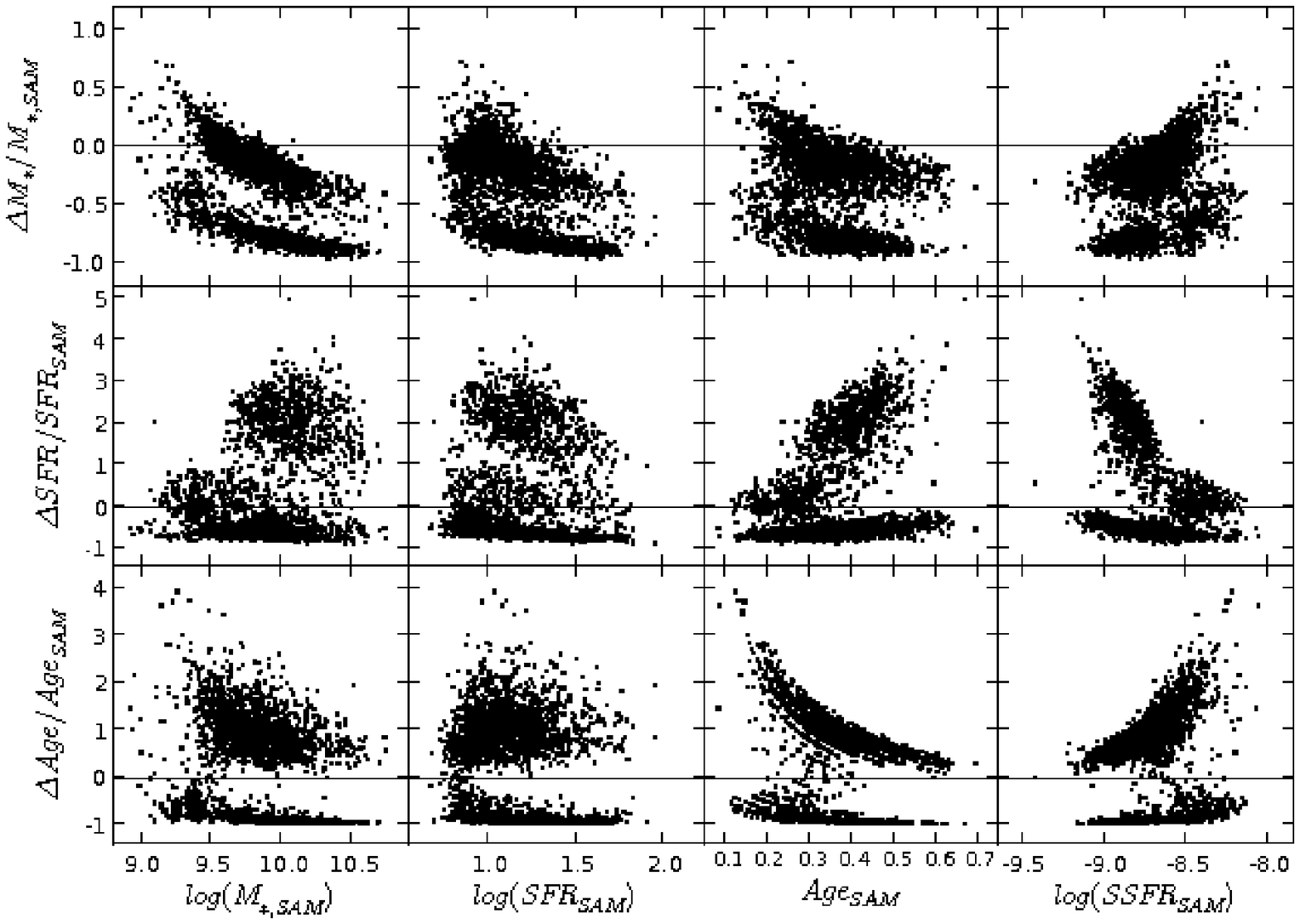}
\caption{Relative errors of stellar masses, SFRs, and mean ages vs. 
intrinsic stellar masses, SFRs, mean ages, and SSFRs for B-dropout LBGs. 
$ACS$/$ISAAC$/$IRAC$ passbands are used and redshifts are allowed to vary. \label{bersamzfree}}
\end{figure}

\begin{figure}
\plotone{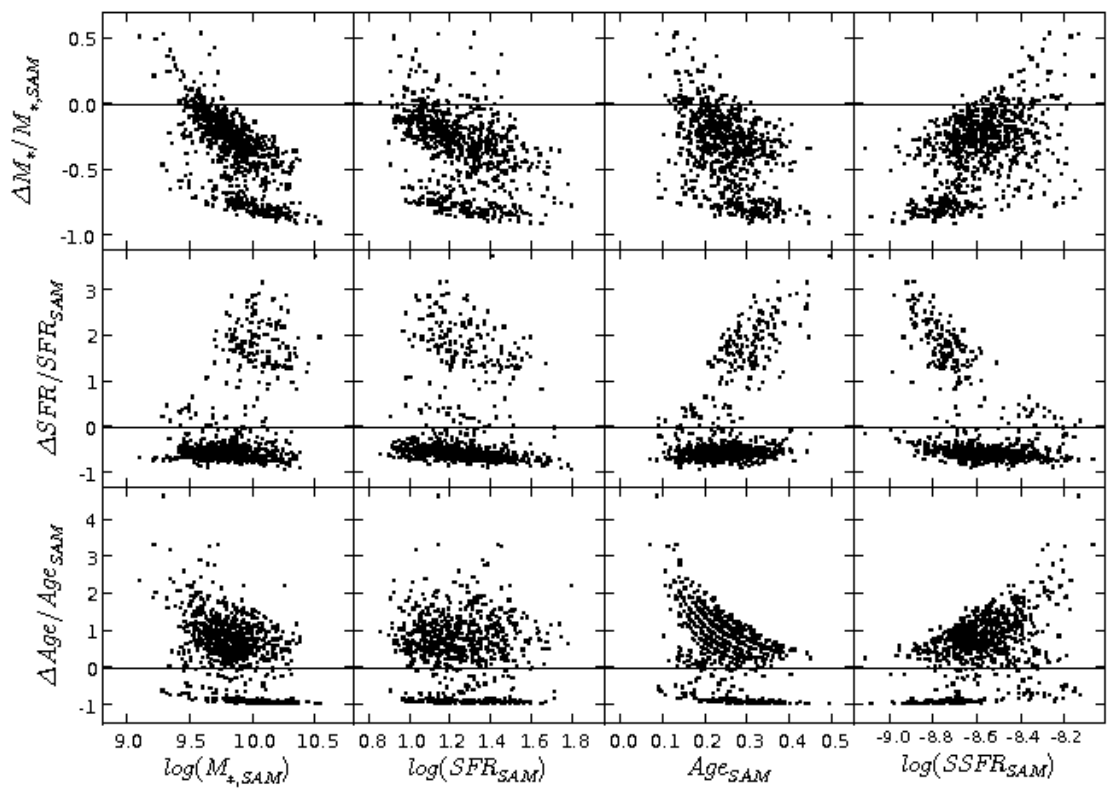}
\caption{Relative errors of stellar masses, SFRs, and mean ages vs. 
intrinsic stellar masses, SFRs, mean ages, and SSFRs for V-dropout LBGs. 
$ACS$/$ISAAC$/$IRAC$ passbands are used and redshifts are allowed to vary. 
One object with $\Delta M_{*} / M_{*,SAM}$ larger than 2.5 is excluded 
from the figures in the top row for visual clarity. \label{versamzfree}}
\end{figure}

\begin{figure}
\plotone{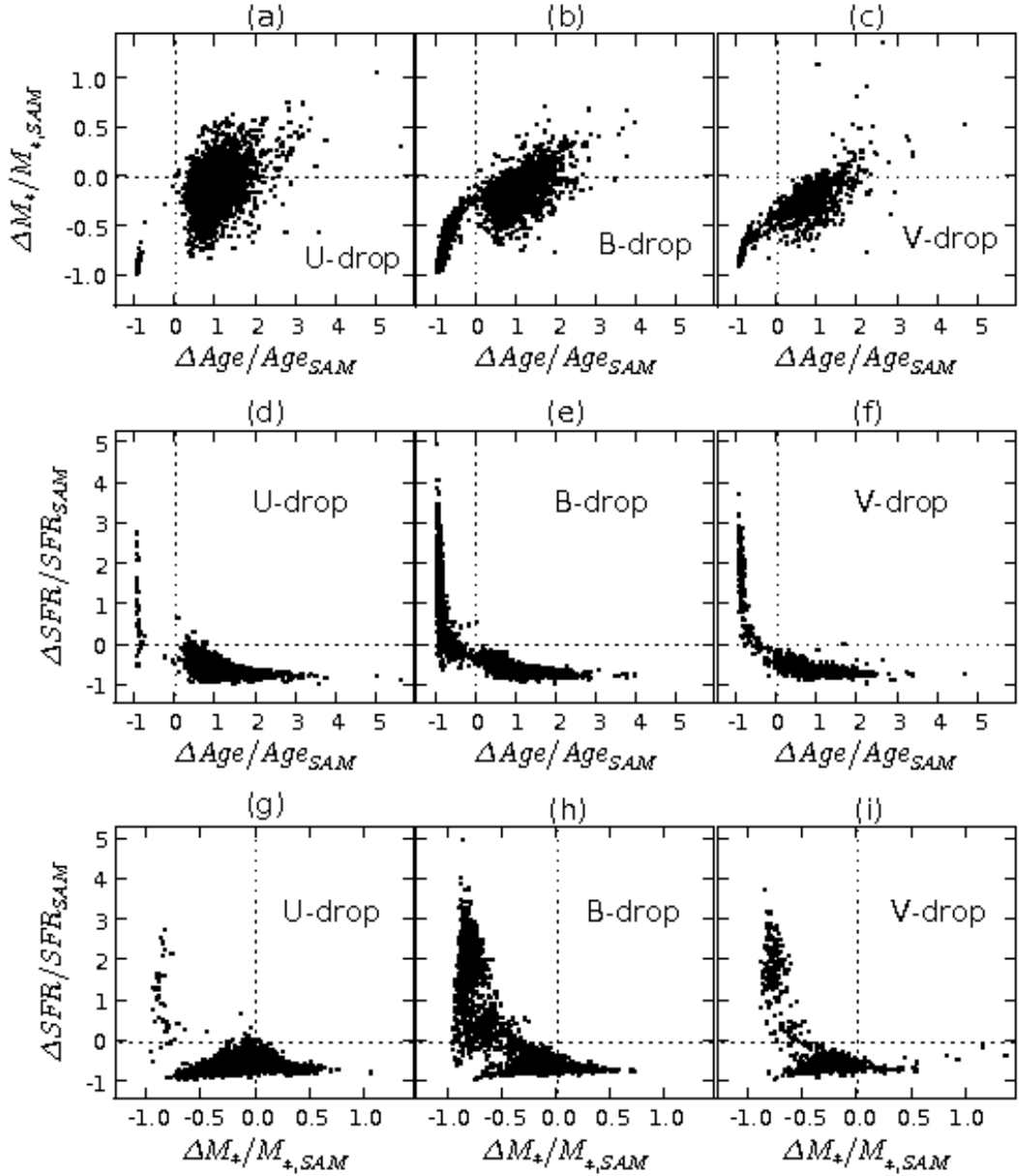}
\caption{Correlations among relative errors of stellar masses, SFRs, and 
mean ages for U-,B-,V-dropout galaxies.
$ACS$/$ISAAC$/$IRAC$ passbands are used and redshifts are allowed 
to vary. \label{errvserrzfree}}
\end{figure}

Figures \ref{uersamzfree}-\ref{versamzfree}, and figure 
\ref{errvserrzfree} are similar plots with figures 
\ref{uersamzfix}-\ref{versamzfix}, and \ref{errvserr}, 
but when redshift is an additional free parameter in the 
SED-fitting procedure.
These figures show trends similar to the redshift-fixed case 
except that subpopulations are more evident, especially 
for B- and V-dropouts.
The increased ambiguity due to the lack of redshift information 
evidently enlarges the subsets of galaxies that 
behave distinctly from the majority of galaxies in the SED-fitting. 
However, for the majority of galaxies, the lack of redshift 
information does not significantly affect the SED-fitting results, 
which is not surprising given the relatively small mean redshift 
errors in SED-fitting.

There is a subset of galaxies whose stellar masses are severely 
underestimated, SFRs are severely overestimated, and ages are 
severely underestimated.
These are similar trends with those shown by a subpopulation 
of B-dropouts when redshift is fixed. However, the number of galaxies 
which belong to this subpopulation substantially increases when 
the redshift is allowed to float as an additional free parameter.
For these galaxies, redshift is underestimated in the SED fitting. 
For U-dropouts, this sub-population is not as significant as for 
B- or V-dropouts.
            
Also, there are galaxies with high SSFRs/young ages which act 
differently if we perform the SED-fitting without fixing redshift. 
For these galaxies, ages are greatly overestimated, stellar masses 
are overestimated, and SFRs are greatly underestimated when we fix 
redshift in the SED-fitting (i.e. these are the galaxies with type-2 
SFHs). 
When we vary redshift freely as a free parameter, redshifts are 
slightly underestimated and ages/stellar masses are underestimated. 
The SFRs are similar to or slightly higher than the intrinsic values.

\subsection{Characteristics of Sub-populations in the Fitted Distributions} \label{subpop}

Figures \ref{bmgclrzfix}, \ref{bersamzfix}, and (b), (e), and (h) 
of figure \ref{errvserr} reveal the presence of a subpopulation 
of B-dropout galaxies whose behavior in the SED-fitting is 
distinct from the majority of galaxies.
For these galaxies, the ages are underestimated, the SFRs are 
overestimated, and the stellar masses are more severely 
underestimated than other galaxies. 
What makes the behavior of the galaxies in this subpopulation 
different from the majority of galaxies?

\begin{figure}
\plotone{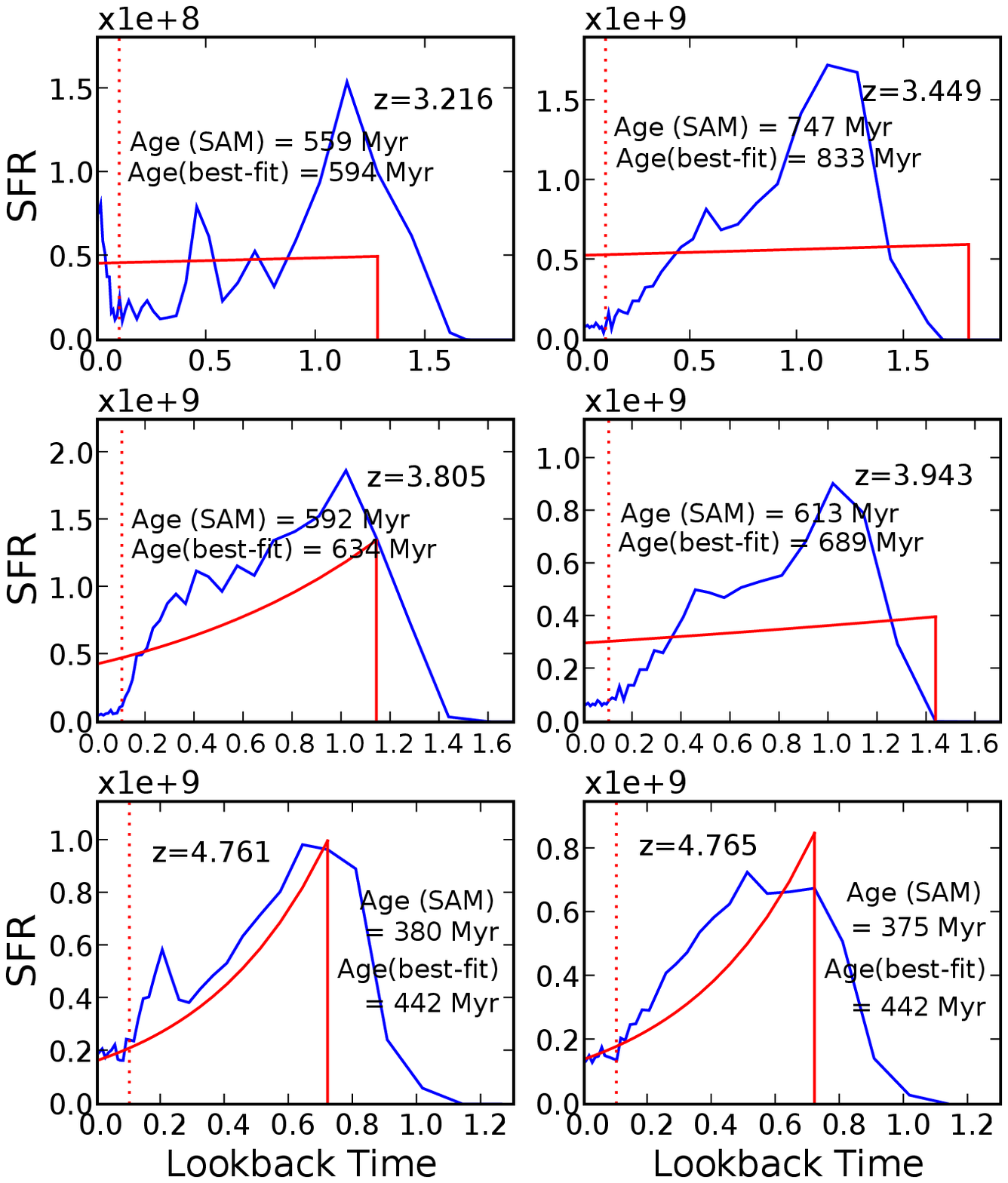}
\caption{Intrinsic star formation histories of typical dropout galaxies among the 
galaxies with the smallest overestimation of mean stellar population ages (blue line) 
along with star formation histories of the best-fit BC03 template (red line). 
These galaxies are the ones with type-1 SFH (see text). 
The red, dotted vertical line shows the point where lookback time is 100 $Myr$. 
SFR are measured over past 100 $Myr$ timespan in this work. 
Lookback time is given in $Gyr$\label{sfhrdagsm}} 
\end{figure}

\begin{figure}
\plotone{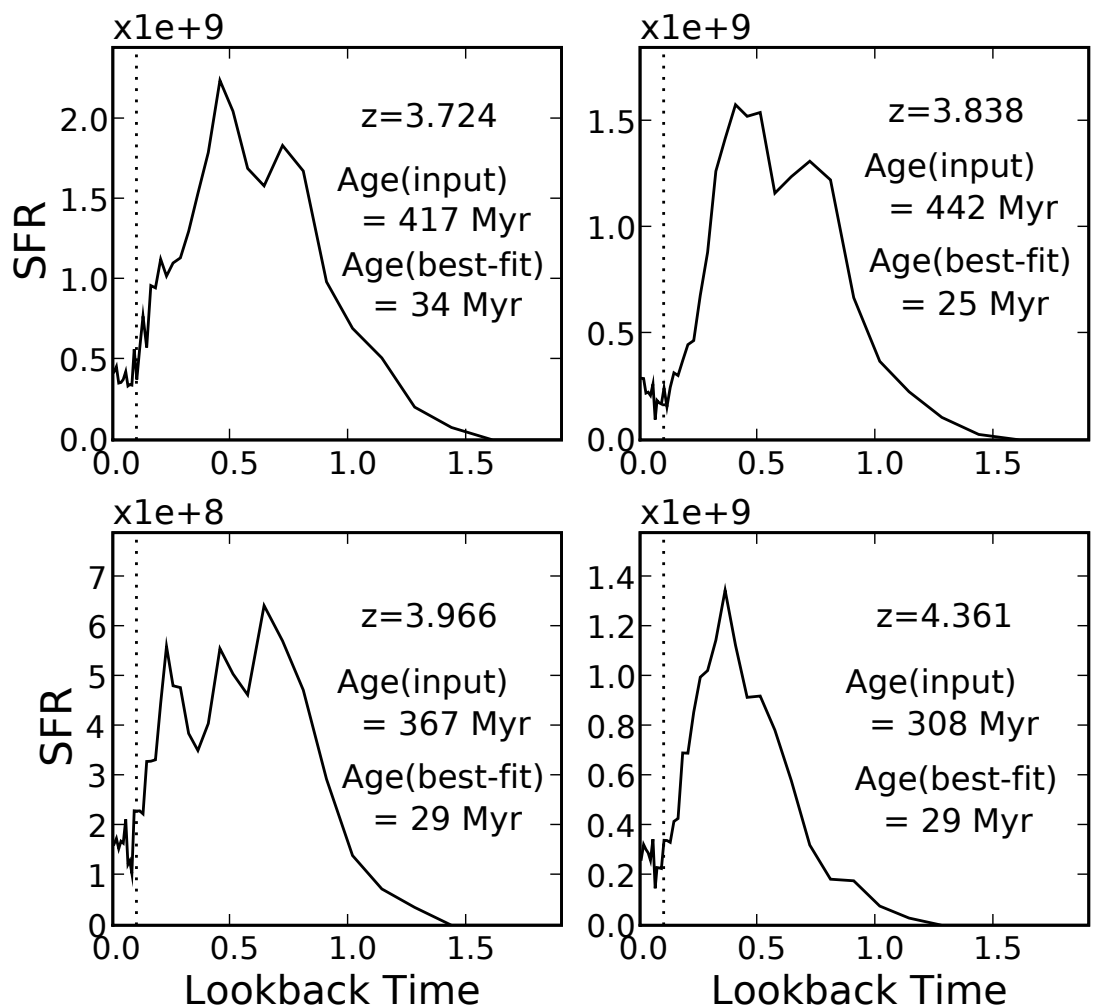}
\caption{Star formation histories of typical B-dropout galaxies with 
underestimated mean stellar population ages (type-3 SFH, see text). 
The dotted vertical line shows the point where lookback time is 100 $Myr$. 
SFR are measured over past 100 $Myr$ timespan in this work. 
Lookback time is given in $Gyr$. \label{sfhrdagneg}} 
\end{figure}

Because one of the main origins for biases in SED fitting is 
the SFH difference between the SAM galaxies and the BC03 templates, 
it is plausible that the SFHs of these galaxies are distinct 
from others.
Figure \ref{sfhrdagneg} shows SFHs of typical model galaxies in 
this subpopulation ('type-3' SFHs from now on). 
Generally, they have small SSFR values like those shown in 
figure \ref{sfhrdagsm}. 
However, the SF activity in figure \ref{sfhrdagneg} shows a 
slower increase and more rapid decrease with a peak at later 
time compared with the galaxies shown in figure \ref{sfhrdagsm}.
The gradual decreases of the SFRs shown in figure 
\ref{sfhrdagsm} are not significantly different from 
the exponentially decreasing SFRs assumed in the BC03 model, 
which makes the relative age errors small for these galaxies. 
On the other hand, for galaxies shown in figure \ref{sfhrdagneg}, 
the strong SF activity, which occurred relatively recently, 
dominates SEDs. Combined with the age-extinction degeneracy, 
this causes the mean ages to be severely underestimated, 
distinguishing behaviors of these galaxies in the SED-fitting.
The purple crosses in figure \ref{subagessfr} represent 
galaxies in this sub-population. They are not clearly 
distinguished from other galaxies with the type-1 SFH in this 
age-SSFR domain, but have, on average, slightly younger ages 
than the type-1 SFH galaxies with similar SSFR.  

Of course, the SFHs of all galaxies are not clearly divided 
into typical examples shown in figures \ref{sfhrdaglg}, 
\ref{sfhrdagsm}, or \ref{sfhrdagneg}. 
For example, when only $ACS$ and $IRAC$ fluxes are used 
(i.e. if smaller number of passbands are used; see 
\S~\ref{woisaac}), more galaxies behave similarly to 
galaxies with SFH type-3, making the bimodal distributions 
of SFRs and ages more prominent.

\section{Effects of Parameter Changes on the Results of the SED-fitting} \label{parachange}

In the following sections, we investigate how the bias in the 
SED fitting behaves as we change some of the conditions in the 
SED-fitting procedure, such as the range of e-folding time of 
star formation history, $\tau$, combinations of broad passbands 
used, or the assumed SFH. 
 
\subsection{Two-Component Fitting} \label{2comp}

Here, we try to allow more complex star formation histories in 
the SED-fitting, by using the two-component templates instead of 
single component ones. Some studies tried this method to constrain 
the hidden mass in old stellar population or for better estimation 
of total stellar mass.
To construct the two-component stellar population templates, various 
combinations of simple SFHs have been tried in the literature, 
including: (1) combining a maximally old, instantaneous burst 
with a more moderately decreasing SFH \citep[e.g.][]{pap01,sha05}, 
or (2) adding a secondary young bursty SFH component to an old, slowly 
decreasing SFH \citep[e.g.][]{kau03,dro05,poz07}.
Also, different authors have used different ways of fitting two 
components: 
(1) fitting with a young SF component first, then fitting the 
residual SED with an old component \citep{sha05}, or 
(2) constructing the combined templates with various ratios 
between a young and an old simple SFH templates \citep{kau03,poz07}.      

\subsubsection{Slowly Decreasing Star Formation Histories with a Secondary Burst} \label{2cset1}

Here, we constructed the two-component SFH templates by adding 
(maximally) old, very slowly varying SFH templates ($\tau$ = 
15 Gyr), and younger, more bursty templates ($\tau$ = 0.2 Gyr). 
Old components are assumed to start forming at $z_{f}$ = 10. 
The star formation activity of the secondary burst is constrained 
to be initiated at least 200 $Myr$ later than that of the old 
component, and at most 500 $Myr$ earlier than the observed time, 
with the percentage of young templates varying between 5 $\%$ 
to 95 $\%$ (with 5 $\%$ step size). 
This construction is expected to reflect better the SFHs of some 
galaxies in the semi-analytic models used in this study (for example, 
galaxies with type-2 SFHs shown in figure \ref{sfhrdaglg}), and using 
this type of composite templates ought to reduce the systematic bias 
for galaxies with type-2 SFHs. 
The SFHs constructed in this way can mimic the SFHs of galaxies that 
experienced the secondary star formation due to merger/interaction.

\begin{figure}
\plotone{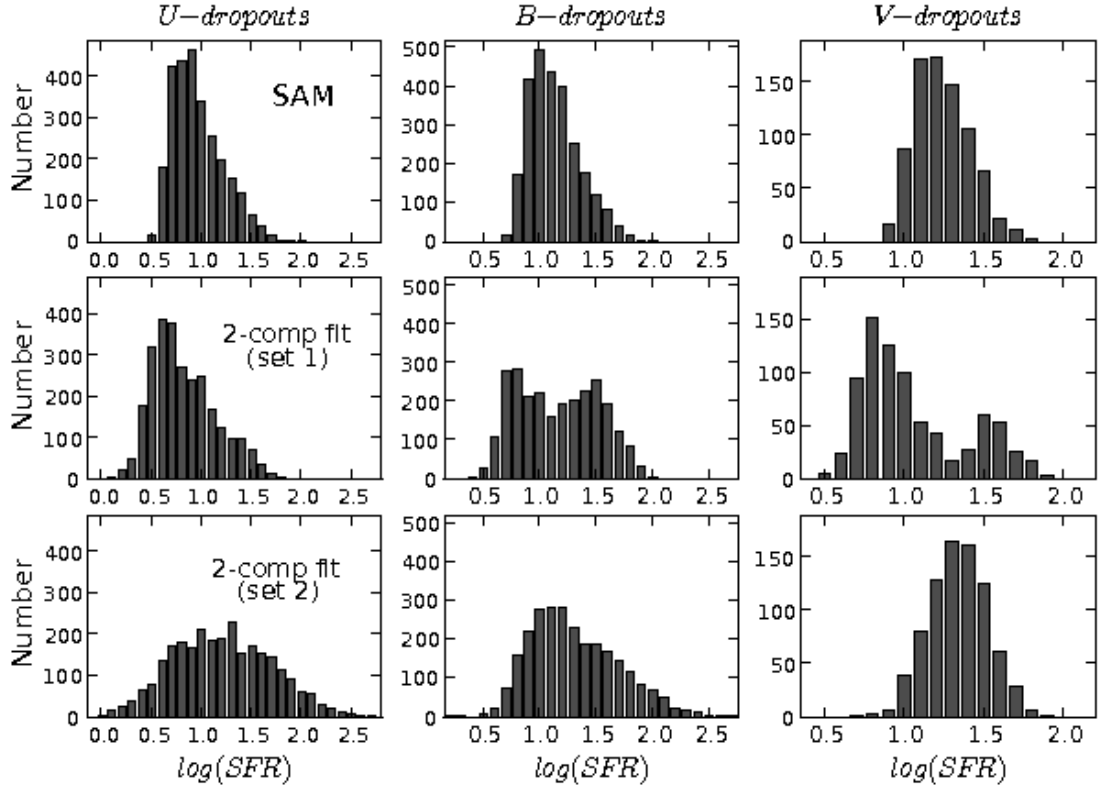}
\caption{SFR distributions of U-dropouts (left column), B-dropouts 
(middle column), and V-dropouts (right column). Figures in the top row show 
intrinsic SFR distributions. The middle row is for SFRs from the two-component 
fitting with a secondary young burst added to a main component (\S~\ref{2cset1}), 
and the bottom row is for SFRs from the two-component fitting with a maximally old burst 
component plus a younger, long-lasting component (\S~\ref{2cset2}). SFRs are given 
in $M_{\sun}$ $yr^{-1}$. \label{2csfrdst}}
\end{figure}

\begin{figure}
\plotone{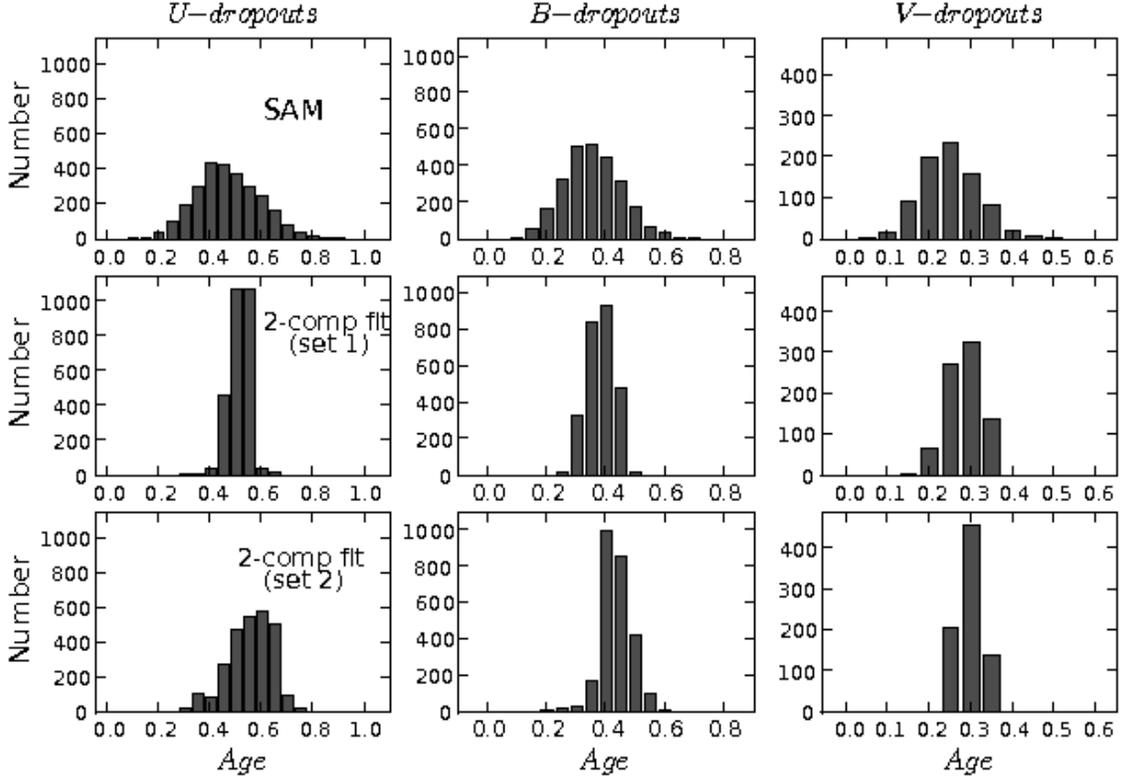}
\caption{Mass-weighted stellar-population mean age distributions of U-dropouts 
(left column), B-dropouts (middle column), and V-dropouts (right column). 
Figures in the top row show intrinsic age distributions. The middle row is for ages 
from the two-component fitting with secondary a young burst added to a main component 
(\S~\ref{2cset1}), and the bottom row is for ages from the two-component fitting with 
a maximally old burst component plus a younger, long-lasting component(\S~\ref{2cset2}). 
Age is given in $Gyr$. \label{2cagedst}}
\end{figure}

The trends in the SED-derived distributions of various parameters in 
this type of two-component fitting are: 
(1) the stellar masses are underestimated more severely than in the 
single-component fitting (middle row of figure \ref{2cmsdst}), 
(2) the SFR/age distributions show reduced offsets compared with 
the single-component fitting (middle rows in figures 
\ref{2csfrdst} and \ref{2cagedst}), and 
(3) the bimodalities that existed in the age distributions have 
disappeared, while the bimodalities in the SFR distributions are 
enhanced for B- and V-dropouts (for U-dropouts, the SFR distribution 
becomes much broader). These bimodalities in the SFR distributions are, 
however, not driven by the subpopulation of galaxies with type-3 
SFHs, as can be seen below. 

The first two trends are not unexpected since we are adding a young 
stellar population with a low mass-to-light ratio to the SED templates. 
This young component makes the best-fit ages younger, thereby reducing 
bias in the mean ages, while its lower mass-to-light ratio makes the 
best-fit stellar masses smaller, thereby increasing bias. 
Younger ages, and thus increased dust extinctions, lead to higher SFRs, 
and decrease bias in the SFRs.

\begin{figure}
\plotone{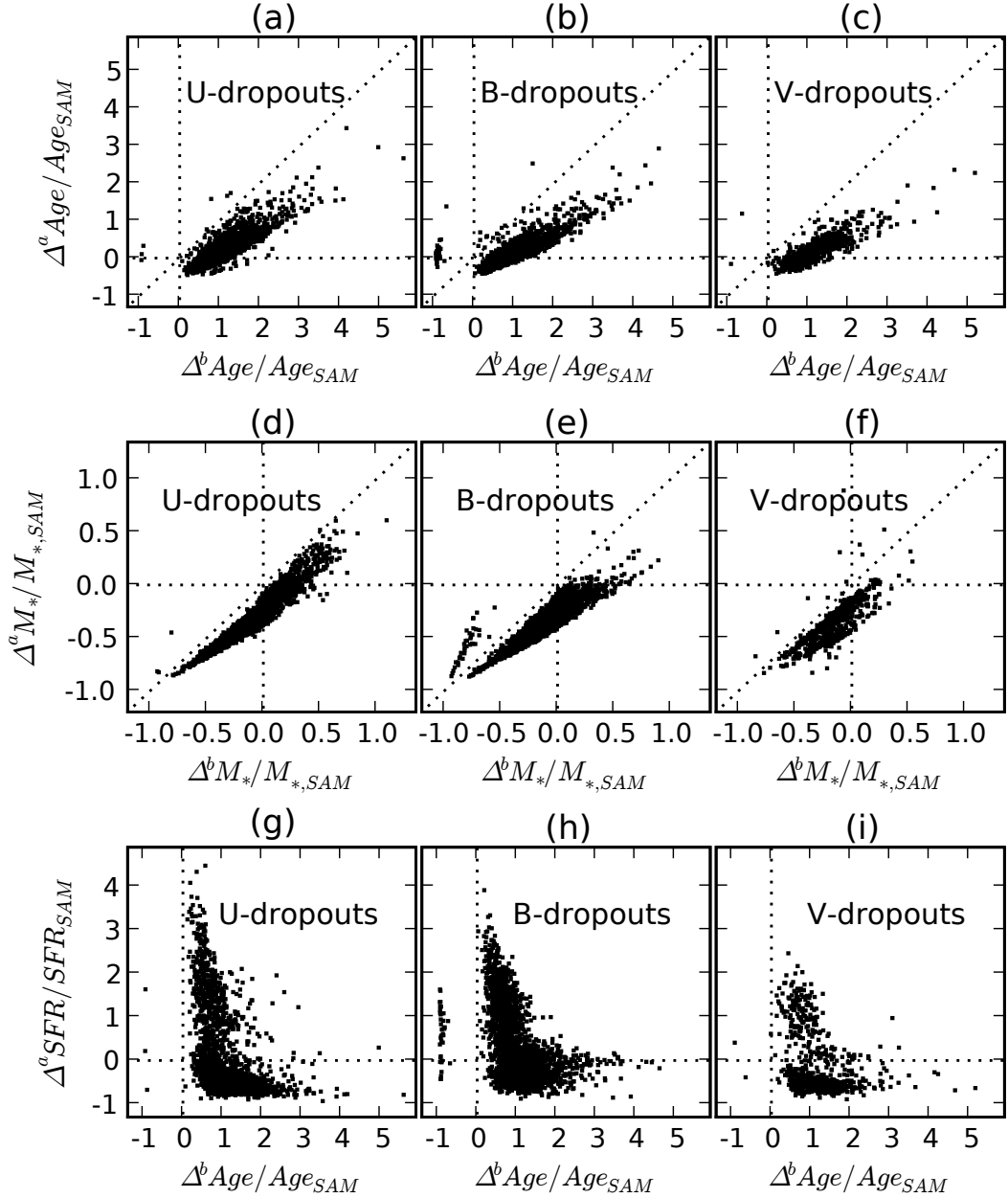}
\caption{Correlations between relative errors in two-component fitting with a 
secondary young burst (\S~\ref{2cset1}) and relative errors in single-component 
fitting. Figures in the top row show correlations between relative age errors 
in two-component fitting and relative age errors in the single-component 
fitting. The middle row is for correlations between relative stellar mass errors in 
the two-component fitting and relative stellar mass errors in the single-component 
fitting. The bottom row is for correlations between relative SFR errors in 
the two-component fitting and relative age errors in the single-component fitting.
$\Delta^{a} value$ represents `\it{(value from two-component fitting) - 
(intrinsic value)}\rm{'}, and $\Delta^{b} value$ represents `\it{(value from 
single-component fitting) - (intrinsic value)}\rm{'}. \label{rd2cs1vsrd1c}}
\end{figure}

Figure \ref{rd2cs1vsrd1c} -- showing the correlation between the 
relative errors arising in the two-component fitting and the relative 
errors in the single-component fitting -- reveals behaviors of galaxies 
with different SFH types in the two-component fitting performed in 
this section.
Galaxies whose ages are greatly overestimated in the single-component 
fitting (with type-2 SFHs) show significantly reduced age-overestimation 
in the two-component fitting.
This improvement in the age estimation is expected.
In the single-component fitting, the ages are greatly overestimated for 
the SFH type-2 galaxies because the young component is ignored. 
By adding a young component in the templates, ages are better fitted for 
this type of galaxies improving the age estimation. 
For these galaxies, the stellar masses are generally overestimated in 
the single-component fitting. Improving the age estimates also improves 
the stellar-mass estimates

However, when the age is relatively well constrained in the 
single-component fitting (i.e. for galaxies with type-1 SFHs), an 
added young component causes the underestimation of ages, and therefore 
more severe underestimation of the stellar mass which is already 
underestimated in the single-component fitting. The age underestimation of 
these galaxies couples with the age-extinction degeneracy, leading to the 
overestimation of the SFRs shown in (g), (h), and (i) in figure \ref{rd2cs1vsrd1c}. 
This causes the bimodalities in the SFR distributions shown in middle row of 
figure \ref{2csfrdst}.

Interestingly, for a small number of galaxies, the ages and the stellar 
masses derived in the two-component fitting are higher than the ones 
derived in the single-component fitting.
These galaxies are the ones with SFH type-3 and are manifested as a 
subpopulation in figures \ref{bmgclrzfix} and \ref{bersamzfix}. 
Through the single-component fitting, ages and stellar masses of 
these galaxies are greatly underestimated, and the SFRs are severely overestimated 
due to the large age underestimation. The higher values of the ages 
and stellar masses derived through the two-component fitting reduce the 
errors in the age and stellar mass for these galaxies.
For these galaxies with the type-3 SFHs, the best-fit $t$s are small in 
the single-component fitting. So, the actual effect of the two-component 
fitting performed in this section is to add an old component for these galaxies, 
while for the other galaxies (with the type-1 or type-2 SFHs), the the effect of 
the two-component fitting is to add a young component. 
This makes the best-fit ages from the two-component fitting older than the ones 
derived from the single-component fitting, the best-fit masses higher (due to 
higher mass-to-light ratios), and the best-fit SFRs lower for galaxies with type-3 
SFHs.
Older best-fit ages of the type-3 SFH galaxies derived in the two-component fitting 
removes the bimodalities shown in the SED-derived age distributions in the 
single-component SED-fitting. 

The age distributions derived in the two-component fitting are 
narrower than the intrinsic distributions (middle row of figure 
\ref{2cagedst}). The ages tend to be underestimated for older galaxies 
(possibly with SFHs of type-1 or type-2), while ages are more 
likely overestimated for younger galaxies (probably SFH type-3 
galaxies). 
The SFR distributions derived in the two-component fitting are more 
extended toward the high SFRs than in the single-component fitting for 
U-dropouts (bringing the SFR distributions closer to the intrinsic 
distributions). 
For B- and V-dropouts, the bimodality in the SFR distribution is 
enhanced, i.e. the SFRs are overestimated for more galaxies.   

In summary, the two-component fitting performed in this section 
reduces bias in the SFR and age distributions, but increases the offsets 
in the stellar mass distributions. 
However, the detailed investigation reveals that the changes of behavior 
in the two-component fitting compared with the case of the single-component 
fitting are different for galaxies with different types of SFH. 
Errors in the estimation of ages and stellar masses are reduced for galaxies 
with SFHs type-2 or type-3, while the stellar mass errors increase for 
galaxies with type-1 SFHs.

\subsubsection{Maximally Old Burst Combined with Slowly Varying Younger Components} \label{2cset2}

In the previous section, we experimented with two-component fitting 
by adding a young, burst-like ($\tau = 0.2$ Gyr) component to a more 
continuously varying ($\tau = 15$ Gyr) old component. 
Such two-component templates are expected to match better the type-2 
SFHs, and turned out to give better age estimates for the galaxies with 
type-2 as well as type-3 SFHs.

In this section, we perform another type of two-component fitting in 
an attempt to give a better constraint on the hidden old stellar mass. 
To achieve this we construct the two-component SFH templates in a similar 
way done as in \S~\ref{2cset1}, but exchange the roles of a 
$\tau = 0.2$ Gyr component and a $\tau = 15$ Gyr component. 
We add an old, $\tau = 0.2$ Gyr component formed at $z_{f} = 10$ 
and younger $\tau = 15$ Gyr components with various ages. 
The star formation activity of the young components is constrained 
to start at least 200 Myr later than an old burst, to make the 
two-component templates clearly distinguished from the 
single-component template. 
By adding an old, bursty component to a more continuously varying 
SFH component, we can expect that this method will give us higher 
mass than the single-component fitting. 

\begin{figure}
\plotone{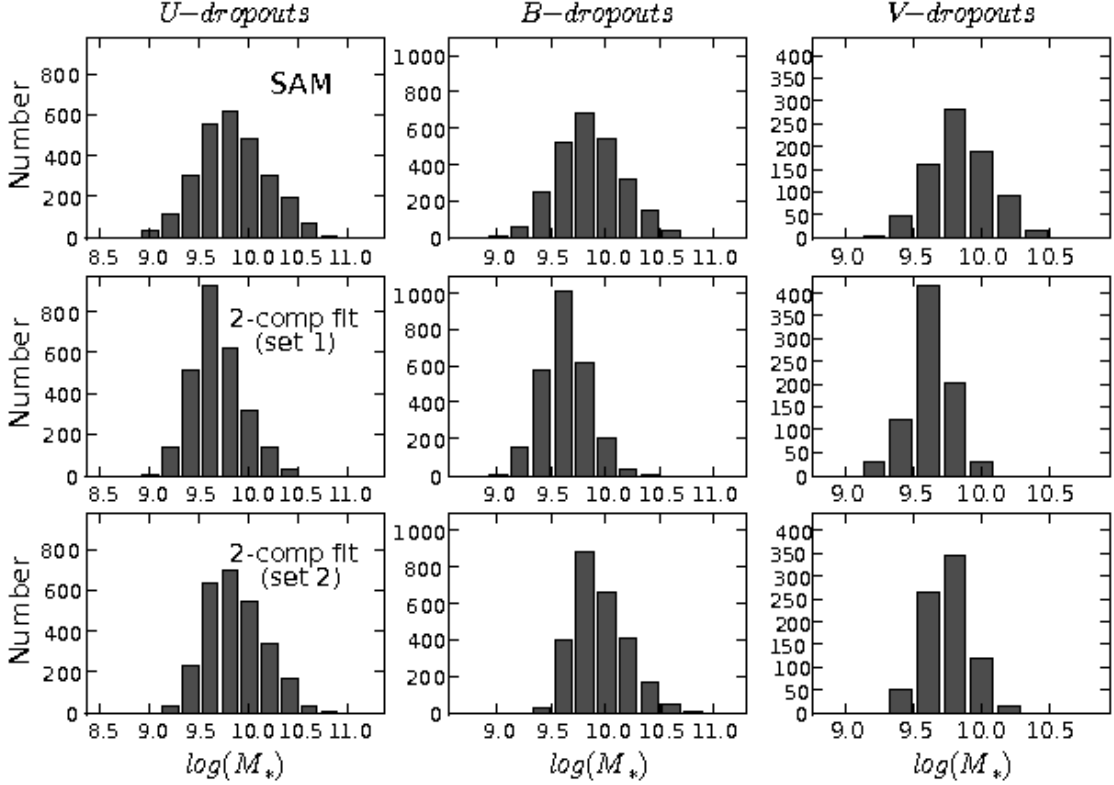}
\caption{Stellar mass distributions of U-dropouts (left column), B-dropouts 
(middle column), and V-dropouts (right column). Figures in the top row show 
the distributions of intrinsic stellar masses. The middle row is for stellar masses 
from the two-component fitting with a secondary young burst added to a main component 
(as investigated in \S~\ref{2cset1}), and the bottom row is for stellar masses 
from the two-component fitting with a maximally old, burst component plus a younger, 
long-lasting component (\S~\ref{2cset2}). Stellar masses are given in 
$M_{\sun}$. \label{2cmsdst}}
\end{figure}

The bottom row of figure \ref{2cmsdst} shows that the stellar mass
distributions are moved toward higher values than the 
ones derived in \S~\ref{2cset1} (middle row of figure \ref{2cmsdst}) 
and also than the ones from the single-component fitting. 
Compared with the intrinsic distributions, the derived stellar mass 
distributions from the two-component fitting with an additional old, 
burst-like component are slightly crowded at high stellar mass for 
U- and B-dropouts.
The bimodalities both in the SFRs and ages have disappeared (bottom 
rows of figures \ref{2csfrdst} and \ref{2cagedst}).

The mean values of stellar masses are higher by $\sim 19$ $\%$, 
54 $\%$, and 1 $\%$ than the values from the single-component 
fitting for U-, B-, and V-dropouts, respectively. 
For individual galaxies, the stellar mass from the two-component 
fitting with an old, burst-like component can be as large as 
several times of the stellar mass from the single-component fitting. 
For a few galaxies (mostly with type-3 SFHs), the stellar mass 
from the two-component fitting with an old burst can reach $\sim 4-9$ 
times of the stellar masses from the single-component fitting. 
However, the relatively small increase in the mean values of the 
best-fit stellar mass (especially for V-dropouts) indicates that 
the young component dominates the SEDs even in the two-component 
fitting performed in this section.

The dominance of the young component can be seen by the fact that 
the mean values of the best-fit ages derived in the two-component 
fitting are much younger (by $\sim 39$ $\%$, 35 $\%$, and 42 $\%$ 
for U-, B-, and V-dropouts) than the mean values derived in the 
single-component fitting. 
(The star-formation time scale of the young component in the 
two-component fitting performed in this section is fixed as 
$\tau = 15$ Gyr, and this leads to younger best-fit ages -- see
 \S~\ref{tau15}, below.)

Even though the two-component fitting performed in this section gives 
higher values of the mean stellar mass, it is not always true that 
stellar masses derived from the single-component fitting and this kind 
of two-component fitting bracket the true, intrinsic stellar mass.
For some galaxies with small stellar mass (with $log$ 
($M_{*} / M_{\sun}$) $\lesssim$ 9.5), even the single-component 
fitting overestimates the stellar mass. 
On the other hand, the stellar mass derived through the two-component 
fitting with an old, burst-like component often remain smaller than 
the intrinsic value for some massive galaxies.

If we were to limit the fractional contribution of the young 
component to smaller values than allowed here -- as done, for example, 
in \citet{kau03} -- the stellar masses derived in the two-component 
fitting would become higher than the ones derived in this section. 
Also, we can derive higher stellar masses from the two-component 
fitting: 
(1) by fitting the old component first, then fitting the younger 
component to the residual fluxes (i.e. forcing the contribution 
from the old component to increase), or 
(2) by setting the formation redshift ($z_{f}$) of the old 
component higher (i.e. increasing the mass-to-light ratio of the 
old stellar component). 
However, even with these more extreme settings, it is still possible 
that the derived stellar masses for some very massive galaxies will 
be smaller than the intrinsic ones.  

In summary, it is not universally true that the single-component fitting 
and the two-component fitting (with an old, burst component added on 
more continuous SFH components) bracket the true stellar mass.

\subsection{Effects of Wavelength Coverage} \label{bands}

The main results presented in \S~4 are based on the analysis 
using broadband photometric information from observed-frame 
optical through MIR range -- i.e. $ACS$ $B_{435}$- to 
$z_{850}$-bands, $ISAAC$ $J$- to $Ks$-bands, $\&$ $IRAC$ 
3.6 $\mu m$ through 8.0 $\mu m$.
However, not all the observed LBGs have photometric data with 
this wavelength coverage. 
Before the $Spitzer$ era, the majority of the observed 
photometric data only covers up to the observed-frame NIR range. 
Thus, it is interesting and important to examine how the results 
vary as we use different combinations of passbands in the SED-fitting. 

\subsubsection{SED-fitting without IRAC Data} \label{woirac} 
    
Several authors investigated the effects of inclusion of IRAC 
photometry (of wavelength coverage of $\sim$ 3-10 $\mu m$) 
in constraining the properties of high-redshift 
galaxies \citep{lab05,sha05,wuy07,els08}. 
\citet{sha05} analysed $z \sim 2$ star-forming galaxies with 
and without IRAC photometric data. 
They reported that the SED-derived stellar mass distribution 
shows little change with the inclusion of IRAC data, 
while including IRAC data can reduce errors in the stellar mass 
estimation for individual galaxy. 
Investigating 13 $z \sim 2-3$, red ($J_s - K_s > 2.3$) galaxies, 
\citet{lab05} showed that the best-fit ages are younger without 
IRAC data for dusty star forming galaxies, while there is little 
change for old, dead galaxies.
\citet{wuy07} studied $2 < z < 3.5$, $K$-selected galaxies, and 
showed that inclusion of IRAC data does not change the overall 
distributions of stellar masses and ages.
Analyzing $0 < z < 5$ observed galaxies, \citet{els08} showed that 
the mean stellar masses increase when derived omitting $Spitzer$ data. 
The discrepancy is maximum at $z \sim 3.5$ with 
$log (M_{U-4}/M_{U-K}) \sim -0.5$, and decreases with redshift 
at $z > 3.5$. (At $z \leq 3.0$ or $z \geq 4.0$, 
$log (M_{U-4}/M_{U-K}) \leq -0.3$.)\footnote{$M_{U-4}$ and $M_{U-K}$ 
  are stellar masses that are derived with and without IRAC data, 
  respectively.}
In addition, they reported that despite this overall trend, the 
stellar masses and mean ages decrease without IRAC data for some 
very young (faint) galaxies. 
Thus, there is no clear consensus in the literature on the 
benefits of including IRAC.

The effects of omitting IRAC data are shown in the third rows of  
figures \ref{udropdsts}, \ref{bdropdsts}, and \ref{vdropdsts}).
Without IRAC data, the stellar mass estimates shift to lower values 
for U- and B-dropouts, but are virtually unchanged for V-dropouts. 
Conversely, the SFR estimates are virtually unchanged for U- and 
B-dropouts, but shift to lower values (increasing the discrepancy 
with the SAMs) for V-dropouts.
The age distributions shift to younger ages for the U- and 
B-dropouts (which actually brings them into better agreement with 
the intrinsic distributions) while for V-dropouts, the age 
distribution is only slightly changed.
 
Evidently, removing the IRAC photometry for the U- and B-dropouts 
increases the dominance of the younger stellar populations due to 
the shorter wavelengths.
This drives the best-fit ages to lower values, resulting in lower 
stellar masses for a given amount of stellar light. 
The stellar masses are reduced by $\sim 57$ $\%$ and $\sim 48$ $\%$ for 
U- and B-dropouts, respectively, relative to the results from fits, 
in which the IRAC data are included. 
 
The age and stellar mass decreases are largest for the galaxies 
with the high SSFRs and young ages, i.e. galaxies with type-2 SFHs 
(similar with the ones shown in figure \ref{sfhrdaglg}).
These galaxies have roughly two components of stellar populations 
-- an `old', slowly varying component and a `young', burst-like 
component. In the SED-fitting with IRAC photometry included, the 
`old' component dominates the SED, resulting in much older 
best-fit ages than the intrinsic ages. 
However, without IRAC photometry, the SEDs cover only up to the 
rest-frame $\sim 4000-5000$ $\rm{\AA}$. Due to the resulting 
shortage of information at long wavelengths, the `young' component 
comes to dominate the SED. This moves the best-fit ages younger, 
even younger than the intrinsic ages in extreme cases.   

The changes of the best-fit age and stellar mass for V-dropouts are 
much smaller ($\sim 13$ $\%$ and $\sim 4.8$ $\%$ of decreases, 
respectively). This is presumably due to the generally younger ages 
of V-dropouts.
As can be seen in figures \ref{sfhrdaglg} and \ref{sfhrdagsm}, 
V-dropouts, on average, started forming stars more recently than U- 
and B-dropouts. 
Therefore, the proportion of the old stellar population hidden without 
IRAC data is much smaller for V-dropouts than U- and B-dropouts.

Interestingly, the bimodalities in the SFR- and age-distributions 
shown (for B-dropouts) in figures \ref{sfrdstzfix} and \ref{agedstzfix} 
disappear when we exclude IRAC photometry.
For the type-3 SFHs shown in figure \ref{sfhrdagneg}, the SFRs have 
lower values and the ages (and stellar masses) have higher values 
than the ones derived including IRAC bands. 
The behaviors of these galaxies are thus opposite to the majority of B-dropouts.

What makes this sub-population of galaxies (with type-3 SFHs) behave 
differently from other galaxies?
The difference of the SFHs between the ones shown in figure \ref{sfhrdagneg} 
(i.e. type-3 SFHs) and the ones shown in figure \ref{sfhrdagsm} (i.e. 
type-1 SFHs) becomes significant when the lookback time is larger than 
$\sim 400-500$ Myr.
With only ACS and ISAAC photometry, which covers only up to the rest-frame 
$\sim 4000-5000$ $\rm{\AA}$, the SFHs at early time are hard to constrain. 
Therefore, the SED-fitting without IRAC data cannot discriminate between 
SFH type-1 (figure \ref{sfhrdagsm}) and type-3 (figure \ref{sfhrdagneg}), 
resulting in the disappearance of the bimodalities in the SFR and age 
distributions.   

\begin{figure}
\plotone{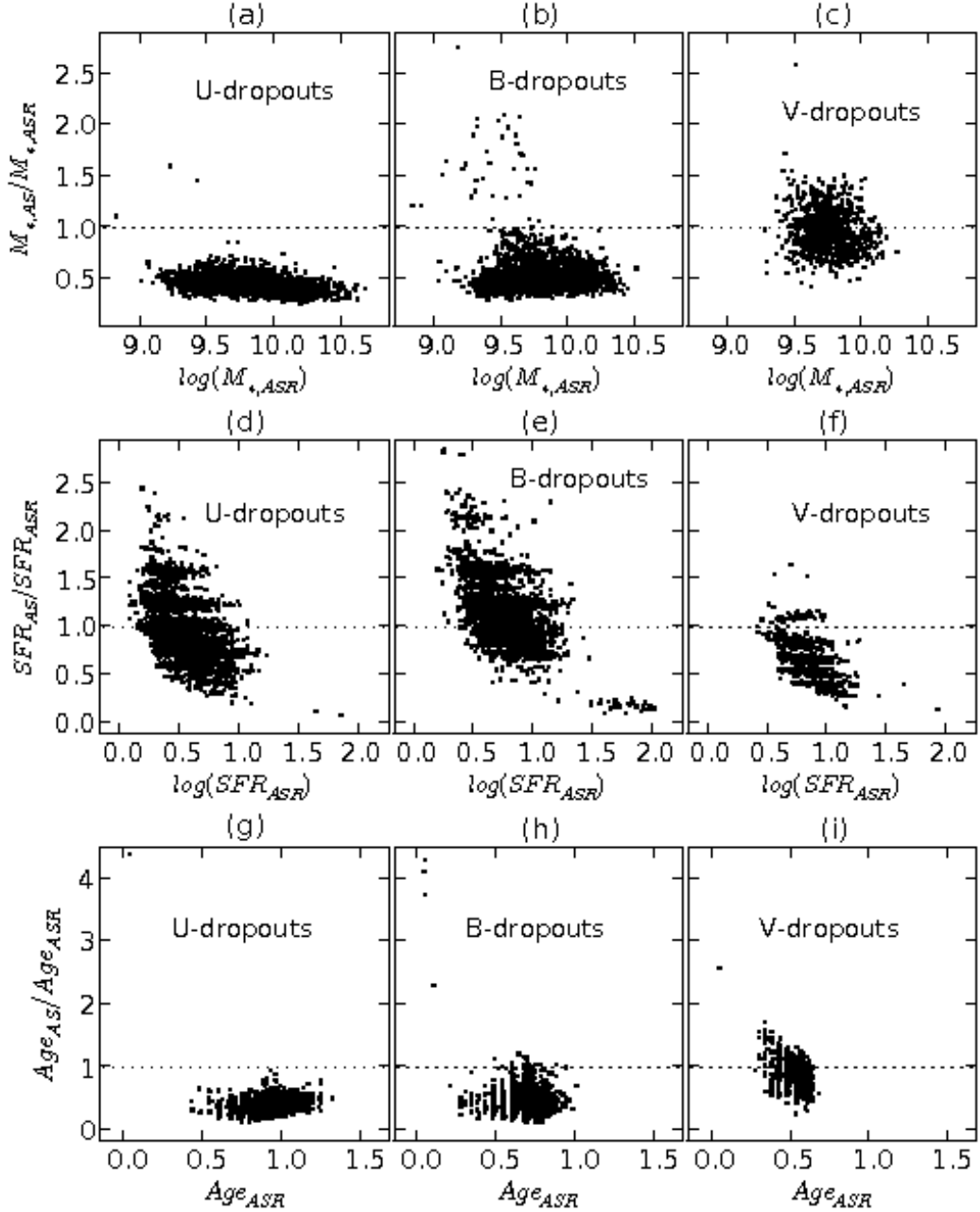}
\caption{Ratio of the best-fit stellar masses ((a)-(c)), SFRs ((d)-(f)), and mean 
ages ((g)-(i)) with and without IRAC photometry for U-dropouts ((a), (d), and 
(g)), B-dropouts ((b), (e), and (h)), and V-dropouts ((c), (f), and (i)) dependent 
on the best-fit values with IRAC photometry. 
`AS' stands for values without IRAC photometry, and `ASR' for values with IRAC 
photometry. Objects with $Age_{AS} / Age_{ASR} > 4.5$ are excluded from the figures in 
the bottom row for visual clarity. \label{as121}}
\end{figure}

Figure \ref{as121} shows the ratios of the best-fit stellar masses, SFRs, 
and ages with and without IRAC photometry for U-, B-, and V-dropouts.
Without IRAC photometry, the stellar masses are underestimated 
(compared with when IRAC data are included) for most U-dropouts 
(figure \ref{as121}-(a)), and for the majority of B-dropouts 
(figure \ref{as121}-(b)). For the subpopulation of galaxies in 
B-dropouts, the best-fit stellar masses without IRAC data are 
larger than the ones with IRAC data. 
For V-dropouts (figure \ref{as121}-(c)), the stellar masses are 
underestimated for some galaxies, and overestimated for other 
galaxies, making the distributions with and without IRAC data similar.

In summary, the effect of removing the IRAC data depends on the redshift 
and/or SFHs. 
This redshift and SFH dependence can explain the apparent disagreement 
between the different previous investigations, since the samples 
included galaxies with different redshift ranges and also different types 
of galaxies with possibly different SFHs.

\subsubsection{SED-fitting without ISAAC data} \label{woisaac}

Next, we perform the SED-fitting with only $ACS$ and $IRAC$ photometry, 
excluding the $J$, $H$, and $Ks$-band $ISAAC$ photometry. 
The effect of omitting $ISAAC$ photometry is insignificant 
for U-dropouts and for the majority of B- and V-dropouts. 
As can be seen by comparing the second and the fourth rows 
from the top in figure \ref{udropdsts}, the distributions of 
best-fit stellar masses, SFRs, and ages show little change 
without $ISAAC$ data for U-dropouts. 
The mean stellar mass and mean age increase by $1.9 \%$ and $2.4 \%$ 
compared with the values derived using $ACS+ISAAC+IRAC$ photometry.

For B- and V-dropouts, the bimodalities in the SFR/age distributions 
become more prominent as can be seen in the fourth rows of 
figures \ref{bdropdsts} and \ref{vdropdsts}. 
For some galaxies whose age offsets are small when the SED-fitting is done 
with full photometry, the best-fit ages become much younger if we use only 
$ACS$ and $IRAC$ photometry. This age underestimation leads to the stellar 
mass underestimation and SFR overestimation as explained in 
\S~\ref{intrinsic}, and the extinction overestimation due to the well-known 
age-extinction degeneracy enhances the SFR overestimation further.
These galaxies skew the mean values of stellar mass and mean age of total sample 
lower by about 10 $\%$ for B- and V-dropouts, even though the majority of 
galaxies show little change.

\subsection{Effects of $\tau$-range Used in SED-fitting} \label{parachoice}

The changes of the allowed ranges of parameters, such as $\tau$, $t$, or 
metallicity, in the SED-fitting would affect the derived values of physical 
parameters, as well. Here, we focus on the effects of the different range
of $\tau$s used in the SED-fitting on the estimation of physical parameters 
of LBGs. This investigation is beneficial for the comparison with previous 
works done using the SED-fitting methods with various $\tau$ ranges as well 
as for better understanding biases of the SED-fitting methods.

\subsubsection{SED-fitting with $\tau \leq 1.0$ Gyr Templates} \label{taule1}

First, we limit the $\tau$ range to $\leq 1.0$ Gyr during the SED-fitting.
Through this experiment, we can look into the biases arising due to the 
usage of `not-long-enough' $\tau$ values in the SED-fitting.

As expected, the enforced smaller $\tau$ values cause smaller best-fit $t$, 
to match $t$/$\tau$ values, compared with the case when the full range of $\tau$ 
is allowed from 0.2 Gyr to 15 Gyr. This bias systematically makes the 
best-fit ages to be younger.
As explained in \S~\ref{intrinsic}, this age-underestimation leads to the 
mass-underestimation, increasing differences between the intrinsic- 
and SED-derived stellar masses.
The effects of the limited $\tau$ values on the SFR estimation is complicated 
due to the age-extinction degeneracy.
The offsets due to the limited $\tau$ values as $\leq 1.0$ Gyr are 
greatest for B-dropouts because of the severely enhanced bimodalities.   
Relative changes of the mean stellar masses and ages when we limit 
$\tau$ range to $\leq 1.0$ Gyr, compared with the mean values derived 
with the full range of $\tau$s, are shown in table.~\ref{tauofftab}.

\begin{figure}
\plotone{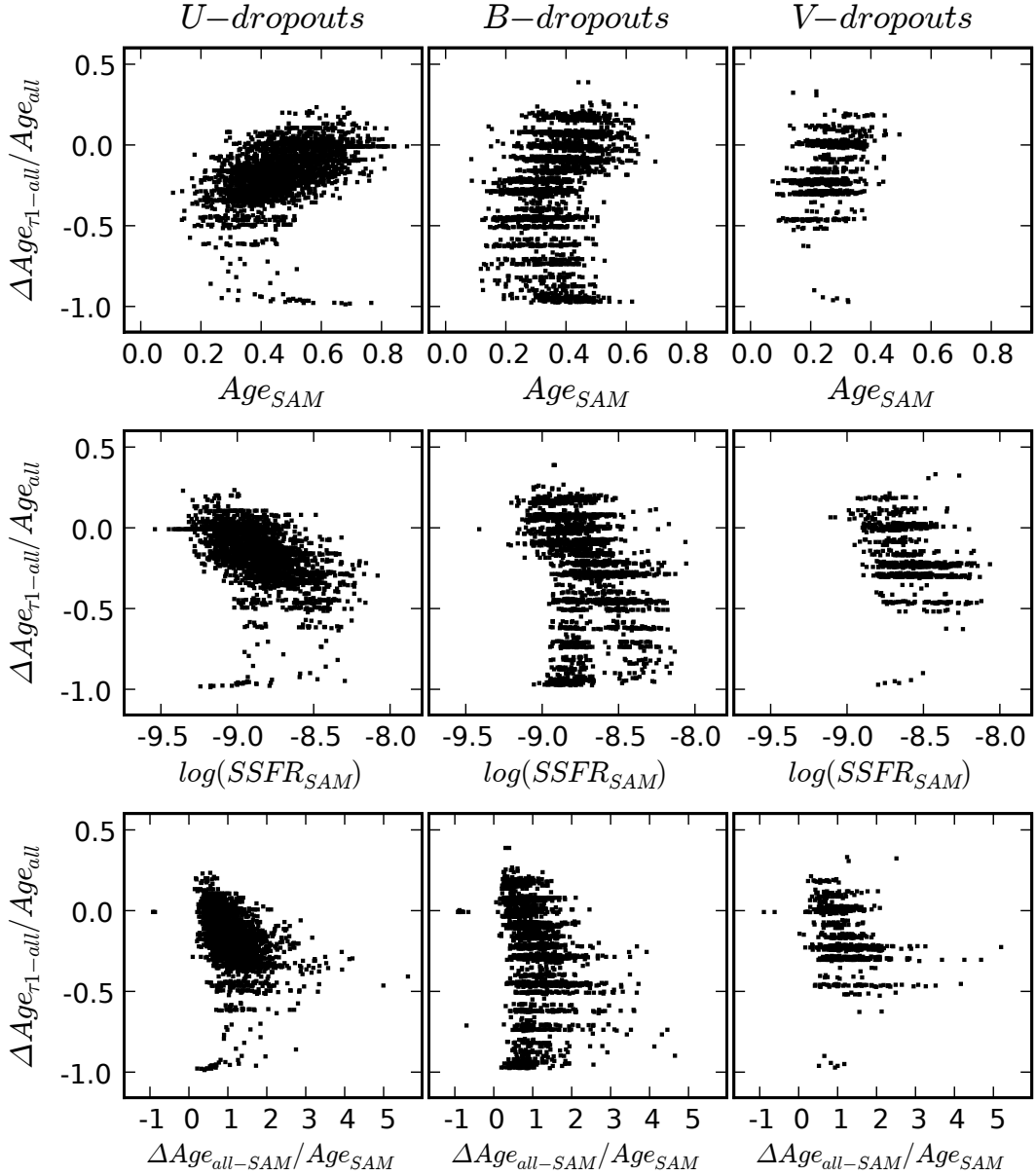}
\caption{$(Age_{\tau 1} - Age_{all})/Age_{all}$  
as a function of intrinsic age (top row), intrinsic specific SFR (middle row), 
and $(Age_{all} - Age_{SAM})/Age_{SAM}$ (bottom row) for U-dropouts (left column), 
B-dropouts (middle column), and V-dropouts (right column). 
Here, $Age_{\tau 1}$ and $Age_{all}$ are the best-fit mean ages derived when we 
limit $\tau$ as $\leq 1.0$ $Gyr$ and when we allow $\tau$ to vary from $0.2$ $Gyr$ 
to $15.0$ $Gyr$, respectively. $Age_{SAM}$ is an intrinsic age, and $SSFR_{SAM}$ 
is an intrinsic SSFR. \label{taule1rd}}
\end{figure}

As can be seen in figure \ref{taule1rd}, the lowered best-fit values of 
ages/stellar masses are largely driven by galaxies with preferentially 
young intrinsic ages (top row) and/or high intrinsic SSFRs (middle row).
These are mostly the galaxies with SFH type-2, whose ages are severely 
overestimated when we fit the SEDs with the full range of $\tau$s (from 0.2 
Gyr to 15 Gyr) (bottom row of figure \ref{taule1rd}). 

The type-2 star formation histories have two main components. 
One component is a relatively low level of long-lasting SF activity, which 
corresponds to large $\tau$ values. The other is a relatively young, 
strong SF activity, which is more likely represented by small $\tau$. 
When we fit SEDs with full range of $\tau$s, best-fit models tend to be 
determined by the underlying, long-last SFH component giving the severely 
overestimated ages to these galaxies (see \S~\ref{ageest}).
In the case when only small values of $\tau$ ($\leq 1.0$ Gyr) are 
allowed in the SED-fitting procedure, the best-fit models are more likely 
determined by the young SF component with relatively small star formation 
time scale. 
The best-fit $t$ values then are much smaller than the ones derived 
utilizing the full range of $\tau$s, resulting in much younger best-fit 
mean ages and much smaller best-fit stellar masses.

Another significant feature is the more prominent bimodalities in the 
distributions of best-fit SFRs and ages. Galaxies in the smaller 
sub-population have very young best-fit ages and very high best-fit 
SFRs compared with the remaining galaxies. 
Figure \ref{taule1rd} shows the increase in the number of galaxies 
which belong to this sub-population as an effect of limitation on 
the allowed $\tau$ values.
They are the galaxies with small intrinsic SSFRs (middle row) as galaxies 
with type-1 or type-3 SFHs. The discrepancies between the intrinsic- 
and best-fit ages are very small when the full range of $\tau$s is used in 
the SED-fitting (bottom row), which means they behave like the galaxies with 
type-1 SFH. However, when we restrict $\tau$ as $\leq 1.0$ Gyr, their 
best-fit ages become much younger and join the sub-population. 
The bottom row of figure \ref{taule1rd} also shows that galaxies which 
are in this subpopulation when $\tau$ has the full range remain in 
the subpopulation when $\tau$ is restricted to be $\leq 1.0$ Gyr.

In summary, if we limit  $\tau$ to be $\leq 1.0$ Gyr in the SED-fitting, 
the overall trends are: 
(1) the best-fit ages (and hence the best-fit stellar masses) become smaller 
for the galaxies with type-2 SFHs. 
(2) a larger number of galaxies joins the subpopulation that has much 
smaller best-fit ages and stellar masses than the intrinsic SAMs values. 
These trends result in (1) an overall downward shift of the age/stellar 
mass distributions (increasing the discrepancies between the intrinsic- 
and SED-derived stellar mass distributions) and (2) more prominent 
bimodalities in the SFR/age distributions.

\subsubsection{SED-fitting with $\tau$=15 Gyr Templates} \label{tau15}

\begin{figure}
\plotone{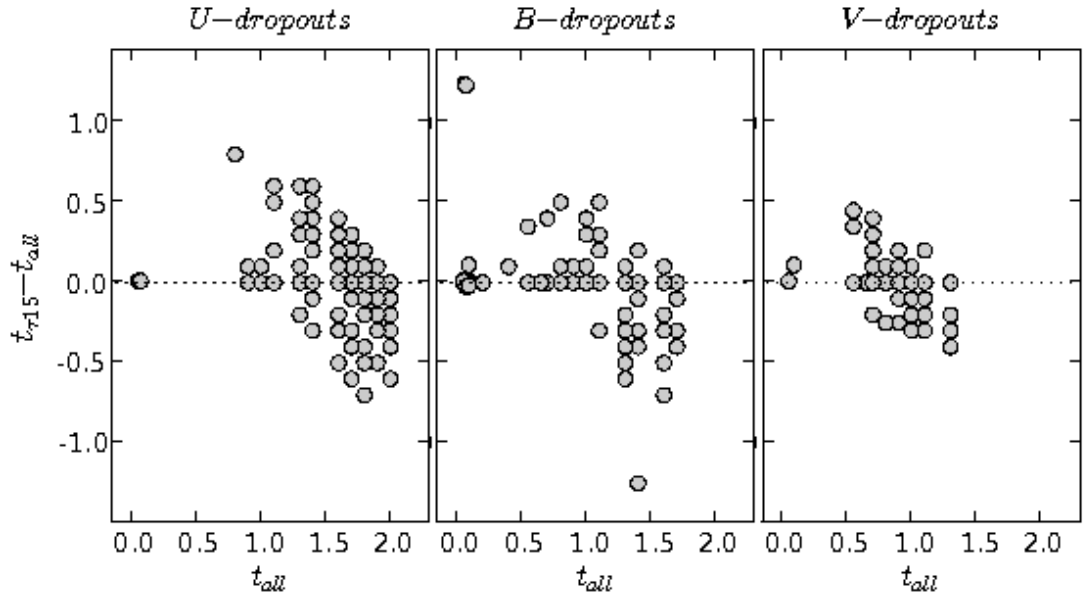}
\caption{Best-fit $t_{all}$ dependent discrepancies in best-fit $t$ 
($t_{\tau 15} - t_{all}$). $t_{all}$ is the best-fit $t$ obtained 
when we allow the full range of $\tau$, and $t_{\tau 15}$ is the best-fit $t$ 
derived if we use single value of $\tau$ ($= 15.0$ $Gyr$) in the SED-fitting. 
$t_{\tau 15}$ and $t_{all}$ are in $Gyr$. \label{tau15dtt}}
\end{figure}

Next, we hold $\tau$ fixed at 15 Gyr, which is equivalent to assigning 
a constant star formation rate, considering the age of the universe at 
redshifts $\sim 3-5$.
Figure \ref{tau15dtt} reveals that the best-fit values of $t_{\tau 15}$ 
generally increase for galaxies with small best-fit $t_{all}$ derived with 
full range of $\tau$s, but decrease if the best-fit $t_{all}$s are large. 
Here, $t_{\tau 15}$ refers the best-fit $t$ derived if we fix $\tau$ at 15 
Gyr, and $t_{all}$ is the best-fit $t$ obtained when we allow the full 
range of $\tau$.
This behavior is caused by the restriction on $t$ to be younger than the 
age of the universe at each redshift 
If the best-fit $t$ is already large for the full range of $\tau$s, there 
is no room to increase $t$ to match the red color of these old galaxies. 
Instead, the fitted value of the extinction increases and the fitted age 
generally decreases.

Also, with the larger value of $\tau$, larger $t$ does not always results in 
older mean ages while smaller $t$ always makes mean age younger, since mean age 
is a function of $\tau$ as well as $t$. 
Therefore, the mean ages (and the stellar masses also as a result) are slightly 
underestimated overall. 
Galaxies whose best-fit $t$s are very small ($\leq 0.2$ Gyr) show little 
differences in best-fit $t$ with or without the $\tau = 15$ Gyr restriction. 

Table.~\ref{tauofftab} shows the relative changes in the mean values of stellar 
masses and ages when we set $\tau = 15$ Gyr compared with the case when we 
allow the full range of $\tau$.

\subsection{SED Models with Extreme SFHs from BC03 Model} \label{test}

We further test what would be the results of the SED-fitting for the galaxies 
with the extreme star formation histories (SFHs).
To do this, here, we construct three types of toy models from the BC03 model, 
replacing SAM galaxies. The parameter settings used in these toy models are
summarized in table.~\ref{toymodel}. 
The aims of each toy model are to examine the biases which arise: 
(1) when we use shorter $\tau$s than real in the SED fitting (toy model 1), 
(2) when we use longer $\tau$s than real (toy model 2), and 
(3) when we try the single-component fitting for the galaxies with clearly 
distinct, two generations of star formation.

\subsubsection{Effects of SED-fitting Using Too Small $\tau$s} \label{test15}

First, in the case (1), the toy model SEDs have a very long SF time scale 
($\tau = 15.0$ Gyr). 
By restricting $\tau$ for the SED fitting not to exceed 1.0 Gyr, we can re-examine 
more transparently (because we compare the same BC03 models) what would happen if 
we fit the SEDs with $\tau$ values much shorter than the actual SF time scales 
of (model or real) galaxies. 

As expected, the best-fit mean ages are underestimated. To match colors (or SED 
shapes) of `$\tau = 15$ Gyr' samples with much shorter $\tau$s, the SED templates 
with smaller $t$s are found as the best-fit templates, which causes the mean ages 
are systematically underestimated. The amount of the age underestimation increases 
with age. 
The relative age error reaches up to $\sim$ 30 $\%$ underestimation for the 
oldest galaxies (with $t = 1.0$ Gyr).

The systematic underestimation of ages leads to systematic mass underestimation, 
since the given amount of light from galaxy is attributed to younger (more massive) 
stars with lower mass-to-light ratios. 
The relative mass underestimation increases as the relative age underestimation 
increases. For the oldest galaxies whose relative age underestimation reaches 
up to $\sim$ 30 $\%$, relative mass underestimations are $\sim$ 9 - 19 $\%$. 
The `Younger-than-input' best-fit ages result in higher average SFRs, whose 
relative overestimation spans from $\sim$ 0 through 9 $\%$.

For the samples with $t$ as small as 0.01 $Gyr$, all the physical parameters, such as 
stellar masses, SFRs, and mean ages, are well recovered through the SED-fitting. 
This reflects the fact that the effects of SFH difference are insignificant for the 
very young galaxies.

In summary, if one tries to fit galaxies' SEDs with relatively short $\tau$s, 
the resultant best-fit stellar masses and mean ages can be underestimations of 
the true values especially for galaxies which have very extended SFHs for 
relatively long timescales. These trends confirm the speculation of 
\S~\ref{taule1}. The SFRs are overestimated, but the relative errors are not 
as large as those of stellar masses/mean ages.

\subsubsection{Effects of SED-fitting Using Too Large $\tau$s} \label{test02}

In the case (2), the input toy models have shorter $\tau$ than the values allowed 
in the SED-fitting procedure. This highlights the biases that can arise in the 
SED-fitting if galaxies have much shorter SF time-scales (probably burst-like) 
than the $\tau$ values allowed in the SED fitting.

The direction of bias in the best-fit mean ages is divided into two regimes 
depending on the actual age (or $t$) of each galaxy.
For the model galaxies with small enough $t$ (i.e. $t$ = 0.1, 0.2 Gyr) compared 
with the age of the universe at corresponding redshift (which is $z$ $\sim$ 4, 
in this case), the best-fit $t$s are larger than the input $t$s to match the SEDs 
with longer $\tau$s than the input ($\tau$ = 0.2 Gyr), leading to the overestimation 
of the mean ages by amount of $\sim$ 30 $\%$ for the galaxies with the input $t$ = 0.1 
Gyr, and $\sim$ 50 $\%$ for the ones with the input $t$ = 0.2 Gyr.

For the similar reason as in the case (1) (but, in the opposite direction), 
the age overestimation results in the mass overestimation by attributing light to 
older stars with higher mass-to-light ratios. 
The mass overestimation is about $\sim$ 11-13 $\%$. 
The `older-than-input' best-fit ages make the SFRs underestimated by $\sim$ 
22-34 $\%$.

However, for the model galaxies with relatively large values of $t$, which are 
not much shorter than the age of the universe, there is not so much room for 
$t$ to increase.
So, for the toy model galaxies with input $t$ = 1.0 and 1.3 Gyr, the best-fit 
$t$s are only slightly larger than the input $t$s, and the mean ages are younger 
than the input due to larger $\tau$s than input. 
Here, the stellar masses are overestimated even for the underestimated mean ages, 
mainly because of the more extended SFHs.
The important difference between the models with large input $t$s and small input 
$t$s is that the derived dust extinctions, parameterized as $E(B-V)$, are greatly 
overestimated to compensate the `younger-than-input' mean ages for the SED models 
with large input $t$s. 

During the SED-fitting procedure, the best-fit $t$s generally move in the 
direction which `correct' the difference in $\tau$s -- `smaller-than-input' 
$t$s for the long input $\tau$ models, and `larger-than-input' $t$s for 
the short input $\tau$ models. 
However, for the toy model galaxies with small $\tau$s and large $t$s, the 
SED-fitting cannot overcome the $\tau$ difference by adjusting the best-fit
$t$s due to the restriction that $t$ be smaller than the age of the universe.
Instead, the best-fit $E(B-V)$ starts to be overestimated by the amount of 
$\Delta E(B-V)$ $\sim$ 0.25-0.3 for the input-$t$ = 1.0 Gyr models and $\sim$ 
0.45 for the input-$t$ = 1.3 Gyr models. 
This large bias, in turn, results in a large bias toward overestimated SFRs. 
The best-fit SFRs are about $\sim$ 13-17 times 
of the input SFRs for input-$t$ = 1.0 Gyr models, and reach up to $\sim$ 70-110 
times of the input SFRs for the input-$t$ = 1.3 Gyr models. 
The extinction overestimation also leads to the larger mass overestimation, but 
the effects are not as dramatic as in the SFR estimation.
This example illustrates how the well-known `age-extinction' degeneracy affects 
the results of the SED-fittings, especially for the galaxies with extreme SFHs 
and/or for the case when parameter (for example, $\tau$) space allowed during 
the SED-fitting is not sufficiently large.

In summary, if one tries to fit galaxies' SEDs with very long $\tau$s only, 
the resultant stellar masses are generally overestimated. If galaxies have 
sufficiently young ages (compared with the age of the universe at the redshifts 
where galaxies reside) the mean ages are overestimated, as the best-fit $t$s 
to be much larger than the input $t$s to compensate the $\tau$ difference. 
However, for old galaxies whose input $t$s are compatible to the age of the 
universe at corresponding redshift, the direction of bias in the age estimation 
is that the mean ages are underestimated (even though the best-fit $t$s are 
slightly larger than the input $t$s). Instead, the dust extinctions are 
greatly overestimated, which makes the best-fit SFRs erroneously high (up to 
two orders of magnitude).

\subsubsection{Effects of Single Component SED-fitting for Galaxies with Two Generations of Star Formation} \label{test2c}

Lastly, in the case (3), the model galaxies have two clearly distinguished 
generations of star formation with $t$ = 0.1 $\&$ 1.0 Gyr.
Both of the components are set to have $\tau$ = 0.2 Gyr.
When we try to derive the best-fit physical parameters of these two component 
model galaxies via the single-component SED-fitting, both the mean ages and 
stellar masses are underestimated, as the result of the older stellar 
population being at least partially ignored.

The age underestimation is minimal ($\sim$ 10 $\%$) for the toy model galaxies 
in which the young-to-old ratio is smallest (i.e. $young$ component fraction 
$\sim$ 0.1).
The age underestimation increases as the proportion of young component increases 
(up to $\sim$ 77-88 $\%$ for the ones with $young$/$old$ $\sim$ 1.0).
This implies the best-fit parameters tend to be determined by the stellar 
component whose fractional occupation is large.
However, as the $young$ component fraction increases further to $young$/$old$ 
= 2.6, the age underestimation decreases and becomes similar to that for the 
galaxies with $young$/$old$ = 0.52, because the mean ages of these galaxies (with 
$young$/$old$ = 2.6) are already small enough to be greatly underestimated.

As stated in previous sections, the age underestimation propagates to the mass 
underestimation. 
The mass underestimation is minimal ($\sim 9-11$ $\%$) for the model galaxies 
with $young$/$old$ = 0.1 due to the lowest age underestimation, and for the 
ones with $young$/$old$ = 2.6 because they have the smallest portion of their 
mass in the $old$ component. 
This means the mass underestimation depends on two factors -- the degree of age 
underestimation (i.e. the amount of the mis-interpretation of the mass-to-light 
ratio) and the fraction of stellar mass in the $old$ component.
The mass underestimation for galaxies with $young$/$old$ = 0.26, 0.52, and 
1.0, is in the range 21-40 $\%$.

The SFRs are underestimated by $\sim$ 14-43 $\%$, and the discrepancy 
is larger for the galaxies with larger fractional mass in the $old$ component. 
This is because the best-fit parameters are affected by the $old$ component 
while the SFR contribution of the $old$ component is very small ($\sim$ 0.2-4 
$\%$ of the total SFR, depending on the mass fraction of the $old$ component). 
The best-fit $\tau$s are larger than the input (i.e. 0.2 Gyr), and the best-fit 
$t$s are larger than the input $young$ component (i.e. 0.1 Gyr) due to the effects 
of the $old$ component, and this leads to lower SFRs.

In summary, if one tries to fit galaxy's SED with single-component SED-fitting 
when the real SFH has two episodes, the mean ages and the stellar masses are 
generally underestimated indicating the $old$ components are to some extent masked. 
And, the SFRs are underestimated since the $old$ components with very little 
current SF activity pollute the SEDs.

\section{SFR Estimation from Rest Frame Ultra-violet Luminosity} \label{uvsfr}

For galaxies with a roughly constant star-formation rate, the extinction-corrected 
UV luminosity is expected to provide a reasonably good estimate of the 
star-formation rate. This is fortunate, because often the only data available 
for high-redshift galaxies are a few photometric data points in the rest-frame UV. 

The relation between SFR and UV luminosity can be calibrated using spectral 
synthesis models, such as BC03. 
\citet{ken98} notes that the calibrations differ over a range of $\sim 0.3$ dex, 
when converted to a common reference wavelength and IMF, with most of the difference 
reflecting the use of different stellar libraries or different assumptions about 
the star-formation timescale. 
The calibrations usually assume constant or exponentially declining star-formation 
rates.

It is interesting to see how well this technique works for the more varied 
star-formation histories of the semi-analytic models. 
In this case, we are using BC03 and the same IMF for both the calibration and the 
SAM galaxies, so the discrepancies in derived SFRs must be primarily due to 
the different star-formation histories.

Here, we derive SFRs of SAM B-dropout galaxies with the assumption that we do 
not know their redshifts, which is similar to the case when there is no 
spectroscopic redshift information for a color-selected LBG sample.  

The (dust-uncorrected) rest-frame UV ($\lambda_0 = 1500 ~\rm{\AA}$) luminosity 
of each SAM B-dropout galaxy is calculated from the $i_{775}$ band flux as 

\begin{equation} \label{uv2sfr}
L_{\nu,1500} = \frac{4 \pi d^{2}_{L}}{(1+z)} \times f_{\nu,i775}, 
\end{equation} 

assuming all B-dropouts are at $z = 4.0$. 

Here, $f_{\nu,i775}$ and $L_{\nu,1500}$ are specific flux at $i_{775}$ band 
and specific luminosity at rest-frame 1500 $\rm{\AA}$, respectively, and 
$d_{L}$ is luminosity distance at redshift $z$, which is assumed to be 4.0.

There are two major possible sources of systematic biases, which can arise 
due to the assumption that all galaxies are at $z = 4.0$: 
(1) ignoring the bolometric correction -- which is a decreasing function of 
redshift and is small -- a factor of $1.0 \pm 0.2$ in the redshift range 
of B-dropouts ($3.6 \leq z \leq 4.6$) for \citet{cha03} IMF and solar 
metallicity, and (2) ignoring error in luminosity distance ($d_{L}$). 

These two sources of bias act in opposite direction. 
Not including bolometric correction causes the UV luminosity to be slightly 
overestimated at high redshift, since it is a decreasing function of redshift. 
Underestimation of the luminosity distances for galaxies at high redshift 
($z > 4.0$) leads to underestimated UV luminosities. 
The correction factor due to the error in $d_L$ estimation is slightly 
larger at high-redshift than the bolometric correction factor, and ranges 
from 0.8 to 1.4 in the redshift range of B-dropouts.
Together, this will cause rest-frame UV luminosity to be slightly 
underestimated at the high-redshift end of the range.

For all B-dropout galaxies, dust-extinction is assumed to be $E(B-V) = 0.15$, 
which is the same value used for dust-correction in high-redshift LBG 
studies, such as \citet{gia04b} and \citet{saw06}.

Star-formation rates are then calculated using the conversion of \citet{ken98} 
divided by 2.0 to correct for the different assumed IMF, because \citet{ken98} 
uses a \citet{sal55} IMF with a mass range 0.1 to 100 $M_{\odot}$.

\begin{figure}
\plotone{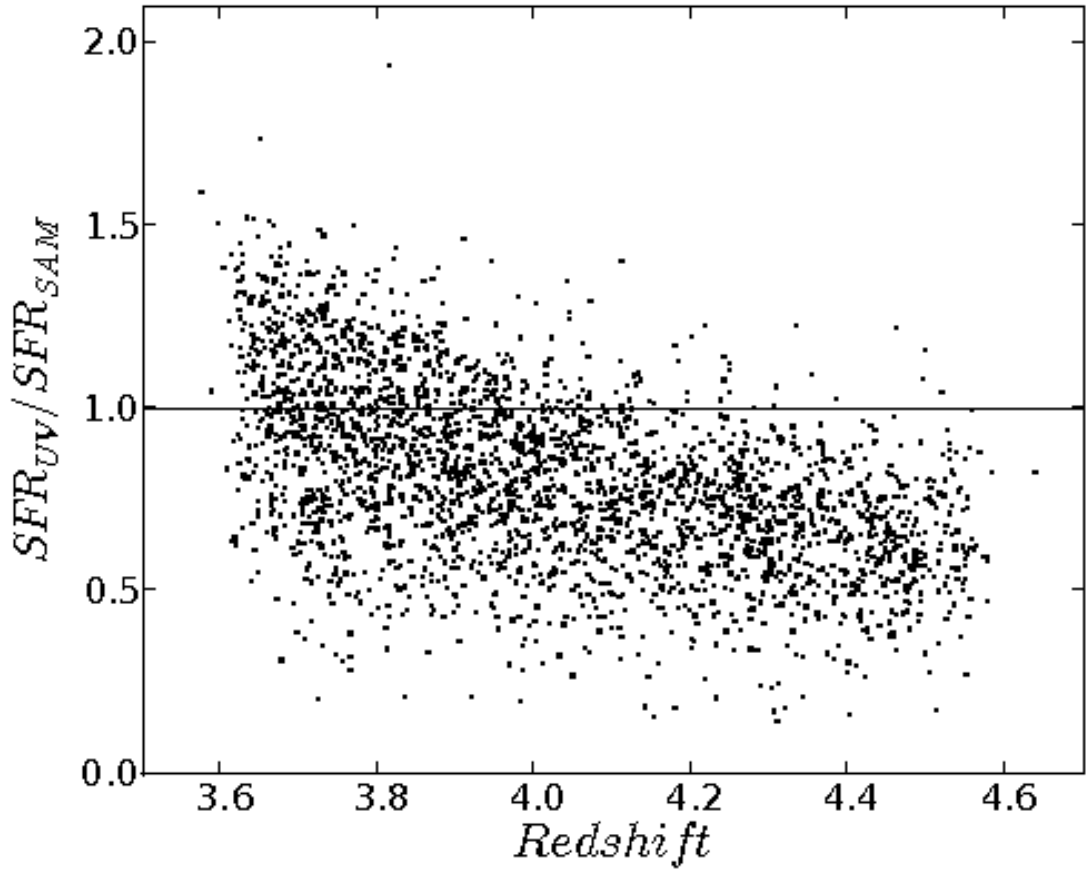}
\caption{Ratio of SFR derived from rest-frame UV luminosity to intrinsic SFR as a 
function of redshift for B-dropouts. 
All B-dropouts are assumed to be at $z = 4.0$, and to be extincted by dust with 
amount of $E(B-V)=0.15$. \label{sfrfromuv}}
\end{figure}

Figure \ref{sfrfromuv} shows the ratio of dust-corrected, UV-derived SFR 
to intrinsic SFR as a function of redshift.
In this figure, we can see the redshift dependent behavior of 
the star-formation rate calibration from the rest-frame UV. 
This behavior is a combined effect of luminosity-distance underestimation, 
which is small, though, as explained above, and the redshift-dependent 
difference in average dust-extinction. 
The large scatter of SFR ratios at a given redshift reflects the variation 
of galaxies' intrinsic dust-extinction and SFH.

Differences in dust extinction can have a large effect. 
For example, small change in assumed value of mean $E(B-V)$ will significantly 
change the derived SFR values -- for a range of $E(B-V)$ from 0.10 to 0.20, 
the dust-correction factor vary from 2.6 to 6.5.

With the assumed dust-extinction of $E(B-V) = 0.15$, mean UV-derived SFR is 
11.099 $M_{\sun} ~yr^{-1}$. 
For comparison, mean values of intrinsic and SED-derived SFRs are 15.650 
$M_{\sun} ~yr^{-1}$ and 6.638 $M_{\sun} ~yr^{-1}$, respectively\footnote{Since 
calibration between UV luminosity and SFR is derived assuming constant SFR over 
100 Myr \citep{ken98}, comparison between UV-derived SFR and intrinsic or 
SED-derived SFR, which is averaged over last 100 Myr, is relevant.}.

\begin{figure}
\plotone{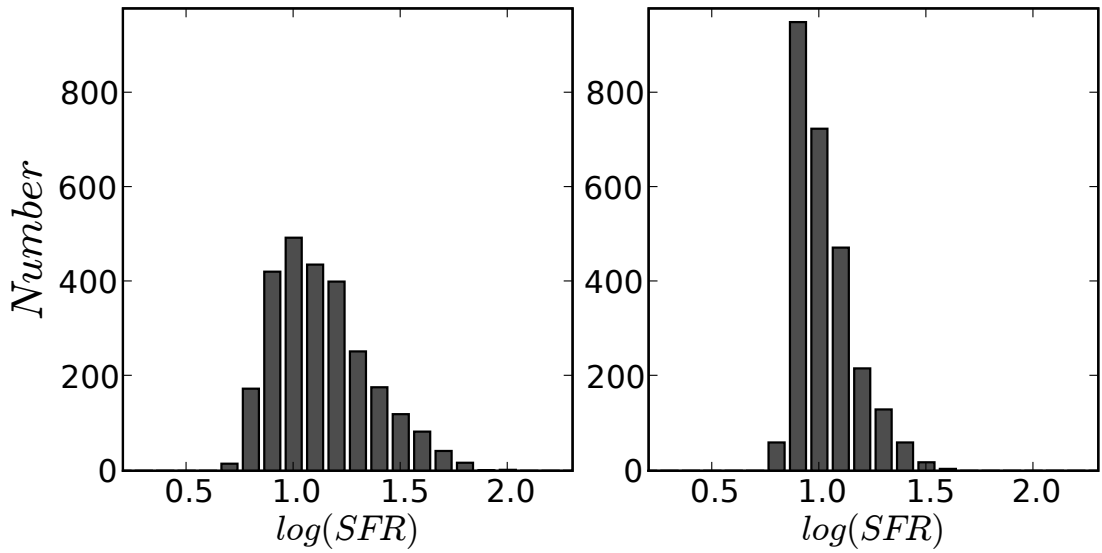}
\caption{Distributions of intrinsic SFRs ($left$), and SFRs derived from rest-frame UV 
luminosity assuming all galaxies are at $z = 4.0$ and have mean dust-extinction of 
$E(B-V) = 0.15$ ($right$) for B-dropout galaxies. \label{uvsfrdst}}
\end{figure}

Figure \ref{uvsfrdst} shows the distributions of intrinsic SFRs ($left$), 
and of SFRs derived by assuming all galaxies are at $z = 4.0$ and assuming  
dust-extinction of $E(B-V) = 0.15$ ($right$).
The distribution of SFRs derived from rest-frame UV luminosity, assuming all 
galaxies are at $z = 4.0$, is much narrower than the intrinsic one.

\section{Discussion} \label{discus}

We have shown in this paper that there are significant biases in the physical 
parameters derived from the SED-fitting of standard $\tau$-model to 
broadband photometry. 
Biases are severe especially for the SFR and mean age, but even the SED-derived 
stellar masses are biased. 
We now address in some detail how the biases in the derivation of these physical 
parameters can affect the investigation of high-$z$ galaxies and the inferred 
galaxy evolution studies.

\subsection{Artificial Age bimodality} \label{agebimod} 

Figures \ref{sfrdstzfix} and \ref{agedstzfix} show that there are bimodalities 
in the SFRs and mean ages derived through the SED-fitting (most clearly shown for 
B-dropout LBGs) while there is no such bimodality in the intrinsic distributions.
These bimodalities are enhanced when we try to fit the SEDs omitting some 
available input information, such as NIR photometry from ISAAC or the 
spectroscopic redshift (figures \ref{udropdsts}, \ref{bdropdsts}, 
and \ref{vdropdsts}).
The main origin of these artificial bimodalities is the mismatch of the SFHs 
between the SAM model galaxies and the templates from the BC03 stellar population 
model as explained in \S~\ref{subpop}.
Such bimodalities can lead to the false interpretation that there are clearly 
distinguished populations among the similarly selected star-forming galaxies.

Recently, \citet{fin08} analysed 14 Lyman-$\alpha$ emitting galaxies 
(LAEs) and found that there is a clear bimodality in their age distribution, 
which are derived through SED-fitting, such that their ages either very young 
($<$ 15 Myr) or old ($>$ 450 Myr). 
Based on this bimodality, they concluded that there are two distinct 
populations of LAEs -- dusty starbursts and evolved galaxies. 
However, according to the results presented in our work, it is possible that 
this age bimodality reported in \citet{fin08} may not be real but an artifact 
which arises in the SED-fitting procedure due to the difference of their SFHs. 
Of course, caution should be applied in interpreting the SED-fitting results 
of the LAEs based on our analysis performed for the LBGs. 
It is still controversial how similar (or how different) the LBGs and LAEs 
are, despite some indications of similarities in their physical parameters 
\citep{lai07} and the possible overlap between LBGs and LAEs \citep{rho08}. 

\citet{sha05} also reported a subset of galaxies in their sample of $z \sim 2$, 
star-forming galaxies with extremely young ages (with $t \leq 10$ Myr). 
The galaxies in this subpopulation are relatively less massive 
($<$log $M_{*}/M_{\sun}$$>$ = 9.64 (calculated from column 7 in table 3 
of \citet{sha05}), while the mean value of total sample is 10.32. 
More dramatic difference between galaxies in this subset and other 
remaining galaxies is shown in the derived SFR distribution. 
Among 72 galaxies, there are 10 galaxies whose SFRs are larger than 200 
$M_{\sun}$ $yr^{-1}$, and among these 10, nine galaxies belong to this 
subset of galaxies with $t \leq 10$ Myr according column 8 in their 
table 3 (one remaining galaxy with high SFR has the best-fit $t = 15$ Myr). 
In contrast, almost half of their sample has very low SFRs 
($\leq 10 M_{\sun}$ $yr^{-1}$).
Based on our analysis, it is plausible that the 10 galaxies with the very 
high SFRs have similar SFHs as shown in figure \ref{sfhrdagneg} (i.e. type-3 SFHs), 
and therefore their ages and stellar masses are underestimated while their 
SFRs are greatly overestimated. 
If so, the ages and SFRs of their total sample would present more continuous 
distributions.

Bimodalities in the inferred ages are also seen among $z \sim 5$ LBGs of 
\citet{ver07} and among 14, $z \sim 5$ LBGs with spectroscopy of \citet{sta07}. 

More interestingly, in the \citet{sha05} sample, there are three galaxies 
whose physical parameters derived through the SED-fitting methods do not agree 
with the indications from their rest-frame UV spectra. 
The SED-fitting results indicate that these are old ($t / \tau$ $>>$ 1) 
galaxies with very low level of SFRs ($\sim$ a few $M_{\sun}$ $yr^{-1}$), 
while there are indications of a young population of stars with active star 
formation in there spectra (see \S~ 4.5 in \citet{sha05} for detail). 
This apparent disagreement can be explained if these galaxies have the star 
formation histories similar with the ones shown in figure \ref{sfhrdaglg} 
(i.e. type-2 SFHs).

\subsection{Possible Descendants of High Redshift Star Forming Galaxies} \label{descen}

There is great interest in connecting LBGs to their possible descendants 
-- i.e. determining whether or not they are progenitors of local massive 
ellipticals \citep[e.g.][]{low97,ade98,saw98,ste98,som01}.
Many properties of LBGs -- including their sizes, morphologies, number 
densities, clustering properties, and physical properties -- are 
closely linked to this issue. 
Therefore, more accurate estimation of LBGs' physical parameters are 
clearly important for addressing this issue properly.  

According to our analysis, the stellar masses derived using the single-component 
SED-fitting methods tend to underestimate the true values, on average, 
by 19-25 $\%$. 
Moreover, bias in the stellar mass estimation seems to strongly depend on 
the stellar mass itself, in a sense that the stellar masses are more severely 
underestimated for more massive galaxies (see figures \ref{uersamzfix}, 
\ref{bersamzfix}, and \ref{versamzfix}).
For some of very massive galaxies ($M_{*} > 10^{10} M_{\sun}$), the best-fit 
stellar masses can be less than half of the intrinsic stellar masses.
Accompanied by the SFR underestimation, which is generally more severe than the 
stellar mass underestimation, this discrepancy can significantly affect the 
discrimination among possible evolutionary descendants of massive LBGs. 
Since the current (or recent) SFR is a measure of the possible additional stellar 
mass which can be added to LBGs during their evolution to the lower redshift, the 
underestimation of the stellar mass and SFR of massive LBGs can plausibly lower 
their possible stellar masses at lower redshift or at $z \sim 0$ significantly. 

Interestingly, studies based on the clustering properties of LBGs speculated 
that massive, high-$z$ LBGs could be the progenitors of local massive 
ellipticals \citep[e.g.][]{ade98,ste98}, while \citet{saw98}, based on the 
SED-fitting analysis, suggested that LBGs would not become sufficiently 
massive at low-$z$ to be massive ellipticals unless experience significant 
number of mergers. 

Biases and uncertainties in the estimation of LBGs' ages -- which are shown not 
only to be large and but also to vary significantly as the fitting parameters 
change in our study -- can also affect understanding of LBG properties and 
the predictions regarding their possible evolutionary paths.
For example, biases in the age estimation would propagate into errors in their 
estimation of the star-formation duty cycle, which would, in turn, affect the 
estimates of the number of galaxies which have similar stellar masses/ages 
with detected LBGs, but are undetected due to their reduced SFRs.

\section{Summary $\&$ Conclusions}

In this paper, we examine how well the widely-used SED-fitting method can 
recover the intrinsic distributions of physical parameters -- 
stellar mass, SFR, and mean age -- of high-redshift, star-forming galaxies. 
To this end, we construct model high-redshift galaxies from the semi-analytic 
models of galaxy formation, make a photometric catalog via the BC03 synthesis 
model, and select LBGs through the appropriate color selection criteria based on 
their broadband colors.
Then, we perform SED-fitting analysis, comparing the photometric SEDs of 
these model galaxies with various galaxy spectral templates from the BC03 
stellar population model to derive the distributions of best-fit stellar 
masses, SFRs, and mean ages. 
We use this test to explore (rather exhaustively) the errors and biases that 
arise in such SED fitting and the underlying causes of these errors and biases.

Here are the summary of the main results of this work.

1. When we fix the redshift to the given value in the SAM catalog and use 
$ACS$/$ISAAC$/$IRAC$ passbands, the SED-fitting method reproduces relatively well 
the input distributions of stellar masses with a minor tendency to underestimate the 
stellar masses and with substantial scatter for individual galaxies.   
The mean stellar masses are underestimated by about 19$\sim$25 $\%$, which is due 
to the fact that the old generations of stars can be hidden by the current 
generation of star formation. 
The distributions of SFRs and mean ages show larger offsets than the stellar mass 
distributions. The SFRs are systematically underestimated and the mean ages are 
systematically overestimated, and these trends mainly reflect the difference in the 
SFHs predicted by the semi-analytic models and assumed in the simple galaxy templates 
used in the SED fitting. The well-known `age-extinction degeneracy' plays an important 
role in biasing the derived SFRs.

2. When we use redshift as an additional free parameter, the discrepancy between 
the intrinsic- and SED-derived stellar mass distributions increase (i.e., the overall 
stellar mass underestimation becomes worse), while the bimodalities which appear in the 
SFR $\&$ mean age distributions become more significant. 
The distributions of offsets of individual galaxies indicate that there exist 
sub-population(s) of LBGs whose behaviors are distinct from the majority of LBGs in the 
SED-fitting. The SED-fitting generally underestimates the redshift slightly. 

3. The age overestimates are clearly related to 
the intrinsic age and specific star formation rate of each galaxy.  
The overestimation of mean ages is worse for galaxies with younger ages and higher SSFRs. 
Inspection of the SFHs of individual SAM galaxy confirms that the main origin of the bias 
in the age estimation is the difference of assumed SFHs in SAM galaxies and the 
simple galaxy templates used in the SED fitting.
This bias in the age-estimation propagates into the stellar mass and SFR estimations, in 
the sense that the age-overestimation leads to the mass-overestimation and SFR-underestimation. 
The SFR-underestimation is further enhanced by the `age-extinction degeneracy'.
Another source of biases is the dominance of the current generation of star formation 
over the old generation(s) of star formation in the SED-fitting. 
This causes both of the stellar masses and SFRs to be underestimated.

4. We perform two types of two-component SED-fitting: (1) adding a young, bursty 
component to an old component with long-lasting SFH, and (2) combining an old, bursty 
component with a younger, long-lasting component. 
The changes of behaviors in these two types of two-component fitting depend on 
galaxies' SFHs.
Generally, compared with the best-fit values in the single-component fitting, the 
best-fit stellar masses are generally smaller in the two-component fitting with a young, 
bursty component embedded in an older, long-lasting component, while they become larger in 
the two-component fitting with an old burst combined with a younger, long-lasting component. 
For the majority of galaxies (mainly, with type-1/type-2 SFHs), the best-fit ages become 
younger and the best-fit SFRs become higher in both types of the two-component fitting. 
The behavior of galaxies with type-3 SFH is opposite: the best-fit ages are older and 
the best-fit SFRs are smaller than the values derived in the single-component fitting. 

5. We perform the SED-fitting with different combinations of passbands -- by 
omitting $IRAC$ data or $ISAAC$ data.
If we fit the galaxy SEDs with $ACS$ and $ISAAC$ data only omitting $IRAC$ data, 
the derived distributions of stellar masses, SFRs and mean ages are significantly affected. 
The detailed behaviors of change in the SED-fittings with and without IRAC data strongly 
depend on galaxy's redshift and SFH.
Alternatively, if we fit the SEDs of LBGs with $ACS$ and $IRAC$ data only, the derived 
distributions of stellar masses, SFRs, and means age do not show significant changes except 
for the enhanced bimodalities in the SFR/age distributions for B-/V-dropouts. 
This indicates that the significant changes occur only for a small fraction of galaxies, 
while the effects of $ISAAC$ data in the SED-fitting are insignificant for the majority of 
LBGs.
These experiments demonstrate the usefulness of the observed-frame MIR data from $IRAC$ 
in constraining physical properties of the high-$z$, star-forming galaxies. 
 
6. When the allowed range of $\tau$ (star formation $e$-folding timescale) is limited to 
be insufficiently short or long, biases in the SED-fitting increase, in general, compared 
with the case when sufficiently broad range of $\tau$ is allowed (from 0.2 Gyr to 15.0 Gyr 
in this study). 
The mean values of the best-fit stellar mass and age are smaller than the values derived 
with the full range of $\tau$ used. 
The age shifts are larger when we limit $\tau$ to be small ($\leq$ 1.0 Gyr) than 
when we use very large value of $\tau$ (= 15.0 Gyr). 
Detailed behaviors of change for individual galaxy depend on the SFH. 

7. The experiments with the SED templates constructed from the BC03 model (\S~\ref{test}) 
isolate the effect of limiting the allowed range of $\tau$. 
If $\tau$ is restricted to be $\tau \leq 1$ Gyr, both the stellar mass and mean age 
are underestimated while the SFR is overestimated. 
If only very long values of $\tau$ are used, the stellar mass is overestimated for the 
majority of model galaxies (from BC03 model) as a result of the mis-assignment of light 
to older stars (with consequently higher mass-to-light ratio).
The biases in the SFR and age estimation depend on the age of galaxy. 
If the age (more exactly, value of $t$, i.e. time since the onset of star formation) 
is long, compatible to the age of the universe, the mean age is underestimated and 
$E(B-V)$ is greatly overestimated leading to the severe overestimation of SFR. 
Otherwise, the mean age is overestimated and SFR is underestimated. 
In the case when one tries the single-component SED-fitting for the galaxies with 
two clearly distinguished generations of star formation (resembling repeated-burst 
models), the stellar mass, SFR, and mean age are all underestimated.  

8. Star-formation rates estimated from the UV luminosity alone may be less biased 
than those estimated from SED-fitting, provided one has a reasonable estimate of $E(B-V)$. 
This is mainly due to the fact that the results are less subject to the degeneracy 
between age and dust-extinction when we use only rest-frame UV photometry. 
However, the bias depends on galaxies' redshift, and more significantly on the 
estimation of mean dust-extinction, which is challenging. 
A relatively small change in $E(B-V)$ would result in large bias in UV-derived SFRs. 

9. We show that biases arising in the SED-fitting procedure can affect studies of 
the high-redshift, star-forming galaxies. 
The different directions and amounts of biases depending on galaxy's SFH can produce 
the artificial bimodalities in the age or SFR distributions. 
This can affect the interpretation of the properties and nature of these galaxies.
Also, the stellar mass underestimation for massive LBGs, combined with the SFR 
underestimation, can affect the interpretation of possible evolutionary paths for 
these massive LBGs.

In conclusion, we show that single-component SED-fitting generally slightly 
underestimates the LBGs' stellar mass distributions, while the SFR distributions are 
significantly underestimated and the age distributions are significantly overestimated. 
The main causes of these biases are:
(1) the difference of assumed SFHs between in the SAM galaxies and simple templates used 
in the SED-fitting, and 
(2) the effects of the current generation of star formation masking the previous 
generation(s) of stellar population. The well-known `age-extinction degeneracy' (or 
`age-extinction-redshift degeneracy' in the case when redshifts are allowed to vary 
freely as an additional free parameter during the SED-fitting procedure) plays an additional 
role, mainly in the estimation of SFR distributions. Consequently, the directions and amounts 
of the biases in the SED-fitting strongly depend on galaxy's star formation history (SFH). 
If we change various inputs or fitting parameters in SED-fitting, such as the range of $\tau$ 
-- the e-folding timescale of star formation, combinations of passbands used, or assumed SFHs, 
the derived distributions of best-fit stellar masses, SFRs and ages can change dramatically. 
Due to the compensating causes of biases, the best-fit stellar mass distributions are more 
stable against these changes than the SFR/age distributions.
Moreover, the behaviors of individual galaxy in various settings of the SED-fitting 
strongly depend on galaxy's SFH.

These biases arising in the SED-fitting can have significant effects in the context of 
the galaxy formation/evolution studies as well as cosmological studies. 
Besides the effects discussed in \S~\ref{discus}, biases in the estimation of stellar mass and 
SFR (which are dependent on the various input settings in the SED-fitting, on available passbands 
and more importantly on individual galaxy's SFH) can affect, for example, the estimation of the 
stellar mass function as well as the global density of the stellar mass and SFR, which are 
important probes in galaxy formation/evolution. The blind comparison among various works -- which 
were done with different input settings in the SED-fitting, with the different sets of available 
photometric data (e.g. works done before or after $Spitzer$-era), or with differently-selected galaxy 
samples -- can also misleads us. 
Therefore, appropriate caution should be applied to the estimates of physical parameters of high-redshift, 
star-forming galaxies through the SED-fitting, and also to the interpretation in the context of galaxy 
formation/evolution based on (the comparisons of) the derived results.

\acknowledgments

We acknowledge useful discussions of the issues in this paper with our many colleagues on the GOODS, 
FIDEL, and COSMOS teams, including Casey Papovich. 
SL thanks Myungshin Im for useful suggestions applied in this work.
This work is supported in part by the Spitzer Space Telescope Legacy Science Program, which was 
provided by NASA, contract 1224666 issued by the JPL, Caltech, under NASA contract 1407, and in part 
by HST program number GO-9822, which was provided by NASA through a grant from the Space Telescope Science 
Institute, which is operated by the Association of Universities for Research in Astronomy, Incorporated, 
under NASA contract NAS5-26555.

\clearpage

\clearpage

\clearpage

\clearpage

\clearpage

\clearpage

\clearpage

\clearpage

\clearpage

\clearpage

\clearpage

\clearpage

\clearpage

\clearpage

\clearpage

\clearpage

\clearpage

\clearpage

\clearpage

\clearpage

\clearpage

\clearpage

\clearpage

\clearpage

\clearpage

\clearpage

\clearpage

\clearpage

\clearpage

\clearpage

\clearpage

\clearpage

\clearpage

\clearpage

\clearpage

\begin{deluxetable}{crrccc}
\tablecolumns{6}
\tablewidth{0pc}
\tablecaption{Fitting Parameters \label{paraset}}
\tablehead{
\colhead{IMF}   & \colhead{$\tau$ (Gyr)}   & \colhead{$t$ (Gyr)}   & \colhead{metallicity ($Z_{\sun}$)}   &
\colhead{internal extinction}   & \colhead{IGM extinction}}
\startdata
Chabrier & 0.2-15.0 & 0.01-2.3\tablenotemark{a} & 0.2, 0.4, 1.0 & Calzetti\tablenotemark{b} & Madau \\
\enddata

\tablenotetext{a}{$t$ is limited to be smaller than the age of the universe at each galaxy's redshift.}
\tablenotetext{b}{0.000 $\leq$ $E(B-V)$ $\leq$ 0.950 with ${\rm \Delta} E(B-V)$ = 0.025}

\end{deluxetable}

\clearpage

\begin{deluxetable}{crrrr} 
\tablecolumns{5} 
\tablewidth{0pc} 
\tablecaption{Mean Values of Stellar Population Parameter Distributions \label{means}} 
\tablehead{ 
\colhead{}    & \colhead{Redshift}   & \colhead{Stellar Mass ($M_{\sun}$)}   & 
\colhead{SFR ($M_{\sun}$/yr)}   & \colhead{Mean Age (Gyr)}}
\startdata
\sidehead{U-dropouts}
SAMs\tablenotemark{a} & 3.36 & 9.376$\times 10^{9}$ & 11.005 & 0.465 \\
$SED_{z-fix}$\tablenotemark{b} & 3.36 & 7.594$\times 10^{9}$ & 3.887 & 0.915 \\
$SED_{z-free}$\tablenotemark{c} & 3.27 & 7.038$\times 10^{9}$ & 4.682 & 0.870 \\
\sidehead{B-dropouts}
SAMs\tablenotemark{a} & 4.02 & 8.888$\times 10^{9}$ & 15.650 & 0.353 \\
$SED_{z-fix}$\tablenotemark{b} & 4.02 & 6.680$\times 10^{9}$ & 6.638 & 0.667 \\
$SED_{z-free}$\tablenotemark{c} & 3.88 & 4.327$\times 10^{9}$ & 19.891 & 0.420 \\
\sidehead{V-dropouts}
SAMs\tablenotemark{a} & 4.97 & 7.946$\times 10^{9}$ & 18.920 & 0.249 \\
$SED_{z-fix}$\tablenotemark{b} & 4.97 & 5.983$\times 10^{9}$ & 7.287 & 0.508 \\
$SED_{z-free}$\tablenotemark{c} & 4.88 & 4.496$\times 10^{9}$ & 16.905 & 0.344 \\
\enddata 

\tablenotetext{a}{Intrinsic values from SAM catalogs}
\tablenotetext{b}{Values derived in the SED-fitting with redshifts fixed as input values}
\tablenotetext{c}{Values derived in the SED-fitting with redshifts allowed to vary freely}

\end{deluxetable} 

\clearpage

\begin{deluxetable}{crrrr} 
\tablecolumns{5} 
\tablewidth{0pc} 
\tablecaption{Means and Standard Deviations of Relative Offsets\tablenotemark{a} for various Stellar Population Parameters \label{offsets}} 
\tablehead{ 
\colhead{}    & \colhead{Redshift}   & \colhead{Stellar Mass}   & 
\colhead{SFR}   & \colhead{Mean Age}}
\startdata
\sidehead{U-dropouts}
$SED_{z-fix}$\tablenotemark{b} & \nodata & -0.039$\pm$0.232 & -0.585$\pm$0.185 & 1.075$\pm$0.510 \\
$SED_{z-free}$\tablenotemark{c} & -0.021$\pm$0.015 & -0.108$\pm$0.230 & -0.530$\pm$0.258 & 0.955$\pm$0.502 \\
\sidehead{B-dropouts}
$SED_{z-fix}$\tablenotemark{b} & \nodata & -0.120$\pm$0.228 & -0.558$\pm$0.288 & 1.009$\pm$0.586 \\
$SED_{z-free}$\tablenotemark{c} & -0.030$\pm$0.031 & -0.371$\pm$0.355 & 0.212$\pm$1.181 & 0.273$\pm$1.012 \\
\sidehead{V-dropouts}
$SED_{z-fix}$\tablenotemark{b} & \nodata & -0.169$\pm$0.179 & -0.597$\pm$0.133 & 1.162$\pm$0.589 \\
$SED_{z-free}$\tablenotemark{c} & -0.014$\pm$0.012 & -0.339$\pm$0.274 & -0.094$\pm$0.964 & 0.487$\pm$0.875 \\
\enddata 

\tablenotetext{a}{Relative offset is defined as ($Value_{SED} - Value_{SAM}$)/($Value_{SAM}$) for stellar 
mass, SFR, and age. For redshift, relative offset is defined as ($z_{SED} - z_{SAM}$)/(1+$z_{SAM}$).}
\tablenotetext{b}{Relative offsets between values derived in SED-fitting with redshifts fixed and intrinsic values}
\tablenotetext{c}{Relative offsets between values derived in SED-fitting with redshifts allowed to vary and intrinsic values}

\end{deluxetable} 

\clearpage 

\begin{deluxetable}{crr}
\tablecolumns{3}
\tablewidth{0pc}
\tablecaption{Relative Changes\tablenotemark{a} of the Mean Stellar Masses and Ages with the Limited $\tau$ Range \label{tauofftab}}
\tablehead{
\colhead{$\tau$s Used for Fitting}   & \colhead{Stellar Mass ($\%$)}   & \colhead{Mean Age ($\%$)}}
\startdata
\sidehead{U-dropouts}
$\tau \leq 1.0$ Gyr & -4.9 & -15 \\
$\tau = 15.0$ Gyr & -8.0 & -12 \\
\sidehead{B-dropouts}
$\tau \leq 1.0$ Gyr & -12 & -26 \\
$\tau = 15.0$ Gyr & -3.4 & -7.4 \\
\sidehead{V-dropouts}
$\tau \leq 1.0$ Gyr & -5.4 & -16 \\
$\tau = 15.0$ Gyr & -3.2 & -6.6 \\
\enddata

\tablenotetext{a}{Relative change is defined as 
($\langle value_{\tau} \rangle -  \langle value_{all} \rangle$)/$\langle value_{all} \rangle$. 
$\langle value_{all} \rangle$ is the mean value of the stellar mass or age derived with the full range of $\tau$.
$\langle value_{\tau} \rangle$ is the mean value of the stellar mass or age derived with the limited $\tau$ range.}

\end{deluxetable}

\clearpage

\begin{deluxetable}{crrc}
\tablecolumns{4}
\tablewidth{0pc}
\tablecaption{Toy Models \label{toymodel}}
\tablehead{
\colhead{}   & \colhead{Model 1}   & \colhead{Model 2}   & \colhead{Model 3}}
\startdata
\sidehead{input}
$\tau$ (Gyr) & 15.0 & 0.2 & 0.2 (young), 0.2 (old) \\
$t$ (Gyr) & 0.01, 0.05, 0.1, 0.5, 1.0 & 0.1, 0.2, 0.5, 1.0, 1.3 & 0.1 (young), 1.0 (old) \\
$E(B-V)$ & 0, 0.05, 0.1, 0.2 & 0, 0.05, 0.1, 0.2 & 0.0, 0.2 \\
Mass Fraction\tablenotemark{a} & \nodata & \nodata & 0.1, 0.26, 0.52, 1.0, 2.6 \\
Number of SEDs & 120 & 120 & 60 \\
\sidehead{used for fitting}
$\tau$ (Gyr) & $\leq 1.0$ & $\geq 8.0$ & $0.2 \leq \tau \leq 15.0$ \\
$t$ (Gyr) & 0.01 - 2.3\tablenotemark{b} & 0.01 - 2.3\tablenotemark{b} & 0.01 - 2.3\tablenotemark{b} \\
\enddata

\tablenotetext{a}{Stellar mass fraction between $young$ component and $old$ component 
defined as $M_{young}$/$M_{old}$} 
\tablenotetext{b}{$t$ is limited to be smaller than the age of the universe at each galaxy's redshift.}

\end{deluxetable}

\end{document}